\documentclass[fleqn,usenatbib]{mnras}

\usepackage{newtxtext,newtxmath}
\usepackage[T1]{fontenc}
\usepackage{ae,aecompl}

\usepackage{bm}

\usepackage{mleftright}
\let\left\mleft
\let\right\mright

\usepackage{amsmath}

\usepackage{graphicx}	% Including figure files
\usepackage{subcaption}
\usepackage{siunitx}
\usepackage{mwe}
\newcommand{\bs}[1]{\boldsymbol{#1}}

\newcommand{\Div}{\nabla \cdot }
\newcommand{\Lap}{\nabla^2 }
\newcommand{\bigO}{\mathcal{O}}
\graphicspath{{./Figures/}}
\title[Magneto-Thermal Instability In Galaxy Clusters]{Magneto-Thermal Instability In Galaxy Clusters I: Theory and Two-Dimensional Simulations}

\author[L. M. Perrone et al.]{
Lorenzo M. Perrone,$^{1}$\thanks{E-mail: lmp61@cam.ac.uk}
Henrik Latter$^{1}$
\\
$^{1}$Department of Applied Maths and Theoretical Physics, University of Cambridge, UK\\
}

\date{Accepted XXX. Received YYY; in original form ZZZ}

\pubyear{2020}

\begin{document}
\label{firstpage}
\pagerange{\pageref{firstpage}--\pageref{lastpage}}
\maketitle

\begin{abstract}
	Determining the origin of turbulence in galaxy clusters, and quantifying its transport of heat, is an outstanding problem, with implications for our understanding of their thermodynamic history and structure. As the dilute plasma of the intracluster medium (ICM) is magnetized, heat and momentum travel preferentially along magnetic ﬁeld lines. This anisotropy triggers a class of buoyancy instabilities that destabilize the ICM, and whose turbulent motions can augment or impede heat transport. We focus on the magneto-thermal instability (MTI), which may be active in the periphery of galaxy clusters. We aim to take a fresh look at the problem and construct a general theory that explains the MTI saturation mechanism and provides scalings and estimates for the turbulent kinetic energy, magnetic energy, and heat ﬂux. We simulate MTI turbulence with a Boussinesq code, SNOOPY, which, in contrast to previous work, allows us to perform an extensive sampling of the parameter space. In two dimensions the saturation mechanism involves an inverse cascade carrying kinetic energy from the short MTI injection scales to larger scales, where it is arrested by the stable entropy stratiﬁcation; at a characteristic ‘buoyancy scale’, the energy is dumped into large-scale g-modes, which subsequently dissipate. Consequently, the entropy stratiﬁcation sets an upper limit on the size and strength of turbulent eddies. Meanwhile, the MTI conveys a substantial fraction of heat, despite the tangled geometry of the magnetic ﬁeld. In a companion paper, these results are extended to three-dimensional ﬂows, and compared to real cluster observations.
\end{abstract}

%This is a simple template for authors to write new MNRAS papers.
%The abstract should briefly describe the aims, methods, and main results of the paper.
%It should be a single paragraph not more than 250 words (200 words for Letters).
%No references should appear in the abstract.

\begin{keywords}
galaxies: clusters: intracluster medium -- plasmas -- instabilities -- turbulence
\end{keywords}

%%%%%%%%%%%%%%%%%%%%%%%%%%%%%%%%%%%%%%%%%%%%%%%%%%

%%%%%%%%%%%%%%%%% BODY OF PAPER %%%%%%%%%%%%%%%%%%

\section{Introduction}
Galaxy clusters are both fascinating and astrophysically important. First, they are the largest bound structures in the Universe, and are thus ideally suited to probe the validity of cosmological models \citep{Voit2005,Allen2011}. For instance, they can help constrain the baryonic abundance \citep{Allen2008}, and the properties of dark matter \citep{Bradac2008}. Second, they are filled with a hot and rarefied plasma, the intra-cluster medium (ICM), which, despite its extremely low density, constitue a substantial fraction of the total mass of the cluster \citep{Mohr1999}. The ICM is observable in the X-ray frequency band through bremsstrahlung and line emission \citep[see][for recent reviews]{Rosati2002,Boehringer2010}, and is characterized by significant levels of subsonic turbulence \citep{Schuecker2004,Churazov2012,Zhuravleva2014b}. The turbulent nature of the ICM has implications for the transport of heat and momentum in galaxy clusters \citep{Narayan2001,Zakamska2003,Voigt2004,Zhuravleva2019},
%whose specifics are necessary to explain
%which in turn constitutes a fundamental ingredient 
which in turn is fundamental for several processes ranging from heating by AGN feedback \citep{McNamara2000,Birzan2004} -- and its significance in addressing the long-standing cooling-flow problem \citep{Fabian1994} -- to the propagation of shocks \citep{Fabian2003,Markevitch2002,Markevitch2007}. Therefore, understanding the properties and sources of turbulence in the ICM is essential in order to build a reliable evolutionary model for galaxy clusters.

Turbulence in the ICM is closely tied to the unique properties of its plasma: being very hot and dilute, with temperatures up to $\sim 10 \si{keV}$ and typical densities of $\sim 10^{-3} \si{cm^{-3}}$ \citep{Sarazin1988}, the plasma is strongly magnetized even at relatively large plasma betas \citep[typically $\beta\sim 100$;][]{Vogt2005}, and thus charged particles gyrate around magnetic field lines on a timescale much shorter than the particle collision timescale \citep{Sarazin1988}. As a result, the dynamics of the ICM is highly anisotropic with respect to the direction of the magnetic field, with a greatly reduced efficiency of heat and momentum transfer perpendicular to the field lines \citep{Braginskii1965a}. 
Since the particle mean free path $\lambda_\text{mfp}$ is substantially shorter than the pressure scale-height of the cluster $H$ -- a measure quantified by the Knudsen number Kn $= \lambda_\text{mfp} / H \approx 10^{-2} - 10^{-3}$ \citep{Kunz2012}  -- the dynamics of the ICM can be described by the so-called \textit{Braginskii} or \textit{extended} MHD equations, which differ from the standard MHD equations in that the heat flux -- primarily due to the electrons -- is directed along the local magnetic field, as is the viscous damping of plasma fluctuations -- mainly due to ion-ion collisions \citep[see, e.g.,][for an astrophysical discussion]{Kulsrud2005}.

By altering the microphysics of transport processes, the intrinsic plasma anisotropy also impacts on the large-scale dynamics of galaxy clusters, especially their susceptibility to turbulence. It enables a new class of buoyancy instability, and the usual criteria for plasma stability is transformed. In particular, stability is no longer controlled by the sign of the entropy gradient \citep[Schwarzschild's criterion;][]{Schwarzschild1906}, but rather by the temperature gradient. Depending on its sign, the plasma is either unstable to the \textit{heat-flux buoyancy} instability \citep[HBI;][]{Quataert2008}, or the \textit{magneto-thermal} instability \citep[MTI;][]{Balbus2000,Balbus2001}. A compelling reason to study these instabilities comes from X-ray observations by \textit{XMM-Newton} and \textit{Chandra}, which showed that clusters exhibit an approximately flat or radially increasing temperature in their cores, and then a decreasing temperature towards large radii \citep{Vikhlinin2006,Pratt2007,Leccardi2008}, thus suggesting the ICM to be unstable to the HBI at small radii (in cool-core clusters with positive temperature gradients), and to the MTI in the periphery. These instabilities are considered to be one source of turbulence on intermediate scales in galaxy clusters \citep{McCourt2011} and can act on timescales much shorter than the Hubble time \citep{Parrish2008}.

This paper is the first of two re-examining the dynamics of the MTI, and the properties of the turbulence it drives, via two and three-dimensional numerical simulations in the Boussinesq approximation. Before we outline the philosophy and structure of the paper, we review the theoretical background to the MTI, which will help contextualise and motivate our work. Some readers may choose to skip ahead to Section 1.2.

\subsection{The Magneto-Thermal Instability}

The basic instability mechanism of the MTI is relatively intuitive. Consider the idealized but informative scenario of a plane parallel atmosphere pierced by a horizontal magnetic field and with both its gravitational acceleration and temperature gradient pointing downward (i.e.\ cold plasma sits upon hot plasma). A fluid blob displaced upwards will drag the magnetic field along with it and, because thermal conduction occurs solely along that field, it will remain thermally connected to the warm fluid at lower altitudes. As a consequence, the perturbation will be isothermal rather than adiabatic. But to come into pressure equilibrium with its new surroundings, the blob's density must decrease, because the neighbouring fluid is cooler. Now, being lighter, the blob will rise, initiating instability \citep{Balbus2000,Balbus2001}.

On long wavelengths, the conduction time may approach the buoyancy timescale (if the plasma is stably stratified) and instability will be inhibited; on shorter wavelengths the thermal conduction time is short and isothermality more strictly enforced, thus instability is favoured. On very short wavelength, however, magnetic tension or diffusion, viscous or resistive, will stabilise the MTI. When the long `buoyancy scales' and the very short stable wavelengths are sufficiently well separated, the maximum growth rate (and instability criterion) is independent of thermal conduction. In the periphery of galaxy clusters, the characteristic MTI timescale is $\sim 600 \si{Myr}$ (see Sec.~\ref{sec:galaxy_clusters}), which is much less than the Hubble time. 

The MTI's relationship with standard thermal convection is much the same as the MRI's relationship with centrifugal instability \citep{Balbus2000}. But it also shares some features of double-diffusive thermo-compositional convection \citep[e.g.][]{Garaud2018}, because it must contend with competing stabilising and destabilising gradients (those of entropy and temperature respectively), while making use of fast (thermal) diffusion to negate the stabilising gradient. Either magnetic or viscous diffusion may be considered the slow diffusion, as if too large it obstructs extraction of energy from the temperature gradient.

\subsubsection{Previous Numerical Studies}

The linear evolution of the MTI was verified in the early numerical studies of \citet{Parrish2005,Parrish2007}, who performed local 2D and 3D simulations that captured important aspects of its nonlinear behaviour, such as its convective-like nature at saturation and a significant vertical flux of heat by both conduction and advection. If the boundary conditions allow for a rearrangement of the temperature gradient, the MTI saturates by isothermalising the domain \citep{Parrish2005}. If, on the other hand, a steady heat flux is enforced at the boundaries, the resulting convective motions never settle, leading to a state of sustained turbulence that can drive a magnetic dynamo in 3D \citep{Parrish2007}. Simulations with extended vertical domains (allowing for significant density structure) confirmed some of these results and highlighted that the MTI can drive near-sonic motions \citep{McCourt2011}. Furthermore, it was found that linearly stable equilibria with vertical background fields are in fact \textit{nonlinearly} unstable \citep{McCourt2011}. This precludes the MTI from saturating by simply rearranging the magnetic field geometry into a purely vertical configuration, and also significantly increases the generality of the instability.

To assess whether the MTI can appreciably influence the large-scale structure of galaxy clusters, a number of global simulations have been performed, focusing on the potential impact of anisotropic heat conduction on both the cool core dynamics and in the outer part of the clusters. For instance,  \citet{Parrish2008,Parrish2012a} have shown that, in the outskirts of galaxy clusters, the instability can considerably modify the temperature gradient, and produces a substantial radial heat flux due to magnetic field lines that align preferentially in the radial direction. Moreover, the MTI-induced turbulence supplies a significant contribution to the thermal pressure support at radii larger than $\sim 100 \si{kpc}$ \citep{Parrish2012}.  

Before moving on, we briefly mention another astrophysical application of the MTI: hot accretion flows around black holes \citep[see e.g.,][for a review]{Yuan2014}. Numerical simulations of axisymmetric spherical accretion in such flows find that the MTI can produce a large conductive heat flux and, furthermore, enhances the transfer of angular momentum in the radial direction \citep{Sharma2008,Bu2011}. Contrary to galaxy clusters, however, hot accretion flows are unstable to the magneto-rotational instability  \citep[MRI;][]{Balbus1991}. The extent to which the MTI and the MRI positively or negatively interact has not yet been studied in detail, except for some exploratory investigations by \citet{Foucart2016,Foucart2017}, where a general relativistic framework was used to evolve the Braginskii MHD equations. 

\subsubsection{Outstanding Issues}

It must be acknowledged that the MTI is only one potential source of turbulence in the ICM; merging events, galactic motions, substructure accretion, and AGN feedback supply large-scale forcing that may also initiate a turbulent cascade \citep[e.g.,][]{Markevitch2001, Churazov2003, Fabian2012}.
Moreover, these competing sources might be sufficiently vigorous to overwhelm the MTI; indeed, `cosmological style' simulations of disturbed clusters have failed to find compelling evidence of the MTI \citep{Ruszkowski2010, Ruszkowski2011, ZuHone2013a, Yang2016, Kannan2017}. However, quite apart from the difficulties in interpreting the numerical data, questions may be asked of the simulations' effective numerical resolution (and whether it can capture the characteristic MTI lengthscales) and of the (presumably isotropic)  numerical thermal diffusivity in such codes.

On the very short scales the MTI encounters a separate problem. High-beta, weakly-collisional plasmas exhibit pressure anisotropies and heat fluxes that trigger a range of microscale instabilities that are not appropriately modelled by the Braginskii MHD equations \citep{Schekochihin2005,Schekochihin2006}. 
By introducing an `effective collisionality', one might expect these small-scale processes to diminish the intrinsic anisotropies that the MTI and HBI rely on, thus calling into question their viability. Solar wind measurements \citep{Bale2013} and kinetic numerical simulations \citep{Roberg2016,Riquelme2016,Roberg2018,Komarov2018} suggest that microinstabilities suppress thermal conduction by an order one multiplicative factor. These results could cross over to the plasma in galaxy clusters, but conclusions about fluxes on the scale of $\sim 100 \si{kpc}$ are still not definitive \citep{Komarov2018}. Moreover, some studies have shown that the MTI's saturated state appears to be mostly unaffected, though this forms the basis of ongoing research \citep{Berlok2020,Drake2020}.

\subsection{Motivation for and Plan of the Paper}

Despite the continuing interest in the MTI since its discovery by \citet{Balbus2000}, our theoretical understanding of the physics of the instability is still incomplete. For instance, previous local, compressible simulations of the MTI were limited by the somewhat arbitrary choice of vertical structure pioneered by \citet{Parrish2008}, a `sandwiched' configuration whereby only the middle layer is MTI-unstable and the background thermodynamic variables follow fixed power laws. While this setup allowed the first study of the saturated phase of the instability with local compressible codes, it precluded any exploration of the defining physical parameters in the system. Moreover, it remains unclear to what extent the numerical results so produced are captive to this specific set-up. Ultimately, we have no comprehensive theory for how the saturation amplitude of the MTI, and its associated thermal flux, depend on the key physical parameters (thermal conductivity, buoyancy frequency, temperature gradient, etc.). And this absence makes it difficult to directly apply the theory to observations, and thus to decide on the MTI's prevalence and importance. This paper and its companion aim to partially rectify some of these issues.

The paper is structured as follows. First we revisit the theory of the MTI: in Section~\ref{sec:model} we present the MHD equations with anisotropic heat conduction in the Boussinesq approximation, and we examine the energetics of the MTI, deriving a generic expression for the energy injection rate of the instability and an estimate for the saturated turbulent energies; in Section~\ref{sec:linear_theory} we redo the linear theory of the instability and obtain expressions for the maximum growth rate and for the most unstable wavelength in the asymptotic regime of weak magnetic fields, low viscosity and magnetic resistivity. Finally, in Section~\ref{sec:nonlin_MTI} we look at the nonlinear saturation of single MTI modes and uncover the role of nonlinear g-modes and parasitic instabilities. The second part of the paper is concerned with numerical simulations of the MTI: in Section~\ref{sec:numerical_methods} we describe the numerical methods and the diagnostics that we adopt, while in Section~\ref{sec:numerical_results} we present the results of the first 2D simulations of Boussinesq MTI with the SNOOPY code, focusing on the behaviour of the instability at saturation, and exploring a large region of the parameter space. This allows us to derive new relationships between the turbulent energy levels and the physical parameters, and to confute some previous results. In Section~\ref{sec:conclusions} we summarize and discuss our results. 

\section{Theoretical Preliminaries}

\subsection{Model and Governing Equations}\label{sec:model}

We aim to explore the subsonic and small-scale dynamics of fluctuations within a quasi-spherical cluster characterised by a background density $\rho_0$, pressure $p_0$, and temperature $T_0$. We thus adopt the Boussinesq approximation, and so our model describes a small Cartesian block of cluster gas located at a given spherical radius $R_0$. The appropriate equations are derived in Appendix~\ref{app:boussinesq_derivation}, and include anisotropic thermal conduction and a background radial gradient in both entropy and temperature. They read:
\begin{gather}
\Div \bs u = \Div \bs B = 0 ,\label{eq:div_eq} \\
\left( \partial_t + \bs u \cdot \nabla \right) \bs u = - \frac{\nabla p_{tot}}{\rho_0} - \theta \bs e_z + \left( \bs B \cdot \nabla \right) \bs B + \nu \Lap \bs u , \label{eq:mom_eq} \\
\left( \partial_t + \bs u \cdot \nabla \right) \bs B = \left( \bs B \cdot \nabla \right) \bs u + \eta \Lap \bs B, \label{eq:bfield_eq} \\
\left( \partial_t + \bs u \cdot \nabla \right) \theta = N^2 u_z + \chi \Div \left[ \bs b \left( \bs b \cdot \nabla \right)\theta \right]	+ \chi \omega_T^2 \Div \left( \bs b b_z\right), \label{eq:buoyancy_eq}
\end{gather}
where $p_{tot}$ is the sum of thermal and magnetic pressure; the buoyancy variable is defined as $\theta \equiv g_0  \rho' / \rho_0$, with $g_0$ being the gravitational acceleration and $\rho' / \rho_0$ the fractional density perturbation; $\nu$ and $\eta$ are the fluid viscosity and magnetic resistivity; and $N^2$ and $\omega_T^2$ are the squared Brunt-V\"ais\"al\"a and MTI frequencies, respectively, defined via:
\begin{align}\label{freq}
	N^2 = \frac{g_0}{\gamma} \left. \frac{\partial \ln p \rho^{-\gamma}  }{\partial R} \right|_0   , \qquad
	\omega_T^2 = - g_0  \left. \frac{\partial \ln T}{\partial R} \right|_0 ,
\end{align}
where the subscript $0$ denotes that the quantities are computed at the reference radius $R=R_0$ in the cluster. In Eqs~\eqref{eq:div_eq}--\eqref{eq:buoyancy_eq}, we have rescaled the magnetic field so that it has the dimensions of a velocity, and the thermal diffusivity $\chi$ (in cgs units) absorbs a factor of $(\gamma -1)/\gamma$. Throughout, we focus on the case where both $\omega_T^2$ and $N^2 $ are positive.  
Finally, notice that the magnetic field enters in the anisotropic conduction term in Eq.~\eqref{eq:buoyancy_eq} only through the unit vector $\bs b = \bs B / B$, and thus the buoyancy equation is independent of the magnitude of $\bs B$.

We acknowledge that an accurate description of ICM requires the inclusion of both anisotropic heat conduction and anisotropic viscosity, but in this work have chosen to neglect the latter and instead use a regular isotropic viscosity term. Our decision is driven by both practical considerations, in that the inclusion of Braginskii viscosity would lead to unnecessary complications that would obscure the essence of the MTI, and by results in the existing literature \citep[e.g.][]{Kunz2011,Parrish2012a,Kunz2012}, which suggest that Braginskii viscosity sets a lower cutoff in the physical scale of magnetic fluctuations but otherwise only weakly affects the properties of MTI-driven turbulence.

\subsection{Energetics of the MTI}

We next present the time-evolution equations for the total kinetic ($K$), magnetic ($M$) and potential ($U$) energies of the system, defined by
\begin{align}
    K = \langle u^2 \rangle /2, \quad M = \langle B^2 \rangle /2, \quad U = \langle \theta^2  \rangle /2 N^2,
\end{align}
where the angle brackets denote a box average. The equations are:
\begin{align}
\frac{d}{dt} K &= - \langle \theta u_z \rangle - \nu \langle \lvert \nabla \bs{u} \rvert^2 \rangle - \langle \bs{B} \bs{B} \colon \nabla \bs{u} \rangle,  \label{eq:kin_en_cons} \\
\frac{d}{dt} M &= - \eta \langle \lvert \nabla \bs{B} \rvert^2 \rangle + \langle \bs{B} \bs{B} \colon \nabla \bs{u} \rangle, \label{eq:mag_en_cons} \\ 
\frac{d}{dt} U  &= \langle \theta u_z \rangle  - \frac{\chi}{N^2} \langle \left| \bs b \cdot \nabla \theta \right|^2 \rangle - \frac{\chi}{N^2} \omega_T^2  \langle   b_z \bs b   \cdot \nabla \theta \rangle. \label{eq:th_en_cons}
\end{align}
Combining Eqs.~\eqref{eq:kin_en_cons}--\eqref{eq:th_en_cons} gives an equation for the conservation of the total energy $E = K + M + U$:
\begin{align}
\frac{d E }{dt}&= - \nu \langle \lvert \nabla \bs{u} \rvert^2 \rangle - \eta \langle \lvert \nabla \bs{B} \rvert^2 \rangle - \frac{\chi}{N^2} \langle \left|  \bs b \cdot \nabla \theta \right|^2 \rangle -  \frac{\chi \omega_T^2}{N^2}  \langle  b_z \bs b   \cdot \nabla \theta \rangle  \nonumber \\ 
&= -\epsilon_{\nu} - \epsilon_{\eta} - \epsilon_{\chi}  + \epsilon_I , \label{eq:energy_balance}
\end{align}
where the first three terms on the right-hand side are the viscous,  Ohmic  and thermal dissipation, respectively, while the last term accounts for the energy input, i.e., the forcing. We stress that, despite the negative sign in front of forcing term, the quantity is not negative-definite overall (unlike the dissipative terms) 

\subsubsection{The Energy Injection Rate in Braginskii MHD}

At saturation, the terms on the right-hand side of Eq.~\eqref{eq:energy_balance} balance each other in a statistical sense, i.e. if we look at averages over timescales longer than the fluctuation time $\sim \omega_T$. In addition, by taking the weak-field limit, we expect Ohmic dissipation to be negligible. Similarly, we expect that in the $\mathrm{Pr} \ll 1$ regime 
\citep[appropriate for the hot plasma in the ICM, where $\mathrm{Pr} \simeq 0.02$, assuming Spitzer values;][]{Kunz2012} 
the energy dissipates primarily through thermal diffusion rather than viscous dissipation. This suggests that, at saturation, the last two terms of Eq.~\eqref{eq:energy_balance} roughly balance each other, i.e. $\epsilon_I \approx \epsilon_{\chi}$, which yields
\begin{align}\label{eq:approx_gradT}
\frac{\chi}{N^2} \langle \left| \bs b \cdot \nabla \theta \right|^2 \rangle &\approx \frac{ \chi \omega_T^2}{N^2} \langle \left| b_z \bs b   \cdot \nabla \theta \right| \rangle \nonumber  \\ 
\Rightarrow \langle \nabla \theta \rangle &\sim \omega_T^2.
\end{align}
This allows us to estimate the total energy injection rate:
\begin{equation}\label{eq:energy_input_mti}
\epsilon_I \sim  \frac{\chi \omega_T^4}{N^2},
\end{equation}
where we have used the definition of $\epsilon_I$ in Eq.~\eqref{eq:energy_balance} together with Eq.~\eqref{eq:approx_gradT}. 
Interestingly, these estimates apply to both the saturated state of the MTI and any other turbulent process whose source of energy is the temperature gradient, such as the HBI. 

\subsubsection{Saturation and isothermality}
\label{sec:isothermality}

One of the key results in the preceding subsection is Equation \eqref{eq:approx_gradT}, which is more illuminating when re-expressed as
\begin{equation}
\langle \nabla T' \rangle \sim \left(\frac{dT}{dR} \right)_0,
\end{equation}
using Eq.~\eqref{freq}, the definition of $\theta$, and the linearised equation of state (see Appendix A). This equivalence cannot tell us about the relative signs of the two gradients, but it strongly suggests that, on average, the turbulence comes to steady state by redistributing heat so as to counteract the background temperature gradient or, in other words, to bring the atmosphere to \emph{isothermality}. In the context of the MTI this state corresponds to marginal stability, and is thus a natural endpoint of its evolution. In Paper II, we explore this balance in more detail, and show how it holds scale-by-scale and not only on average. We finish by pointing out that saturation strongly constrains the gradient of $\theta$, but not the amplitude of the thermal energy, nor its characteristic lengthscale.

\subsubsection{Saturated energies and the buoyancy scale}\label{sec:theoretical-scalings}

Imposing further assumptions, we can estimate the total kinetic energy in the box at saturation. As the linear MTI involves both density and velocity perturbations, we assume that a similar fixed fraction of the energy injection goes into kinetic fluctuations (with the remaining going into density fluctuations). Suppose next that kinetic energy is removed at a rate $1/\tau$ by some energy sink, where $\tau$ is its characteristic timescale. As we demonstrate later, this `Epstein drag' process is a reasonable approximation to the complicated spectral dynamics in our simulations, through which the extracted kinetic energy is sent to larger scales by an inverse cascade where an energy sink awaits in the form of large-scale g-modes. If we finally assume that the timescale of this process is $\tau \sim \omega_T^{-1}$, we obtain 
\begin{equation}
K \sim \chi \frac{\omega_T^3}{N^2},
\end{equation}
by balancing kinetic energy injection and removal.  

Furthermore, following \citet{Boffetta2012}, we can derive the characteristic friction scale $l_{B}$ at which the energy removal process dominates and the inverse cascade stalls:
\begin{equation} \label{bscale}
l_{B} \sim  \epsilon_I^{1/2} \tau^{3/2} \sim \chi^{1/2} \omega_T^{1/2} N^{-1}.
\end{equation}
This important length demarcates when motions driven by MTI turbulence (on shorter scales) are supplanted by motions dominated by buoyancy (on longer scales). It thus resembles several key scales in the theory of atmospheric turbulence, in particular the Ozmidov and Obukhov scales, though it is distinct from both. We henceforth refer to is simply as the `buoyancy scale'.

Finally, we expect that an initially uniform mean field $B_0$ will be amplified by the random stretching and folding of turbulent flow. This process, called turbulent induction \citep{Schekochihin2007}, is separate from a magnetic dynamo, and can take place in 2D. By equating the (conserved) magnetic flux at the buoyancy scale ($\sim l_B B_0 $), where the mean field is stretched, with the flux at the resistive scale ($\sim l_{\eta} B_{rms}$, with $l_{\eta} \sim \sqrt{\eta / \omega_T}$ and $B_{rms} \sim M^{1/2}$), where fluctuations are damped, we can estimate the saturated magnetic energy:
\begin{equation}\label{eq:mag_energy_saturation}
    M \sim  \frac{\chi}{\eta} \left( \frac{\omega_T}{N} \right)^2 B_0^2.
\end{equation}

\subsection{MTI Parameters in Galaxy Clusters}\label{sec:galaxy_clusters}

\begin{figure}
	\centering
	\includegraphics[width=0.9\columnwidth]{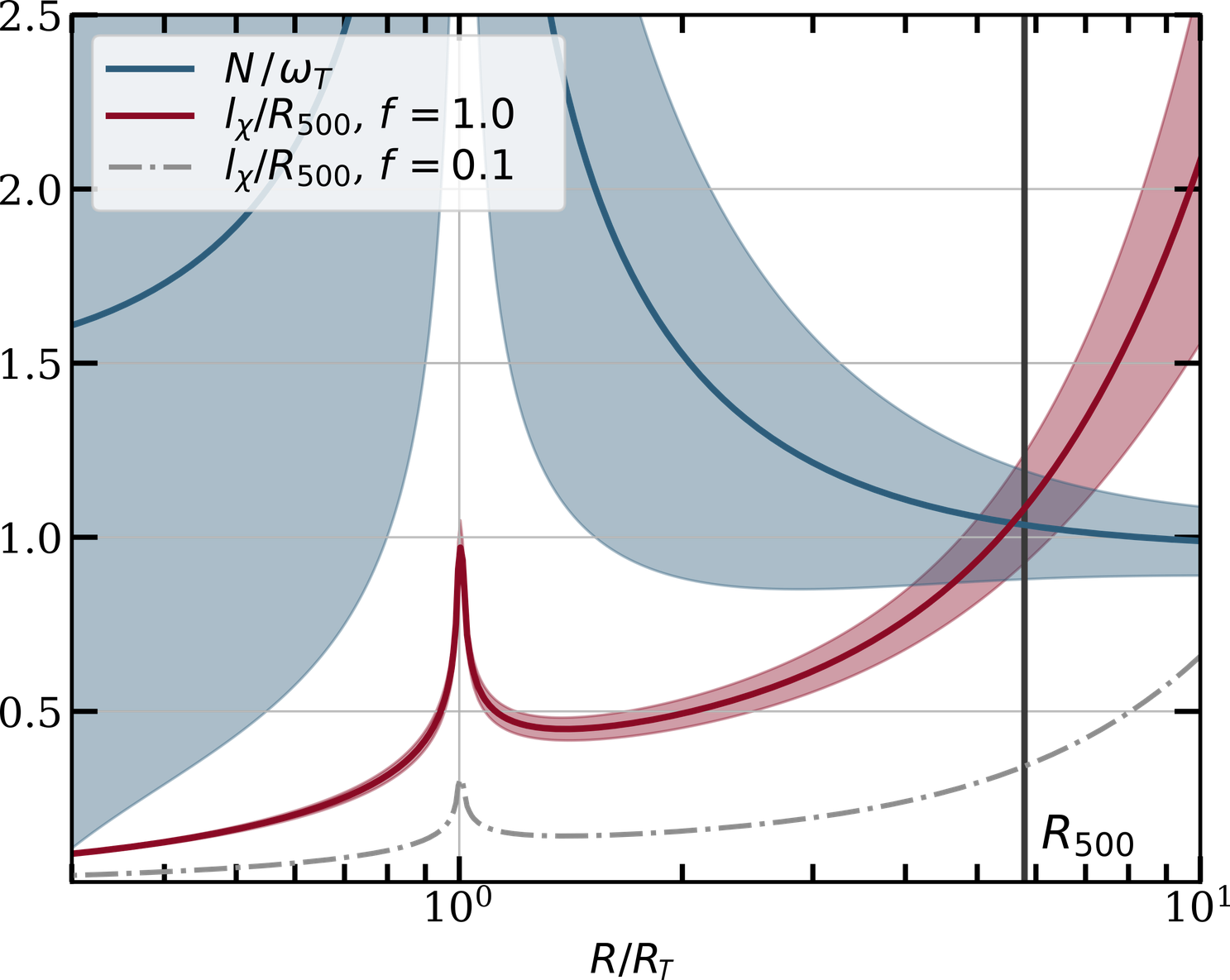}
	\caption{Best-fit radial profile of the ratios $l_{\chi}/R_{500}$ and $N/\omega_T$ as a function of the normalized radius $R/R_T$, where $R_T$ is the radius at which the temperature begins to decrease \citep[data from][]{Ghirardini2019}. The shaded areas represent the standard deviation from the best-fit parameters. We also show the conduction length for a conductivity suppression factor of $f = 0.1$ (grey dash-dotted line). } 
	\label{fig:GC_radial_profiles}
\end{figure}

In this section we estimate typical values of our main parameters in the outskirts of galaxy clusters using available ICM data. We focus on the ratio between the Brunt-V\"ais\"al\"a and MTI frequencies, and on the conduction length defined as $l_{\chi} = \sqrt{\chi /\omega_T}$.

We employ the data analyzed by \citet{Ghirardini2019}, who combine \textit{XMM-Newton} observations and measurements of the Sunyaev-Zel'dovich (SZ) effect in order to characterise the radial profiles of the ICM's density, entropy and temperature. The selected sample of $12$ galaxy clusters was compiled from the first Planck SZ catalogue and comprises low-redshift sources with high signal-to-noise ratio in the SZ signal; of the $12$ systems, $8$ are classified as non-cool-core galaxy clusters. The main reason for our choice is that \citet{Ghirardini2019} supply best-fits of the parametric functions, and thus we can approximate the behaviour of all thermodynamic variables throughout the entire radial range. This is in contrast to previous studies of the MTI \citep[namely][]{Parrish2012a}, which only had access to the best-fits of some of the thermodynamic quantities.  

As we are mostly interested in the periphery of galaxy clusters, we use the best-fit parameters in Table~3 of \citet{Ghirardini2019} for the entire population of clusters, including both cool- and non-cool-core sources. To simplify the computation, we adopted a number of approximations, such as setting the scale factor $h(z) = 1$ and the virialized radius $R_{500} = 1.2\si{Mpc}$, which are reasonable values for the selected sample. 

Fig.~\ref{fig:GC_radial_profiles} shows the ratio of the Brunt-V\"ais\"al\"a and MTI frequencies. The $x$-axis has been rescaled to the radius at which the temperature begins to decrease $R_T$ ($\approx 200\si{kpc}$ for the best-fit parameters of the sample), which we use as a proxy to distinguish between core and periphery. The divergence at $R=R_T$ results from the radial derivative of the temperature going to zero. As we can see, for $R \gtrsim R_T$, the MTI and buoyancy timescales are comparable. In terms of actual values, we find that at $R \lesssim 1\si{Mpc}$, the Brunt-V{\"a}is{\"a}l{\"a} and MTI timescales are both $\sim 600\si{Myr}$, which is much less than the Hubble time and would give the MTI ample time to develop.

To compute the conduction length, we assume the value of thermal diffusivity to be equal to the Spitzer value $\chi_s$ \citep{Spitzer1962} multiplied by a suppression parameter $f \leq 1$ that combines all possible  mechanisms that can reduce the conductivity \citep[tangled magnetic fields, scattering of electrons by micro-scale instabilities, etc.;][]{Roberg2016,Roberg2018,Komarov2018,Drake2020}. Setting $\chi = f \chi_{s}$, with
\begin{align}
	\chi_{s} = 4.98 \times 10^{31} \left( \frac{k_B T_e}{5 \si{keV}}\right)^{5/2} \left( \frac{n_e}{10^{-3}\si{cm^{-3}}}\right)^{-1} \si{cm^2.s^{-1}},
\end{align}
\citep{Kunz2012}, we compute the conduction length $l_{\chi}$ normalized by the outer virialized radius $R_{500}$; it is plotted in Fig.~\ref{fig:GC_radial_profiles} for the maximal value $f = 1$, and for a realistic suppression factor of $f = 0.1$. As is clear, for $R \sim 100\mathrm{s} \, \si{kpc}$ the conduction length is typically smaller than the size of the cluster even in the absence of any suppression of the thermal conductivity.

The radial profiles shown in Fig.~\ref{fig:GC_radial_profiles} have been obtained using the best-fit values for the functional expressions of the thermodynamic variables, and do not provide us with any information about the possible dispersion around the mean profiles. To quantify this dispersion, we vary the best-fit parameters within their margin of error and compute the standard deviation between the resulting modified profiles. In the absence of larger datasets with comparable resolution at large radii, this method quantifies the intrinsic scatter of the sample used in \citet{Ghirardini2019}. The procedure that we adopt is as follows: for each parameter, we sample $N=1000$ values from a gaussian distribution centered around the mean and with standard deviation as in Table 3 of  \citet{Ghirardini2019}, but truncated outside of the range defined by the prior used in the best-fit; we then compute the radial profiles for $N/\omega_T$ and $l_{\chi}$ for each set of parameter so produced, and we rescale each profile by their temperature inversion radius $R_T$ (which depends on the various parameters); finally, we compute the standard deviation of the samples at each radii, which we plot in Fig.~\ref{fig:GC_radial_profiles} with shaded regions. We excluded by visual inspection a small number of samples ($35$ out of $1000$) which showed unphysical features, such as a double peak in the temperature.

\subsection{Linear Theory}\label{sec:linear_theory}

In this subsection, we review a number of standard results on the onset and early evolution of the MTI, in the weak field limit, as these provide important tests for our numerical code. We also account for the effects of fluid viscosity, magnetic resistivity, and magnetic tension, and derive an asymptotic growth rate formula.

\subsubsection{Linearized Boussinesq Equations}

We linearize the Boussinesq MTI equations ~\eqref{eq:div_eq}--\eqref{eq:buoyancy_eq} around an initial equilibrium defined by $\bs u_0 = 0$, $\bs B_0 = B_0 \bs e_x $, and $\theta_0 = 0$, introducing small perturbations denoted by $(\delta \bs u, \delta \bs B, \delta \theta)$. Their governing equations are:
\begin{gather}
\Div  \delta \bs u = \Div  \delta \bs B = 0 , \label{eq:lin_MTI_1}\\
\partial_t \delta \bs u = - \frac{\nabla \delta p_{tot}}{\rho_0} - \delta \theta \bs e_z +   B_0 \partial_{x}  \delta \bs B + \nu \Lap \delta \bs u ,  \label{eq:lin_MTI_2} \\
\partial_t \delta \bs B =  B_0 \partial_{x} \delta \bs u + \eta \Lap \delta \bs B,  \label{eq:lin_MTI_3} \\
\partial_t \delta \theta = N^2 \delta u_z + \chi \partial_x^2 \delta \theta 	+ \chi \omega_T^2 \frac{\partial_x  \delta B_z}{B_0}. \label{eq:lin_MTI_4}
\end{gather}
The perturbed fluid variables take the form $\delta v = \delta \hat{v} \exp \left(i \bs k \cdot \bs x + \sigma t\right)$, where $\sigma$ is a (possibly complex) growth rate, $\bs k$ is a real wavevector with only $x$ and $z$ components. Solvability of Eqs~\eqref{eq:lin_MTI_1}-\eqref{eq:lin_MTI_4} yield the dispersion relation:
\begin{align}\label{eq:mti_disp_rel_full}
&\sigma^3 + \sigma^2 \left( \chi k_x^2 + \eta k^2 + \nu k^2 \right) + \sigma \left[ \frac{N^2 k_x^2}{k^2}  + k_x^2 v_A^2  + \nu k^2 \chi k_x^2  \right. \nonumber \\
&\left.  + \eta k^2 \chi k_x^2 + \nu \eta k^4\right] +   \chi k_x^2 \left( k_x^2 v_A^2 - \frac{ \omega_T^2 k_x^2}{k^2} \right) + \nu \eta k^4 \chi k_x^2 \nonumber \\ 
&      + \eta k_x^2 N^2  = 0.
\end{align}
%
%where the wavevector $k^2=k_z^2 + k_x^2$ is projected along the direction of gravity ($k_z$) and perpendicular to it $(k_x)$. 
Equation \eqref{eq:mti_disp_rel_full} reduces to the dispersion relation obtained by \citet{Balbus2000} when $\eta,\nu \rightarrow 0$ and for $k^2=k_x^2$, and to that of \citet{Quataert2008} when $\eta,\nu \rightarrow 0$ and for a purely horizontal background magnetic field. The MTI eigenmodes are
\begin{equation}\label{eq:mti_eigenmode}
%\delta  \hat{\bs v} = 
\begin{bmatrix}
\delta \hat{u}_x \\\delta \hat{u}_z \\ \delta \hat{B}_x/  B_0 \\ \delta \hat{B}_z/  B_0 \\ \delta \hat{\theta}
\end{bmatrix} =
\begin{bmatrix}
-k_z/k_x \\ 1 \\ - i k_z/ \sigma_\eta \\ i k_x/ \sigma_\eta \\ \left(N^2 - \chi k_x^2 \omega_T^2/\sigma_\eta \right)/(\sigma+\chi k_x^2)
\end{bmatrix} \delta \hat{u}_z,
\end{equation}
where $\sigma_\eta=\sigma + \eta k^2 $.
The dispersion relation can be non-dimensionalized by taking $\omega_T^{-1}$ as the unit of time, and $l_{\chi}=\sqrt{\chi / \omega_T}$ as the unit of length, and then reduces to
\begin{align}\label{mti_disp_rel_full_nondim}
&\tilde{\sigma}^3 + \tilde{\sigma}^2 \xi \left[ k_x^2/k^2  +  q + \mathrm{Pr} \right] + \tilde{\sigma} \left[ \left( \tilde{N}^2  +  \xi \Lambda  +  \xi^2 \left( q + \mathrm{Pr} \right)  \right)k_x^2/k^2  \right. \nonumber \\
&\left.  +  \xi^2 q \mathrm{Pr} \right]  +  \xi^3 q \mathrm{Pr}  k_x^2/k^2  +  \tilde{N}^2 \xi q  k_x^2/k^2 +  \xi k_x^4/k^4  \left( \xi \Lambda - 1 \right) = 0,
\end{align}
where we have defined $\xi = ( k l_{\chi})^2$, $\tilde{\sigma}=\sigma/\omega_T$, $\tilde{N}^2=N^2/\omega_T^2$, $\mathrm{Pr} = \nu/\chi$ is the Prandtl number, $q=\eta/\chi$ is the Robertson number, and $\Lambda = v_A^2/(\chi \omega_T)$ is the Elsasser number. 
By applying the Routh-Hurwitz criterion we obtain the MTI's instability criterion:
\begin{align}\label{eq:instab_criterion}
q  \tilde{N}^2 k^2/k_x^2 + \Lambda \xi + \xi^2 q \mathrm{Pr} k^2/k_x^2 < 1,
\end{align}
which is equivalent to the instability criterion of \citet{Balbus2000} in the limit of vanishing viscosity and resistivity, for which Eq.~\eqref{eq:instab_criterion} reduces to
$k^2 v_A^2 < \omega_T^2.$
Interestingly, the entropy stratification enters in the instability criterion Eq.~\eqref{eq:instab_criterion} only when the resistivity is non-zero, and similarly the viscosity.
We remark that in Eq.~\eqref{eq:instab_criterion} all terms on the left-hand side are positive definite and can thus contribute to stabilizing a given wavelength. 
\begin{figure}
	\centering
	\includegraphics[width=1.0\columnwidth]{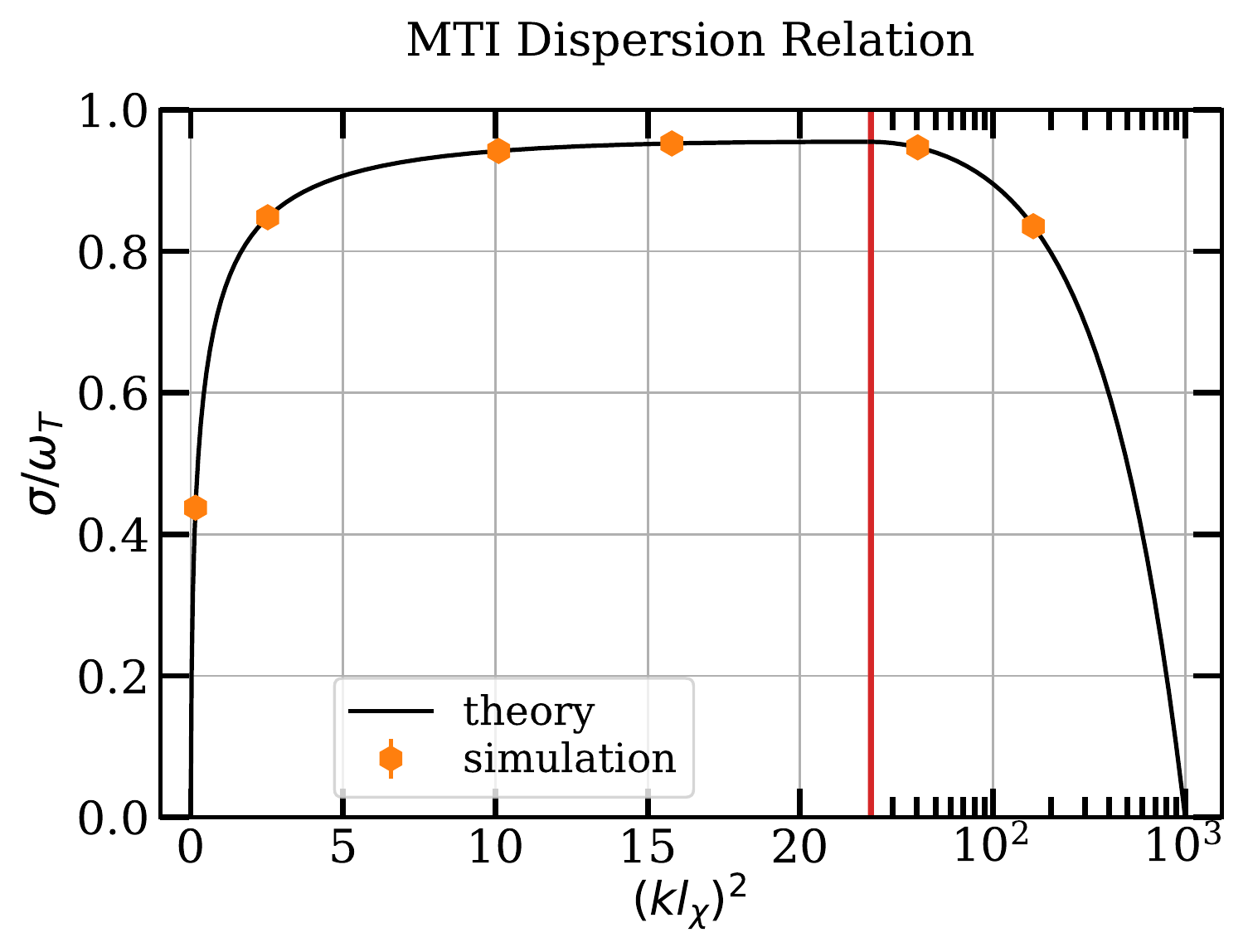}
	\caption{Plot of the unstable root of the MTI dispersion relation for $\mathrm{Pe}=10^3$, $\tilde{N}^2 = 0.1$, $q=\mathrm{Pr}=0.001$, $\Lambda = 10^{-7}$, and with a purely perpendicular perturbation ($k^2=k_x^2$). Note that the $x$-axis scale is linear left of the red vertical line -- which marks the normalized wavenumber at which maximum growth is achieved, $(k_{max}l_{\chi})^2 \approx 22.22$ -- and logarithmic on its right. The orange hexagons represent the numerical growth rates extracted from a series of 2D simulations with SNOOPY with explicit integration of the anisotropic thermal diffusion terms, see Sec.~\ref{sec:onset_2D_mti}. }
	\label{fig:mti_disp_rel}
\end{figure}

In Fig.~\ref{fig:mti_disp_rel} we plot the real (unstable) root of the MTI dispersion relation in Eq.~\eqref{mti_disp_rel_full_nondim}, for a given choice of parameters. As we observe, when dissipative processes like viscosity and resistivity are taken into account (or, alternatively, when magnetic tension is non-negligible), the growth rate increases monotonically from $0$ at $\xi=(kl_{\chi})^2=0$, reaches a maximum at a finite wavenumber $\xi_\text{max}$, and then decreases to zero. This is in contrast to the ideal case -- i.e. with zero viscosity and resistivity, as well as with vanishingly small magnetic fields -- where the unstable root of the dispersion relation asymptotically tends to $\tilde{\sigma} \rightarrow 1$ as $\xi \rightarrow \infty$. For the parameters shown in Fig.~\ref{fig:mti_disp_rel}, the behaviour at small scales is dominated by the viscosity/resistivity, rather than the magnetic tension. Finally, the growth rate goes to zero linearly as $\xi \rightarrow 0$ (albeit with a large slope due to the particular choice of parameters in Fig.~\ref{fig:mti_disp_rel}) .

\subsubsection{Fastest Growing Mode}

Figure \ref{fig:mti_disp_rel} shows that the MTI's grows at or near its maximum rate for a broad range of $k$. This is because at low $q,\mathrm{Pr},\Lambda$ there is a broad range of wavelengths that approximate the ideal case; stable stratification, diffusion, and magnetic tension have little influence on these scales. The dispersion relation hence resembles more a plateau, rather than a peak. 

We can obtain an asymptotic formula for the maximum growth rate in the case of a perpendicular perturbation, and in the limit of weak viscosity, resistivity, and magnetic field: $ 0  \ll \left\{q, \mathrm{Pr}, \Lambda \right\} \ll 1$, and $\tilde{ N}^2 \lesssim 1$. For ease of notation, in the next sections we will drop the tildes, but it is understood that all the quantities are dimensionless. We adopt the small parameter $\epsilon \ll 1$ and set
\begin{align}
 q & \sim \epsilon , \, \, \, \, \, \, \, \, \mathrm{Pr} \sim \epsilon , \, \, \, \, \, \, \, \, \Lambda \sim  \epsilon , \,\,\,\,\,\,\,\, \xi \sim \epsilon^{-1/2}, \\
	\sigma &= \sigma_0 + \sigma_1 \epsilon^{1/2} + \bigO(\epsilon),
\end{align}
where $\sigma_0$ and $\sigma_1$ are $\sim \bigO (1)$. Inserting these in Eq.~\eqref{mti_disp_rel_full_nondim} obtains
\begin{align}
\sigma_{max} &= 1 - \left[ (1+N^2)(q+\mathrm{Pr}+\Lambda) \right]^{1/2}, \label{eq:sigma_max_visc} \\
\xi_{max} &= \left[ \frac{(1+N^2)}{(q+\mathrm{Pr}+\Lambda)}\right]^{1/2}. \label{eq:k_max_visc} 
\end{align}
Notice the interesting result that the Roberts, Prandtl, and Elsasser numbers are additive in both expressions, indicating that though they represent quite different physical effects, they impact on the MTI in exactly the same way.

\subsubsection{Small $k$ Limit}
 
In the long-wavelength limit, i.e. for small $k$, the growing solution of the dispersion relation in Eq.~\eqref{mti_disp_rel_full_nondim} exhibit two different asymptotic behaviours, depending on whether $N^2=0$ (the "adiabatic" case), or $N^2>0$ (the "sub-adiabatic" case).
In the adiabatic case, at leading order in small $\xi$, the growth rate scales as
\begin{align}
\tilde{ \sigma} \approx  \left( k_x^2/k^2  \right)^{2/3} \xi^{1/3},
\end{align}
while in the sub-adiabatic case we have
\begin{align}
\tilde{ \sigma} \approx \left( \frac{k_x^2/k^2 }{\tilde{N}^2}-q \right) \xi.
\end{align}
Both expressions go to zero for $\xi \rightarrow 0$, but when $N^2=0$ the growth rate increases faster. In other words, a stable entropy stratification inhibits the growth of the MTI on long scales, though does not fully stabilise it. 

\subsubsection{Oscillations and g-modes}

Stably stratified atmospheres support buoyancy oscillations/waves, which we refer to as g-modes \citep[see, e.g.,][]{Davidson2013}. These modes survive in the presence of anisotropic heat conduction and, despite being strongly damped, our simulations show that they are continuously excited by the MTI turbulence, thereby acting as an effective sink of turbulent energy. We hence devote some space to review them.

Equations~\eqref{eq:lin_MTI_1}-\eqref{eq:lin_MTI_4} only admit oscillatory solutions above a certain length-scale. When $N^2>0$, these oscillations represents a modified g-mode that reduces to the classical case in the long-wavelength limit. It is worth stressing, however, that when $N^2=0$ damped oscillatory solutions also appear at large scales but are not restored by buoyancy (nor magnetic tension) arising instead from the nature of the thermal conductivity (and are possibly a form of self-oscillation). These waves are of marginal concern to us as the case $N^2=0$ is rather pathological.
 
The critical wavenumber below which such oscillating solutions emerge can be found by setting the discriminant of the cubic polynomial equal to zero. To simplify things, we will only consider the inviscid, perfectly conducting, weak-field limit of Eq.~\eqref{mti_disp_rel_full_nondim}. In this case, the discriminant is a quartic polynomial in $\xi$ whose positive root (the one of interest to us) is
\begin{align}
\xi^2_{c} = \frac{1}{8} \left[ 27+18 \tilde{N}^2 - \tilde{N}^4 + \sqrt{1+\tilde{N}^2} \left( 9 + \tilde{ N}^2 \right)^{3/2} \right]. \label{eq:roots_discriminant}
\end{align}
The expression in Eq.~\eqref{eq:roots_discriminant} reduces to $\xi^2_{c} \approx 9(3+2\tilde{ N}^2)/4$ in the limit of $\tilde{ N}^2 \ll 1$ and to $\xi^2_{c} \approx 8 + 4 \tilde{ N}^2$ for $\tilde{ N}^2 \gg 1$. Thus, the value $k_c = \sqrt{\omega_T \xi_{c}/ \chi }$ at which the transition to occurs is approximately $k_c \approx l_{\chi}^{-1} = \sqrt{\omega_T / \chi }$ with a prefactor of order unity over a large range of $\tilde{ N}^2$. Oscillating modes can only exist for $k< k_c$.

A long-wavelength expansion of the dispersion relation yields
\begin{equation}\label{eq:g-modes-disp-rel}
\sigma_N(\xi) = \pm i \frac{k_{x}}{k} \tilde{N} - \frac{1+\tilde{ N}^2}{2 \tilde{ N}^2} \frac{k_x^2}{k^2} \xi + \bigO(\xi^2)
\end{equation}
where $\sigma_N(\xi)$ is the complex growth rate of the modified g-modes. When $\xi \rightarrow 0$ we retrieve the classical g-mode frequency, as the damping rate is of order $\bigO (\xi)$. If, on the other hand, $\tilde{ N}^2 =0$ we find that no oscillations exist in the limit $\xi \rightarrow 0$, as the two stable roots of the MTI dispersion relation take the form
\begin{equation}
	\sigma(\xi) = \frac{-1 \pm i \sqrt{3}}{2} \left( \frac{k_{x}}{k} \right)^{2/3} \xi^{1/3} - \frac{1}{3}
	\left( \frac{k_{x}}{k} \right)^{2}  \xi + \bigO(\xi^{5/3}),
\end{equation}

\subsection{Nonlinear Theory}\label{sec:nonlin_MTI}
In this section we examine the evolution of the MTI in a reduced 1D model by tracking the nonlinear behaviour of a single MTI mode. Apart from explaining the initial stages of our simulations (and providing additional numerical tests), such an approach can help us understand some of the features of the MTI saturation more broadly.
In particular, we demonstrate how the nonlinearity in the thermal conduction can transform a linear MTI mode into a quasi-steady nonlinear buoyancy oscillation,  discuss the parasitic modes that can destabilise these waves, and how this relates to the impossibility of the MTI developing an `open field' configuration as its main saturation route (in contrast to the HBI's ability to settle on a `closed field' configuration).

\subsubsection{The `Open Field' Solutions}

We consider the development of the MTI when it depends only on $x$, the horizontal direction. We label such solutions `open field' because their associated magnetic field is purely vertical and, when attaining large amplitudes, they transform the system from one dominated by the background horizontal field to one dominated by vertical fields. It might seem natural that this near vertical configuration would be the natural endpoint of the MTI saturation: it maximises the vertical conduction of heat (thus better erasing the destabilising temperature gradient). But this is not the case, as we discuss below. Note that our designation of open field permits the vertical field to change sign across the nodes of the initial sinusoidal profile. 

Unlike the channel modes of the MRI \citep{Goodman1994}, the linear open-field MTI mode is not an exact solution to the nonlinear governing equations. While it is true that all the quadratic  terms vanish identically, such as $\bs u \cdot  \nabla \bs u$, $\bs B \cdot  \nabla \bs B$, etc., the nonlinear terms comprising the anisotropic heat flux do not. Thus the mode cannot grow indefinitely and can saturate at finite amplitude. 

To track the evolution of the 1D open-field MTI solution, we must solve a third order system of time-dependent PDEs. The associated one-dimensional flow defined by $\bs u = u(x,t) \bs e_{z}$, $\bs B = B_0 [ \bs e_{x} + h(x,t) \bs e_{z}]$ and $ \theta = \theta(x,t)$, is governed by
\begin{align}
&\partial_t u = - \theta + \nu \partial_x^2 u , \qquad 
\partial_t h    =  \partial_x u +\eta \partial_x^2 h, \label{eq:non-lin_B} \\
&\partial_t \theta = N^2 u + \chi \frac{\partial}{\partial x} \left[ \frac{\partial_x \theta}{1+h^2} \right] + \chi \omega_T^2\frac{\partial}{\partial x} \left[ \frac{h}{1+h^2} \right] \label{eq:non-lin_buoyancy},
\end{align}
in which we have dropped magnetic tension to simplify the analysis. We now discuss the behaviour of this dynamical system.

\subsubsection{The Nonlinear Timescale}

For small amplitude $u$, $h$, and $\theta$, Eqs \eqref{eq:non-lin_B}-\eqref{eq:non-lin_buoyancy} reproduce the linear analysis of Section 2.4. At next order in the small amplitudes, we find cubic contributions arising from the last two terms in Eq.~\eqref{eq:non-lin_buoyancy}. These become important when the mode achieves an amplitude 
	$ h^{2} (x,t_{\mathrm{NL}}) \sim 1$, which defines the nonlinear time-scale $t_{\mathrm{NL}}$. In the limit of low resistivity and viscosity, and given the exponential growth of low amplitude (linear) modes, we have that $ h \sim k_x u / \sigma$ and thus
	\begin{equation}\label{eq:nonlin_timescale}
		t_{\mathrm{NL}} \sim \sigma^{-1} \ln \frac{\sigma}{  u_0 k_x},
	\end{equation}
	where $u_{0}$ is the initial vertical velocity (at $t=t_0$). Note that $t_{\mathrm{NL}}$ is not necessarily the time at which the MTI saturates, but only the point at which the nonlinear terms in the heat flux become important. In fact, we find numerically that the mode still grows for some period when $t>t_\text{NL}$ albeit at a reduced rate.

\subsubsection{Saturation via Nonlinear g-modes}

At very large amplitude $h\gg 1$, the total heat conduction (comprising its destabilising linear term and all its nonlinear terms) becomes small and potentially subdominant. We may then expect the MTI growth mechanism to fade away once the MTI grows sufficiently large, leaving behind only the linear terms that govern slightly damped g-modes. Given this damping, the system might settle on an amplitude where the weak damping is matched by weak injection of energy from the MTI, this saturated state possibly taking the form of a modified g-mode. 

A separate saturation route minimises any horizontal variation in $h$, thus diminishing the destabilising third term on the right side of Eq.\eqref{eq:non-lin_buoyancy}. This, however, must be achieved while retaining any alternating sign in the solution inherited from the initial sinusoidal mode. We might then expect that the system evolves away from a sinusoidal and towards a square-wave pattern in the magnetic field ($h$), with the other variables adopting piece-wise linear profiles in $x$. Such a situation once again leads to the dominance of the linear terms controlling g-modes. This is not the whole story, however, as across the nodes of this nonlinear oscillation there will be an unbalanced horizontal heat flux. Thus, to leading order, the oscillation will appear as a square wave, with the node encompassed by an internal boundary layer in which diffusion is important. 
	
\begin{figure}
	\centering
	\includegraphics[width=1.0\columnwidth]{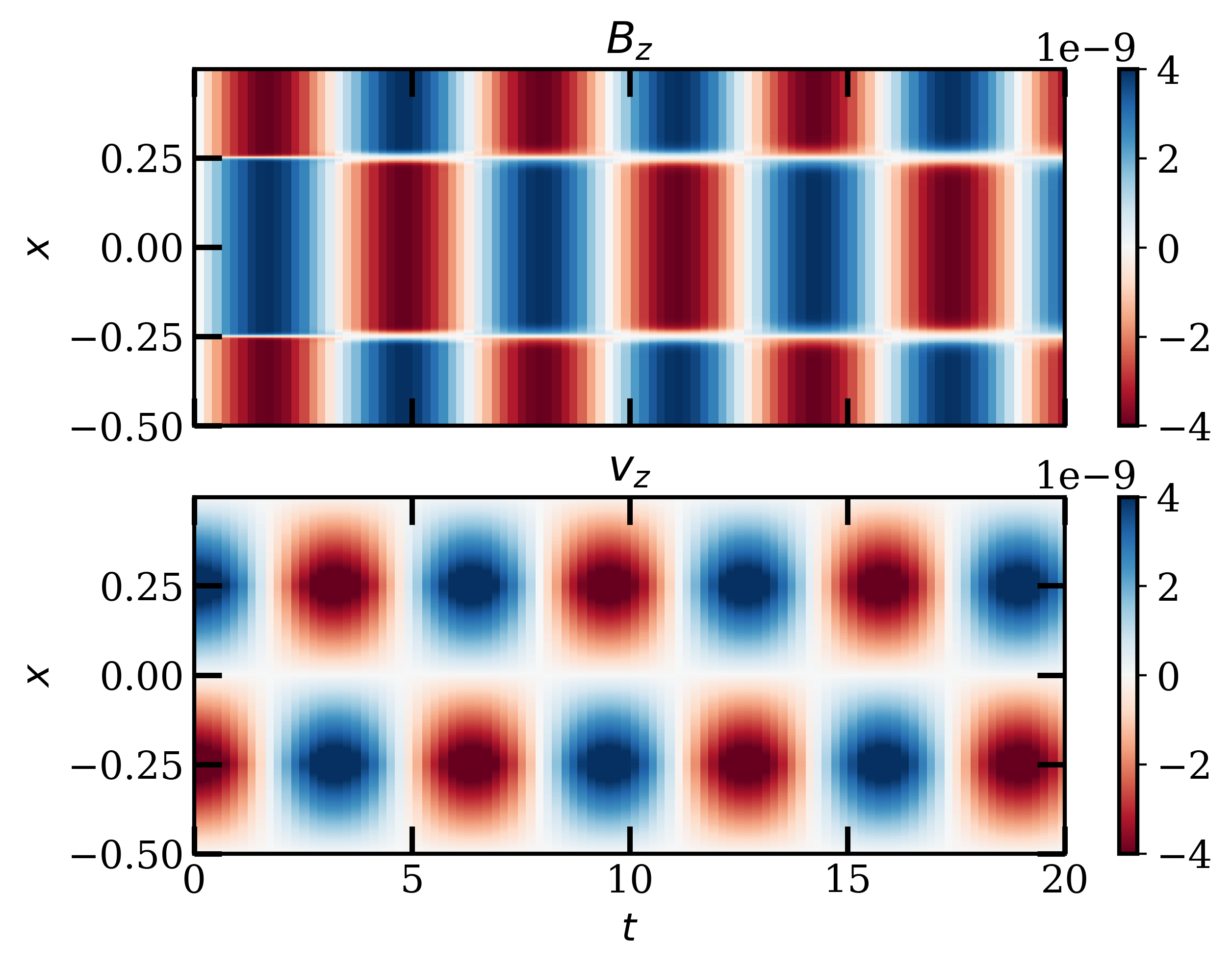}
	\caption{Space-time diagram of a `nonlinear g-mode' solution to Eqs \eqref{eq:non-lin_B}-\eqref{eq:non-lin_buoyancy} obtained with \textsc{SNOOPY} (see Section \ref{snoopy}). Here $N/\omega_T=1$, Pr $=$ Pm $=1$, and $\mathrm{Pe}=10^5$, with the latter sufficiently large to minimise disruption at the internal boundary layers. The wave's oscillation frequency is $\approx N$.}
	\label{fig:mti_nonlinear}
\end{figure}

These expectations have been borne out by solutions calculated numerically, by both an initial value and boundary value approach to the equations. 
For moderate resistivity and viscosity, the system supports a family of steady solutions, but as we reduce the diffusion coefficients this solution branch terminates and we obtain more spatially complicated wave solutions that oscillate with a frequency near $N$. An example of a nonlinear g-mode solution, comprising a square wave in $B_z$ and a triangular wave in $u_z$, is shown in Fig.~\ref{fig:mti_nonlinear}. Though interesting, and potentially offering a toy model for understanding large-amplitude g-modes in the ICM, we do not explore these solutions further. 

\subsubsection{Parasitic Instabilities}\label{sec:parasitic_instabilities}
In addition to the nonlinearities intrinsic to the MTI equations, the exponential growth phase of instability can also be cut short by the development of parasitic instabilities once we permit the system to vary with $z$. In particular, the $k_z=0$ mode of the MTI develops strong vertical shear as it grows exponentially: if small-amplitude perpendicular perturbations are present in the system, a parasitic Kelvin-Helmholtz instability will be triggered that can disrupt the MTI mode. These dynamics cannot be captured by Eq.~\eqref{eq:non-lin_B}--\eqref{eq:non-lin_buoyancy} because the dependence on $z$ is neglected. 

The time at which the MTI is disrupted by these secondary modes can be estimated by equating the velocity amplitude of  the parasite mode $p$ with that of the MTI mode $m $. Following \citet{Latter2016}, we assume that the amplitude of the parasitic mode grows as $dp/dt \sim s(t)p$, where $s(t)$ is a time-dependent growth rate, proportional to the amplitude of the MTI mode. More specifically, we set the growth rate at time $t_0$ to be proportional to the shear rate of the MTI mode, i.e. $s\sim \delta u_z k_x$, in which $\delta u_z$ is growing exponentially with growth rate $\sigma$. By solving the differential equation, one can obtain an analytical expression for the time of disruption $t_{\mathrm{KH}}$ in terms of the second real branch of the Lambert $W$ function:
	\begin{align}\label{eq:KH_timescale}
		t_{\mathrm{KH}} =  - \frac{1}{\sigma} W_{-1}\left( \zeta \exp \left[ -\zeta - \log \frac{m_0}{p_0} \right]\right) - \frac{1}{\sigma} \left( \zeta + \log \frac{u_0}{p_0} \right)
	\end{align}
	where $\zeta = u_0 k_x / \sigma$, and $p_0$ is the initial velocity amplitude of the parasite.
	
	To compare the nonlinear and Kelvin-Helmholtz timescales, we expand the Lambert function in small values of $\zeta$. When the parasitic mode has the same amplitude of the MTI eigenmode at $t=t_0$, 
	the difference between the two timescales is
	\begin{equation}
		t_{\mathrm{KH}} - t_{\mathrm{NL}} \simeq \sigma^{-1} \ln \left( - \ln  \zeta\right) ,
	\end{equation}
	which is always greater than zero but varies exceptionally weakly with $\zeta$. As a result, the two timescales do not drastically differ from one another, although we note that the nonlinear timescale is always less than the Kelvin-Helmholtz timescale for the same value of the initial amplitude $u_0$.

\section{Numerical Methods}\label{sec:numerical_methods}
The remainder of the paper presents numerical simulations of the MTI at its onset and at saturation. In this section we describe the numerical methods that we have used, alongside our diagnostics.

\begin{table*}  
	\centering 
	\caption{Parameters and results of 2D MTI simulations, averaged over $100 \omega_T^{-1}$, for a total of $10^4$ samples. All runs have been initialized with a horizontal uniform magnetic field of strength $B_0 = 10^{-5}$, except for: runs R0B01e-4, R0B01e-3, which have an initial field of strength $B_0 = 10^{-4}$ and $B_0 = 10^{-3}$, respectively; run R0Bz, where the magnetic field $B_0 = 10^{-5}$ instead points in the $z$ direction; and run R0NNF, where the initial horizontal magnetic field has a sinusoidal profile with wavelength $2\pi/L$ and amplitude $B_0 = \sqrt{2}\times 10^{-5}$. The dagger symbol $^{\dagger}$ means the run has not reached saturation.}  
	\label{tab:table_runs} 
	\begin{tabular}{ccccccccccc}  
		\hline 
		Run 	 &  Res. 	 &  $\mathrm{Pe}$ 	 &  $\tilde{N}^2$ 	 &  $\mathrm{Pr},q$ 	 &  $K$ 	 &  $M$ 	 &  $U$ 	 &  $Q_{adv} $ 	 &  $\langle | b_z | \rangle$ 	 &  $Nu$ 	  \\ 
		\hline 
		R0 	    &  $ (2048)^2 $	   &  $ 6 \times 10^{3} $ 	   &  $ 0.10 $ 	    &  $ 0.006 $ 	    &  $ 1.768(6) \times 10^{-4} $ 	    &  $ 1.26(2) \times 10^{-8} $ 	    &  $ 2.10(2) \times 10^{-4} $ 	    &  $ 8.3(5) \times 10^{-6} $ 	    &  $ 0.638(3) $ 	    &  $ 0.213(7) $ 	    \\ 
		RsP00 	    &  $ (1024)^2 $	   &  $ 6 \times 10^{2} $ 	   &  $ 0.10 $ 	    &  $ 0.0006 $ 	    &  $ 1.94(5) \times 10^{-3} $ 	    &  $ 1.36(7) \times 10^{-7} $ 	    &  $ 1.07(6) \times 10^{-3} $ 	    &  $ 2(3) \times 10^{-5} $ 	    &  $ 0.639(4) $ 	    &  $ 0.19(2) $ 	    \\ 
		RsP01 	    &  $ (1024)^2 $	   &  $ 1 \times 10^{3} $ 	   &  $ 0.10 $ 	    &  $ 0.001 $ 	    &  $ 1.17(4) \times 10^{-3} $ 	    &  $ 8.3(2) \times 10^{-8} $ 	    &  $ 7.6(3) \times 10^{-4} $ 	    &  $ 2(1) \times 10^{-5} $ 	    &  $ 0.641(7) $ 	    &  $ 0.20(2) $ 	    \\ 
		RsP02 	    &  $ (1024)^2 $	   &  $ 6 \times 10^{3} $ 	   &  $ 0.10 $ 	    &  $ 0.006 $ 	    &  $ 1.93(2) \times 10^{-4} $ 	    &  $ 1.57(4) \times 10^{-8} $ 	    &  $ 1.85(2) \times 10^{-4} $ 	    &  $ 6.8(7) \times 10^{-6} $ 	    &  $ 0.642(2) $ 	    &  $ 0.257(8) $ 	    \\ 
		RsP03 	    &  $ (1024)^2 $	   &  $ 1 \times 10^{4} $ 	   &  $ 0.10 $ 	    &  $ 0.01 $ 	    &  $ 1.16(2) \times 10^{-4} $ 	    &  $ 9.5(4) \times 10^{-9} $ 	    &  $ 1.24(1) \times 10^{-4} $ 	    &  $ 6(1) \times 10^{-6} $ 	    &  $ 0.641(2) $ 	    &  $ 0.280(8) $ 	    \\ 
		RsP04 	    &  $ (1024)^2 $	   &  $ 3 \times 10^{4} $ 	   &  $ 0.10 $ 	    &  $ 0.03 $ 	    &  $ 3.63(4) \times 10^{-5} $ 	    &  $ 3.82(9) \times 10^{-9} $ 	    &  $ 5.15(4) \times 10^{-5} $ 	    &  $ 3.14(7) \times 10^{-6} $ 	    &  $ 0.648(2) $ 	    &  $ 0.356(5) $ 	    \\ 
		RsP05 	    &  $ (1024)^2 $	   &  $ 6 \times 10^{4} $ 	   &  $ 0.10 $ 	    &  $ 0.06 $ 	    &  $ 1.74(1) \times 10^{-5} $ 	    &  $ 2.10(5) \times 10^{-9} $ 	    &  $ 2.95(3) \times 10^{-5} $ 	    &  $ 2.17(7) \times 10^{-6} $ 	    &  $ 0.653(2) $ 	    &  $ 0.422(5) $ 	    \\ 
		RsN00 	    &  $ (1024)^2 $	   &  $ 3 \times 10^{3} $ 	   &  $ 0.05 $ 	    &  $ 0.003 $ 	    &  $ 7.26(8) \times 10^{-4} $ 	    &  $ 6.0(2) \times 10^{-8} $ 	    &  $ 6.8(2) \times 10^{-4} $ 	    &  $ 8(3) \times 10^{-6} $ 	    &  $ 0.637(5) $ 	    &  $ 0.24(2) $ 	    \\ 
		RsN01 	    &  $ (1024)^2 $	   &  $ 3 \times 10^{3} $ 	   &  $ 0.10 $ 	    &  $ 0.003 $ 	    &  $ 3.91(3) \times 10^{-4} $ 	    &  $ 2.70(9) \times 10^{-8} $ 	    &  $ 3.18(3) \times 10^{-4} $ 	    &  $ 8(3) \times 10^{-6} $ 	    &  $ 0.637(4) $ 	    &  $ 0.228(8) $ 	    \\ 
		RsN02 	    &  $ (1024)^2 $	   &  $ 3 \times 10^{3} $ 	   &  $ 0.20 $ 	    &  $ 0.003 $ 	    &  $ 2.08(3) \times 10^{-4} $ 	    &  $ 1.36(4) \times 10^{-8} $ 	    &  $ 1.32(4) \times 10^{-4} $ 	    &  $ 9(2) \times 10^{-6} $ 	    &  $ 0.636(3) $ 	    &  $ 0.215(5) $ 	    \\ 
		RsN03 	    &  $ (1024)^2 $	   &  $ 3 \times 10^{3} $ 	   &  $ 0.40 $ 	    &  $ 0.003 $ 	    &  $ 1.04(2) \times 10^{-4} $ 	    &  $ 7.0(1) \times 10^{-9} $ 	    &  $ 5.24(8) \times 10^{-5} $ 	    &  $ 7(1) \times 10^{-6} $ 	    &  $ 0.639(2) $ 	    &  $ 0.206(3) $ 	    \\ 
		RsN04 	    &  $ (1024)^2 $	   &  $ 3 \times 10^{3} $ 	   &  $ 0.60 $ 	    &  $ 0.003 $ 	    &  $ 6.80(6) \times 10^{-5} $ 	    &  $ 4.96(5) \times 10^{-9} $ 	    &  $ 2.93(6) \times 10^{-5} $ 	    &  $ 6.3(5) \times 10^{-6} $ 	    &  $ 0.641(1) $ 	    &  $ 0.203(2) $ 	    \\ 
		R0NNF 	    &  $ (2048)^2 $	   &  $ 6 \times 10^{3} $ 	   &  $ 0.10 $ 	    &  $ 0.006 $ 	    &  $ 1.777(9) \times 10^{-4} $ 	    &  $ 2.5(2) \times 10^{-10} $ 	    &  $ 2.04(3) \times 10^{-4} $ 	    &  $ 8.5(4) \times 10^{-6} $ 	    &  $ 0.637(4) $ 	    &  $ 0.22(3) $ 	    \\ 
		R0Bz 	    &  $ (2048)^2 $	   &  $ 6 \times 10^{3} $ 	   &  $ 0.10 $ 	    &  $ 0.006 $ 	    &  $ 1.77(3) \times 10^{-4} $ 	    &  $ 2.8(2) \times 10^{-8} $ 	    &  $ 2.09(3) \times 10^{-4} $ 	    &  $ 8.5(9) \times 10^{-6} $ 	    &  $ 0.641(3) $ 	    &  $ 0.214(7) $ 	    \\ 
		R0B01e-4 	    &  $ (2048)^2 $	   &  $ 6 \times 10^{3} $ 	   &  $ 0.10 $ 	    &  $ 0.006 $ 	    &  $ 1.66(2) \times 10^{-4} $ 	    &  $ 1.34(5) \times 10^{-6} $ 	    &  $ 2.19(2) \times 10^{-4} $ 	    &  $ 1.02(9) \times 10^{-5} $ 	    &  $ 0.642(3) $ 	    &  $ 0.223(7) $ 	    \\ 
		R0B01e-3 	    &  $ (2048)^2 $	   &  $ 6 \times 10^{3} $ 	   &  $ 0.10 $ 	    &  $ 0.006 $ 	    &  $ 1.36(6) \times 10^{-4} $ 	    &  $ 4.2(1) \times 10^{-4} $ 	    &  $ 4.7(2) \times 10^{-4} $ 	    &  $ 3.7(3) \times 10^{-5} $ 	    &  $ 0.894(4) $ 	    &  $ 0.86(4) $ 	    \\ 
		R0Nlow$^{\dagger}$ 	    &  $ (2048)^2 $	   &  $ 6 \times 10^{3} $ 	   &  $ 10^{-4} $ 	    &  $ 0.006 $ 	    &  $ 7.0(4) \times 10^{-3} $ 	    &  $ 1.5(1) \times 10^{-6} $ 	    &  $ 2.73(2) \times 10^{-1} $ 	    &  $ 1.5(2) \times 10^{-4} $ 	    &  $ 0.685(3) $ 	    &  $ 0.5(2) $ 	    \\ 
		\hline 
	\end{tabular} 
\end{table*}  

\subsection{\textsc{SNOOPY}}
\label{snoopy}
Our simulations have been performed with the latest version of \textsc{SNOOPY} \citep{Lesur2015}, a pseudo-spectral 3D code suitable for incompressible MHD simulations, which we have modified so as to include anisotropic heat conduction. \textsc{SNOOPY} applies a Fourier operator to the fluid equations, thus taking the problem from real space to Fourier space, and advances the Fourier coefficients in time through a three-step Runge-Kutta algorithm. Like all pseudo-spectral codes, nonlinear terms are computed in real space and then transformed back into Fourier space, making use of the fast Fourier transform (FFT). 
For this reason, a standard $2/3$ de-aliasing rule is implemented.  We use OpenMP parallelization to speed up all runs.

We implemented  the anisotropic heat conduction term both with explicit integration at each timestep, and with a super time-stepping (STS) scheme based on \citet{Alexiades1996}, together with symmetric Strang operator splitting of the diffusion operators, including resistivity and viscosity \citep{Strang1968}. This led to a considerable speedup of our simulations (about a factor $\approx 9$ for a run with typical parameters and excluding output) as we focus on the regime where thermal diffusivity is many orders of magnitude larger than either viscosity or magnetic resistivity. Indeed, due to the large disparity of typical parabolic $\Delta t_\text{par}$ and hyperbolic $\Delta t_\text{hyp} $ time-scales, especially for larger values of $\chi$, to ensure numerical stability it is necessary to restrict  $\Delta t_\text{hyp} $ below a certain threshold value. We have therefore introduced a tunable parameter $R_\text{max}$ such that if $\Delta t_\text{hyp} / \Delta t_\text{par} > R_\text{max}$ then the hyperbolic time-step is reduced accordingly to be exactly $R_\text{max}$ times larger than $\Delta t_\text{par}$. We found that $R_\text{max} \approx 100$ strikes a good balance between computational speedup and numerical robustness. This procedure is very similar to what is outlined in the STS implementation of the compressible MHD code PLUTO \citep{Mignone2007}.  A numerical validation of our implementation of the STS algorithm can be found in Appendix \ref{app:STS_benchmark}. Finally, viscous and resistive terms are integrated implicitly.  

\subsubsection{Initial Conditions and Parameters}

The physical variables are non-dimensionalized by taking $\omega_T$ as the unit of time, and $L$, the box size, as the unit of length. Therefore, unless otherwise specified, it is understood that all quantities shown in the next sections are  in units where $L=\omega_T=1$. The physical parameters that characterize our simulations are the Peclet number, defined as $\mathrm{Pe}=L^2 \omega_T / \chi$, the reduced Brunt-V{\"a}is{\"a}l{\"a} frequency squared $\tilde{ N}^2 = N^2 / \omega_T^2$, the Prandtl number $\mathrm{Pr} = \nu / \chi$, and the Robertson number $q=\eta / \chi$. Please be aware that Pe is not the same as the physical Peclet number of a cluster, and the two should not be compared. Finally, we vary the strength of the imposed horizontal magnetic field, $B_0$, recalling that in Eqs \eqref{eq:div_eq}-\eqref{eq:buoyancy_eq} $\bs B$ has units of velocity.

The majority of 2D runs -- with the exception of those where we study the evolution of a single MTI eigenmode -- are initialized with random white noise at all wavenumbers in the velocity field, with amplitude $\lvert \delta u_x \rvert = \lvert \delta u_z \rvert = 10^{-4}$, and with a uniform horizontal magnetic field of strength $B_0 = 10^{-5}$. This ensures that the magnetic field is dynamically unimportant, except indirectly through the anisotropic heat conduction term. We performed most of the simulations using $(1024)^2$ Fourier modes on 2D, doubly-periodic domains with aspect ratio of $1:1$. This we have found sufficient for our purposes. To check the convergence of the results, we also performed a number of higher-resolution runs with $(2048)^2$, and with different aspect ratios, such as $2:1$ or $1:2$. A list of all the runs performed, together with their physical and numerical parameters, can be found in Table~\ref{tab:table_runs}.

As is typical of Boussinesq equations, no natural outer scale appears in Eq.~\eqref{eq:div_eq}--\eqref{eq:buoyancy_eq}.  Boussinesq simulations often produce structures at the box-size that end up dominating the overall dynamics, as seen in simulations of Rayleigh-Benard convection \citep["elevator modes";][]{calzone2006}, or in simulations of the MRI \citep["channel modes";][]{Lesur2007}. Fortunately this is not the case in our MTI simulations, where an \textit{a posteriori} check shows that all physical quantities vary on scales much shorter than $L$, and are thus box-size independent.

\subsubsection{Diagnostics}

One significant advantage of using a (pseudo-)spectral code is that it is straightforward to extract the power spectral densities of the fields. This information is an invaluable aid in understanding the dominant processes at different scales. Consequently, in addition to the usual box-averaged quantities calculated in previous MTI simulations -- total, as well as vertical and horizontal energy levels (indicated by the subscript $z$ and $x$, respectively), average magnetic field orientation, heat fluxes, etc. -- we compute the terms that enter in the scale-by-scale energy balance.

To do so, we first obtain an equation for the time evolution of the energy contained in each Fourier mode: starting from Eq.~\eqref{eq:mom_eq}, we apply a Fourier transform, multiply by $\hat{\bs u}^* (\bs k, t)$, where $\bs k = k_x \bs e_x + k_z \bs e_z$ is the wavevector, and add the resulting equation together with its complex conjugate. Following an analogous procedure with Eqs \eqref{eq:bfield_eq} and\eqref{eq:buoyancy_eq}, we obtain
\begin{align}
\frac{1}{2}\frac{d \lvert \hat{u} (\bs k, t) \rvert^2}{d t} &= \hat{T}_{\bs u}^{\bs u}  - \hat{\Theta}  - \hat{L}  - \nu \hat{D}_{\nu}  ,  \label{eq:spectral_vel}\\
\frac{1}{2}\frac{d \lvert \hat{B} (\bs k, t) \rvert^2}{d t} &= \hat{T}_{\bs B}^{\bs u}  + \hat{L}  - \eta \hat{D}_{\eta}  ,  \label{eq:spectral_mag}\\ 
\frac{1}{2}\frac{d \lvert \hat{\theta} (\bs k, t) \rvert^2}{d t} &= \hat{T}_{\theta}^{\bs u} + N^2 \hat{\Theta}  - \chi \hat{D}_{\chi}  + \chi \omega_T^2 \hat{\mathcal{F}} , \label{eq:spectral_temp}
\end{align}	
where $ \hat{T}_{\bs q}^{\bs u}$ denotes the nonlinear transfer term of quantity $\bs q $ by the velocity field, $\hat{\Theta} $ is the buoyancy force, $\hat{L}$ is the Lorentz force, $\hat{\mathcal{F}}$ is the MTI forcing term, and finally $\hat{D}_{\nu}, \hat{D}_{\chi}$ and $\hat{D}_{\eta} $ are the diffusive sinks due to viscosity, conductivity and resistivity, respectively. Their mathematical definitions can be found in Appendix ~\ref{app:spectral_fluxes}. 

In all the simulations performed, we choose to track the evolution over time of the "shell-integrated" spectra, i.e. by summing over the modes with the same magnitude of the wavevector $k= \lvert \bs k \rvert$. For instance, for the kinetic spectral energy density $E_K(k)$ we have
\begin{equation}
E_K(k,t) dk = \sum_{k < \lvert \bs k' \rvert \leq k + dk} \frac{1}{2} \lvert \hat{u} (\bs k', t) \rvert^2 ,
\end{equation}
and so forth for the other quantities. We remark that this procedure averages over the polar angles, and thus it does not capture any anisotropy present. Consequently, we have complemented these shell-integrated spectra with analogous measures that integrate over either the horizontal or vertical direction only.

In the turbulent state of the MTI, heat can be transported across the domain by both advection of the density perturbations and by conduction along the magnetic field lines. 
In the Boussinesq approximation, the total heat flux can be easily derived by rewriting Eq.~\eqref{eq:buoyancy_eq} in conservation form. As we will be mostly concerned with transport of heat across the domain in the vertical direction, we consider only the $z$ component of the total flux and define the box-averaged advected and conductive heat fluxes to be\footnote{We note that in the Boussinesq equations, the heat flux has units of $\si{m^2.s^{-3}}$.}:
\begin{align}
	Q_\text{adv}  &= -\langle \theta u_z \rangle, \\
	Q_\text{cond} &= \chi \langle  b_z \bs b \cdot \nabla \theta \rangle + \chi \omega_T^2  \langle b_z^2 \rangle.
\end{align}

Additionally, as is customary in studies of Rayleigh-Benard convection, we introduce the Nusselt number Nu, usually defined as the ratio of the total vertical heat flux (due to both conduction and turbulent advection) and the conductive heat flux \citep{Spiegel1971}. In the case of anisotropic conduction, however, there is an ambiguity in the definition of Nu, as the value of the heat flux due to conduction depends on the configuration of the magnetic field. We therefore define a turbulent Nusselt number where the sum of the advective and conductive fluxes is normalized by $Q_0= \chi \omega_T^2$, which is the amount of heat that would flow through the box if heat conduction were solely isotropic or, equivalently, if the box were at rest and had a purely vertical magnetic field. Thus
\begin{equation}\label{eq:nusselt_num}
	\text{Nu} = \frac{Q_\text{adv} + Q_\text{cond}}{Q_{0}}.
\end{equation}
\begin{figure}
	\centering
	\includegraphics[width=0.9\columnwidth]{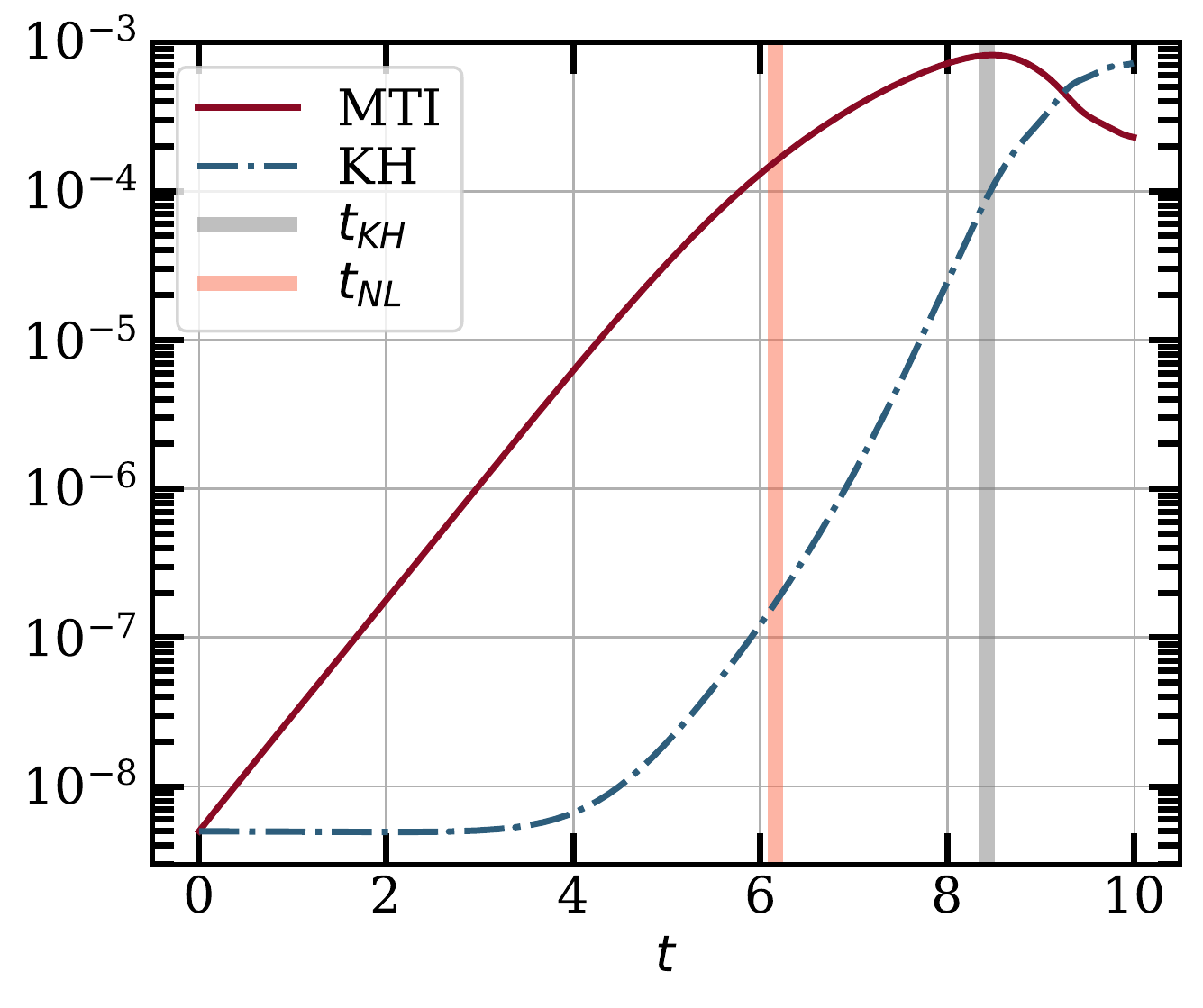}
	\caption{Plot of the kinetic energies of the MTI mode ($K_z$ with $k_x = 36 \times 2 \pi/L$) and of the KH seed perturbation ($K_x$ with $k_z = 10 \times 2 \pi/L$), together with the numerical estimation of the nonlinear and parasitic timescales. }
	\label{fig:parasitic_KH_Lambert}
\end{figure}

\begin{figure}
	\centering
	\includegraphics[width=1.0\columnwidth]{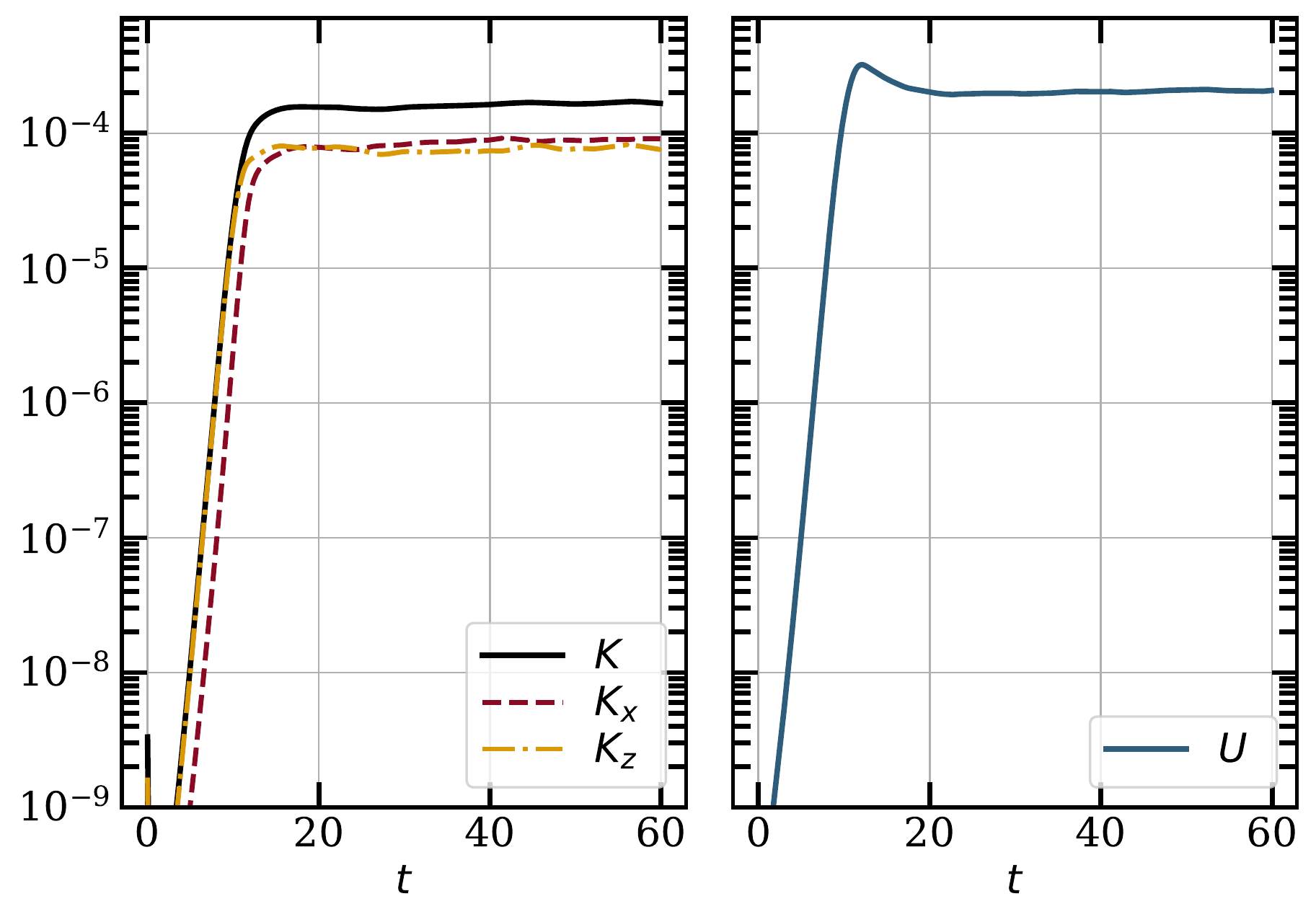}
	\caption{Left panel: total kinetic energy of run R0 (solid), together with its horizontal and vertical components (dashed and dash-dotted, respectively). Right panel: potential energy of run R0. After the exponential growth of the MTI, both the kinetic and potential energy saturate at a similar value.}
	\label{fig:2D_kin_pot_energy}
\end{figure}
\begin{figure}
	\centering
	\includegraphics[width=1.0\columnwidth]{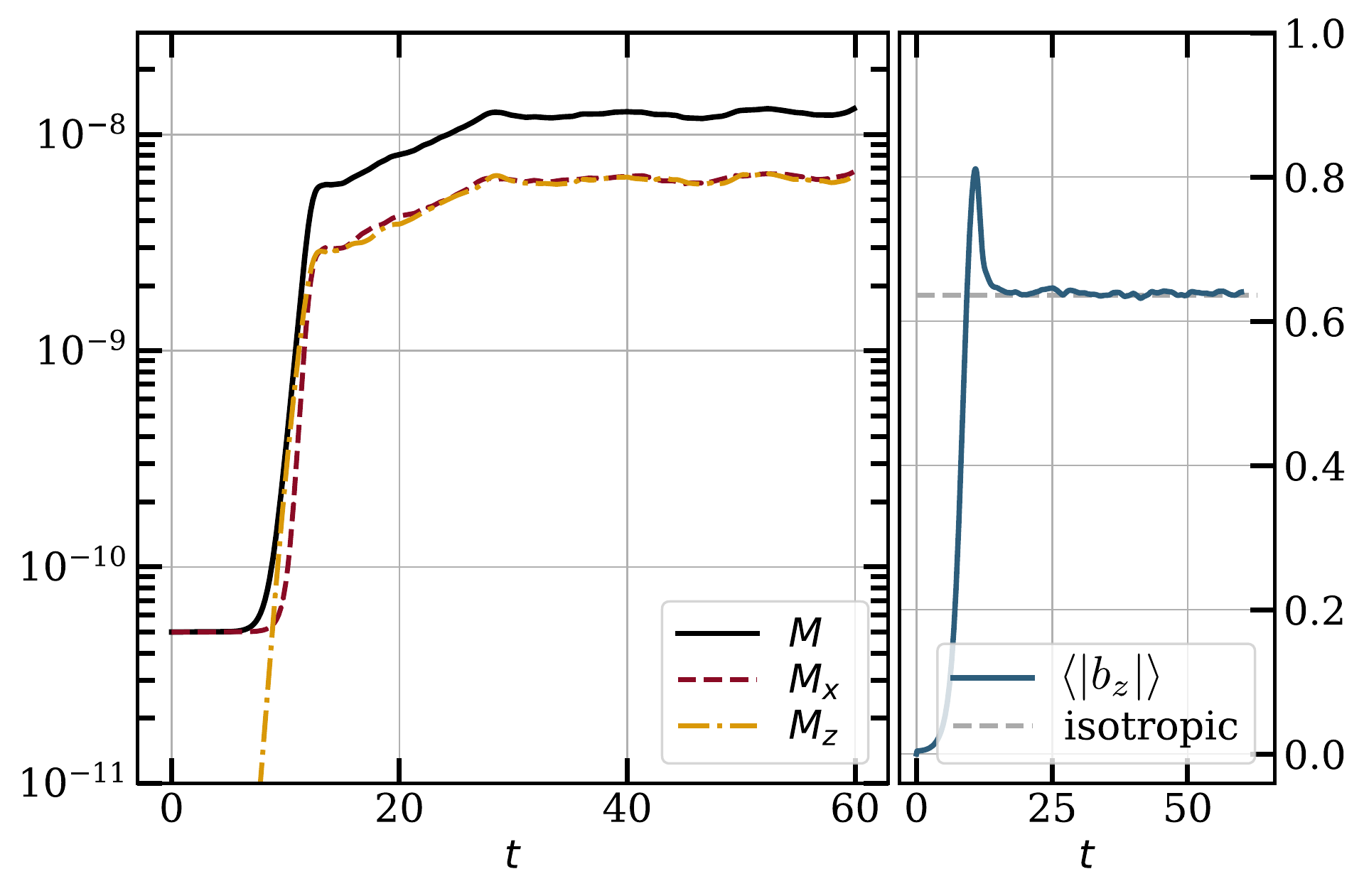}
	\caption{Left panel: R0's total magnetic energy (solid) and its vertical and horizontal components (dashed and dashed-dotted). Right panel: average $b_z$. After the end of the MTI growth phase ($t\approx 15$), the amplitude of the magnetic fields continues to grow before settling on its saturated value (at $t\approx 30$). We attribute this second growth to the turbulent stretching of the field lines (cf. Section \ref{sec:theoretical-scalings}) }
	\label{fig:2D_mag_energy}
\end{figure}
\begin{figure}
	\centering
	\includegraphics[width=0.95\columnwidth]{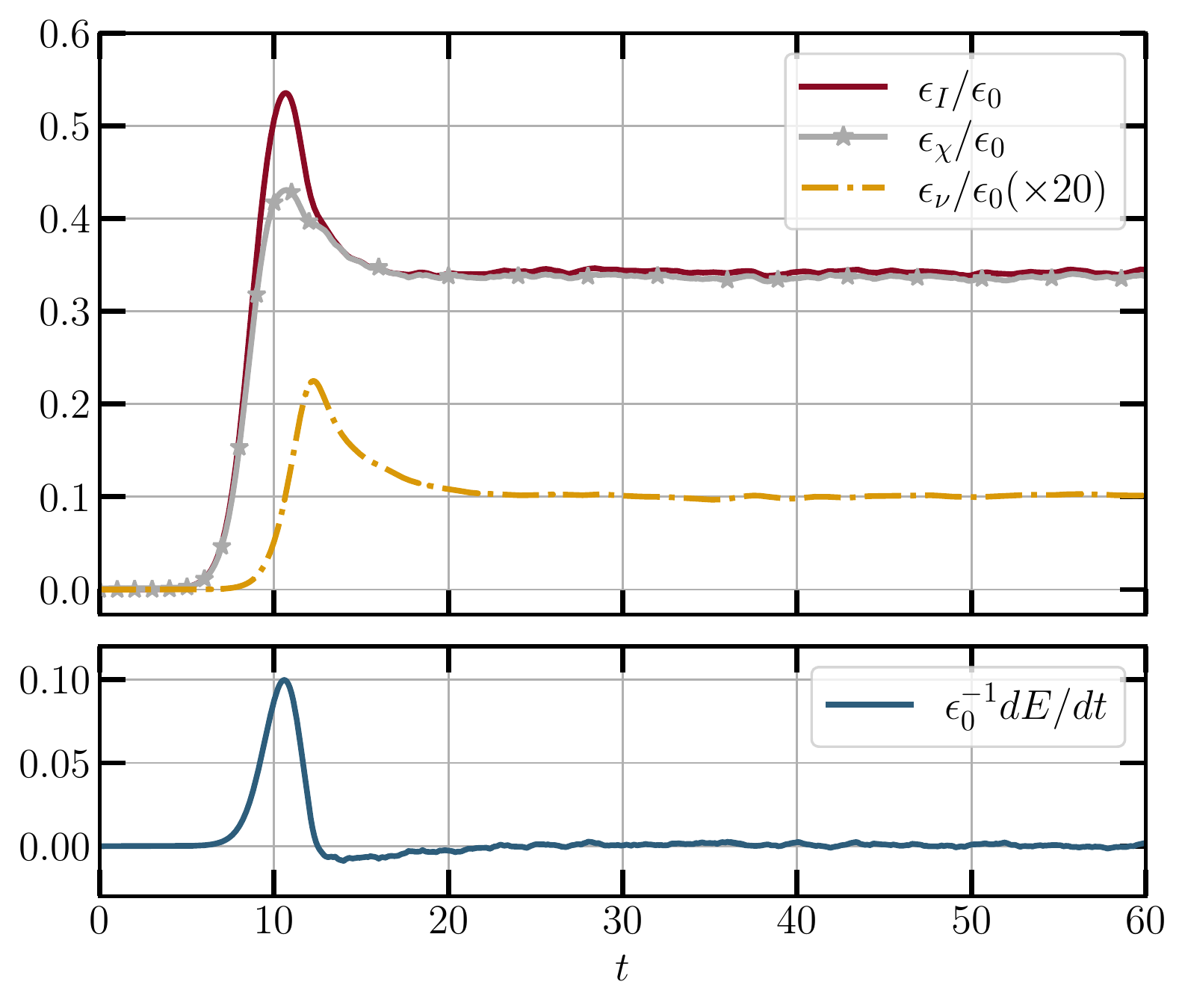}
	\caption{Volume-averaged energy fluxes normalized by $\epsilon_0 \equiv \chi \omega_T^4/N^2$ (top panel), and net balance between energy injection and dissipation normalized by $\epsilon_0$ (bottom panel). 
	Ohmic dissipation is not plotted, as it is about two orders of magnitude less than viscous dissipation. 
}
	\label{fig:en_fluxes}
\end{figure}
\begin{figure*}
	\centering
	\includegraphics[width=1.0\linewidth]{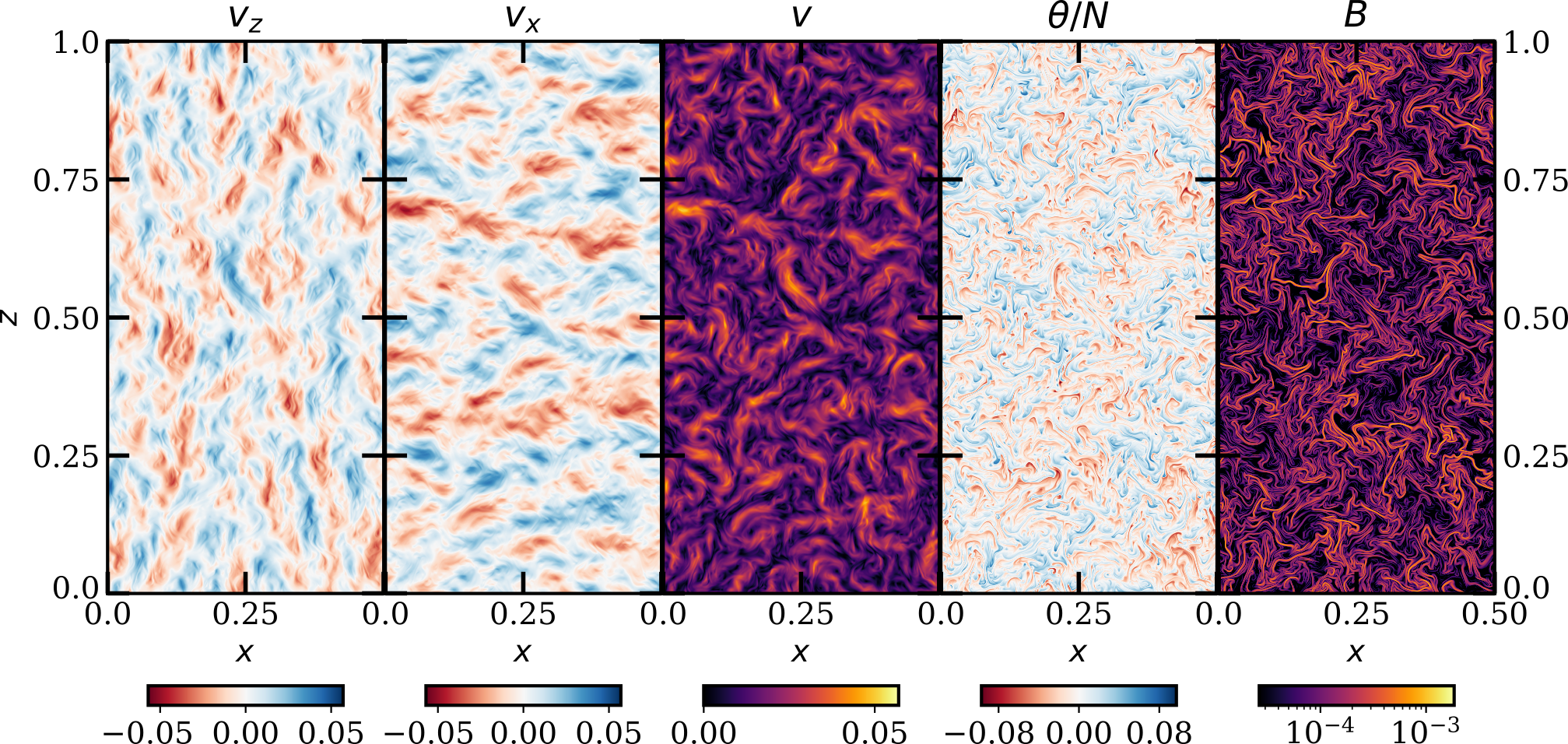}
	\caption{From left to right: vertical velocity, horizontal velocity, magnitude of the velocity vector $v = (v_x^2 + v_z^2)^{1/2}$, density fluctuations and amplitude of the magnetic field $B = (B_x^2 + B_z^2)^{1/2}$. In all five panels, only the left half of the plane is shown. 
	}
	\label{fig:2D_velocity_density}
\end{figure*}

\section{Numerical Results}\label{sec:numerical_results}

\subsection{Onset of MTI}\label{sec:onset_2D_mti}

We first simulate the MTI's exponential growth phase so as to verify our implementation of the anisotropic thermal diffusion term in the SNOOPY code. We seed the exact eigenmodes of the MTI dispersion relation from Eq.~\eqref{eq:mti_eigenmode} with $k_z = 0$ and $k_x=\left\{2,8,16,20,32,64\right\} \times 2 \pi /L$, and we track the time evolution of the kinetic energy over $\sim 2-3$ e-folding times. We then extract the growth rates of the instability through a standard best-fit. All the runs have been performed with a resolution of $(1024)^2$, and with the same physical parameters as in Fig.~\ref{fig:mti_disp_rel}, where we show the numerical growth rates overlaid on the theoretical dispersion relation. Note that the simulations here explicitly integrated the anisotropic term and that all the growth rates shown in Fig.~\ref{fig:mti_disp_rel} have a relative error compared to the theoretical value of $\sim 10^{-5}$.

Having verified that the code reproduces the linear theory, we turn to the estimation of the saturation timescale in Section~\ref{sec:nonlin_MTI}. To compare the nonlinear and the parasitic timescales, we perform a simulation where we add a sinusoidal perturbation to the horizontal velocity on top of an MTI eigenmode: the horizontal perturbation can then act as a seed for the parasitic Kelvin-Helmholtz instability. In Fig.~\ref{fig:parasitic_KH_Lambert} we follow the time evolution of the vertical kinetic energy and the horizontal kinetic energy: as the dominant MTI mode has no horizontal motion, any growth in the horizontal kinetic energy  can then be attributed to the KH parasite. For comparison, the numerical estimates of the nonlinear and KH timescales are represented by the red and grey vertical bars, respectively, using Eqs~\eqref{eq:nonlin_timescale}--\eqref{eq:KH_timescale}. As is clear, while the MTI mode develops, the KH perturbation grows in a super-exponential fashion in accord with the predictions of  Section~\ref{sec:parasitic_instabilities}. However,  by the time the KH perturbation grows to levels comparable to those of the MTI mode, the latter has already entered the nonlinear phase where it continues to grow, albeit at a reduced rate. This confirms that the intrinsic nonlinear timescale of the MTI is indeed shorter than the parasitic timescale, but both are comparable in magnitude. 

Once the amplitude of the parasite mode becomes comparable to the growing MTI mode, the initially ordered, one-dimensional structure of the MTI  breaks down and the system enters a disordered state, which we discuss in the next section. This demonstrates that it is via parasitic instability that the initially one-dimensional MTI generates two-dimensional structure, and ultimately terminates its growth phase.

\subsection{Basic Properties of the Saturated State}

We now choose a fiducial run, which we call R0, and follow its evolution deep into the saturated regime. Run R0 is characterized by $\mathrm{Pe}=6\times 10^3$ and low resistivity and viscosity, with $q=\mathrm{Pr}=0.006$. It has a uniform horizontal magnetic field with strength $B_0 = 10^{-5}$. Further details of the fiducial run are summarized in Tab.~\ref{tab:table_runs}. 

 \subsubsection{Energy evolution}
 
As we can observe in Figs.~\ref{fig:2D_kin_pot_energy}-\ref{fig:2D_mag_energy}, after the initial exponential growth, the kinetic and potential energy level off and the system enters a saturated state. The velocities and the magnetic field are effectively isotropized, with approximate equipartition of energy between their vertical and the horizontal components. This is also visible in Fig.~\ref{fig:2D_mag_energy}, where in the right panel we plot the average vertical component of the unit vector $\bs b$, whose saturated value is close to that expected for isotropic turbulence $\lvert b_{z,iso}\rvert  \approx 0.636$. Because of the imposed net flux, the magnetic energy does not decay with time, but is instead set by a balance between resistive dissipation and the continuous stretching of the large-scale field by the turbulent motions (see discussion at the end of Section \ref{sec:theoretical-scalings}).

\subsubsection{Global Energy Balance}

In Sec.~\ref{sec:model} we argued that in the low Prandtl number, weak-field regime the energy input should be roughly balanced by the energy dissipated thermally. This allowed us to estimate the total energy injection rate at saturation as $\epsilon_I \sim \chi \omega_T^4/N^2$.
To confirm this we plot in Fig.~\ref{fig:en_fluxes} the volume-averaged energy fluxes for run R0 and their balance at saturation. The lower panel reveals that during the initial growth phase energy is added to the system at an increasing rate, peaking when the system enters the nonlinear phase, after which the net energy input rate $d E / dt$ decreases to zero. In the top panel, we find that practically all of the energy injected through the MTI forcing term is dissipated thermally, as viscous dissipation is about two orders of magnitude lower. Furthermore, we can see that the energy injection rate at saturation is consistent with Eq.~\eqref{eq:energy_input_mti}, as $\epsilon_I \approx 0.3 \chi \omega_T^4/N^2$.

\subsubsection{Flow field}

To get a sense of the MTI's saturated state, we show in Fig.~\ref{fig:2D_velocity_density} a representative snapshot of the velocity field and of the density fluctuations of run R0. From visual inspection, we can see that the MTI is characterized by eddies/structures exhibiting a range of lengths less than the box size. Whereas the vertical and horizontal velocity fields express an alignment in the vertical and horizontal direction, respectively, the total magnitude of the velocity fluctuations (middle panel) are approximately isotropic. 

The density fluctuations (fourth panel) peak at a much smaller scale and are elongated in the direction parallel to the local magnetic field (fifth panel), with steep gradients perpendicular to it. Their morphology is thus controlled by two different lengthscales: first, anisotropic conduction isothermalises the gas only along magnetic field lines, with the size of the resulting elongated structures set by a balance between the conduction time and the local eddy turnover time; second, perpendicular to the local magnetic field, the structures' length-scale is determined by the reconnection of adjacent magnetic field lines, which results in the merging of density fluctuations, and thus depends on the magnitude of resistivity.

\subsubsection{Heat Fluxes}

\begin{figure}
	\centering
	\includegraphics[width=1.0\columnwidth]{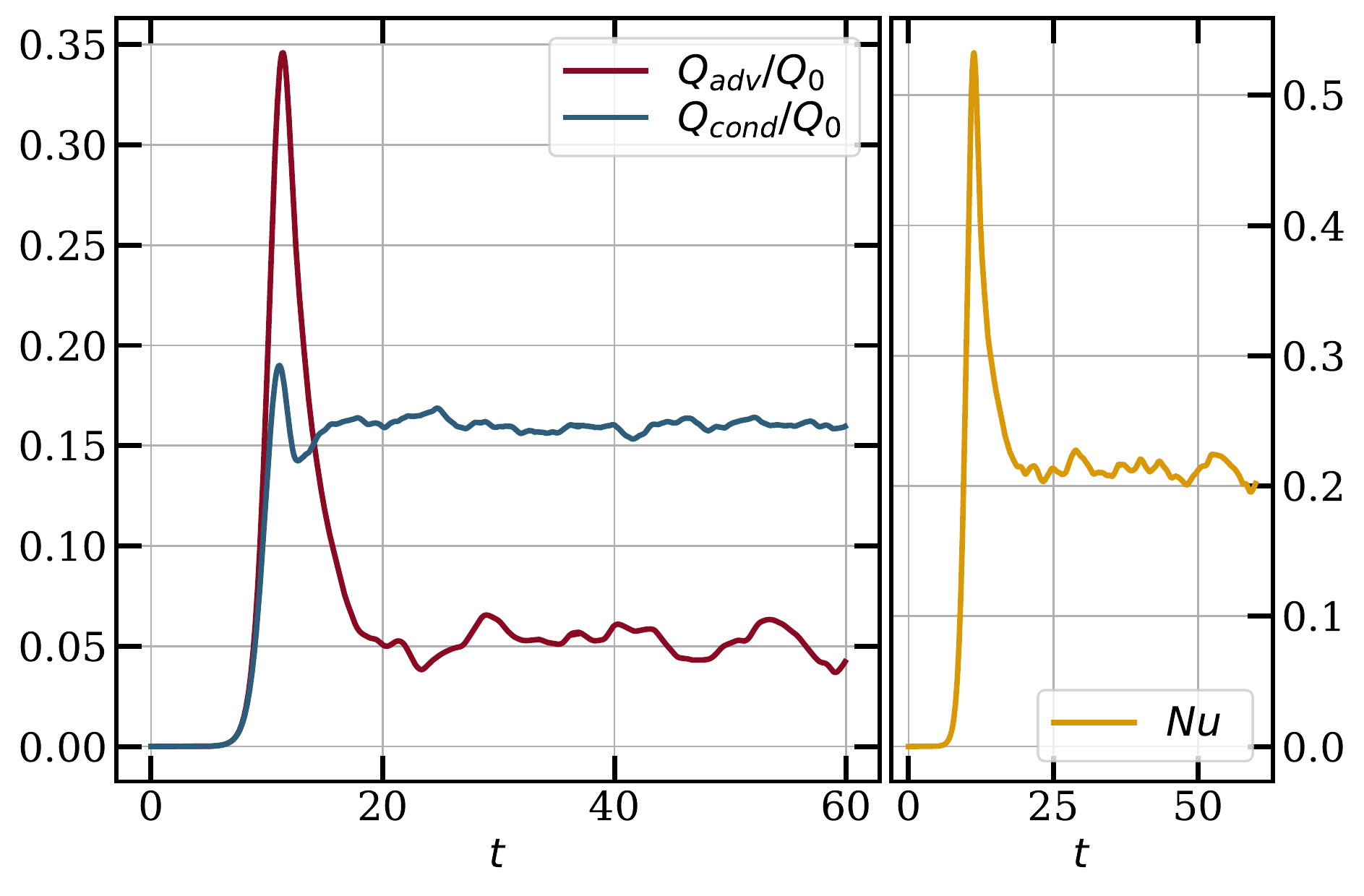}
	\caption{Left panel: advective and conductive heat fluxes normalized by the maximum vertical flux $Q_0= \chi \omega_T^2$. Right panel: Nusselt number. While heat is mostly carried by advection during the late stages of the exponential growth phase, eventually conduction takes over once the system reaches saturation.}
	\label{fig:2D_nusselt}
\end{figure}

In Fig.~\ref{fig:2D_nusselt} we show the time evolution of the advective flux and of the conductive flux for run R0, normalized by $Q_0$. As we can observe from the positivity of $Q_{adv}$, the density fluctuations are anti-correlated with the vertical velocity fluctuations on average. Thus the MTI acts likes standard convection:  lighter (and hotter) fluid elements buoyantly rise, while denser (and colder) fluid elements sink. At saturation, however, the majority of heat is transported by thermal conduction, with turbulent advection making up about $30\%$ of the conductive flux.

In Fig.~\ref{fig:2D_nusselt}, we also plot the Nusselt number, which measures the total efficiency of heat transport relative to the maximum rate possible. As is clear, the MTI drives a substantial fraction (about $20\%$) of the maximum heat flux across the domain $Q_0$, despite the highly tangled magnetic field geometry.

\subsubsection{Power Spectra}
\label{powerspectra}

\begin{figure}
	\centering
	\includegraphics[width=0.9\columnwidth]{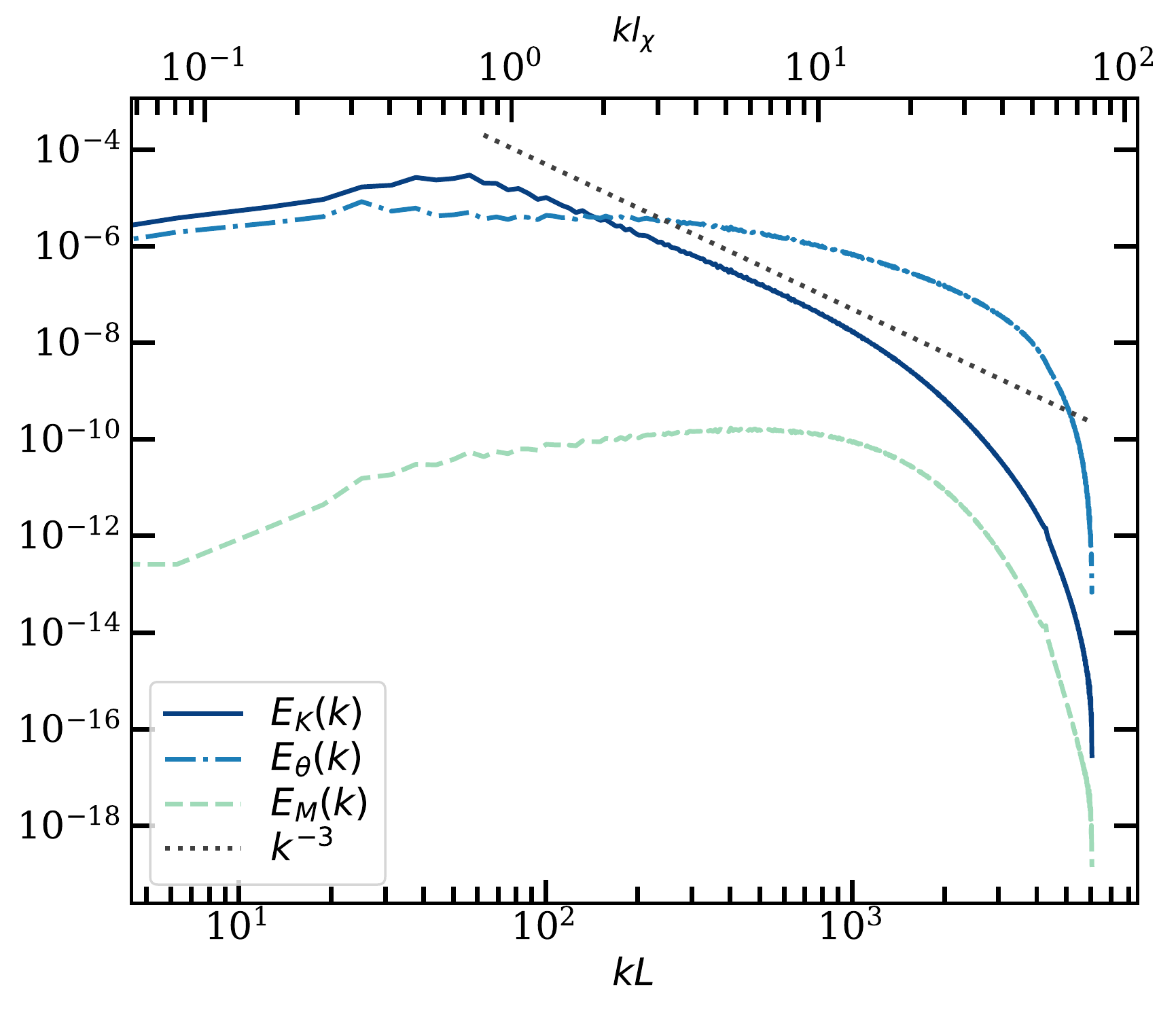}
	\caption{Power spectrum of the kinetic (solid line), density (dash-dotted line), and magnetic (dashed line) fluctuations for run R0 at saturation. For comparison, we also plot the $k^{-3}$ power-law decay in the dotted line. Note that in the bottom and top $x$-axis the wavenumbers $k$ are normalized with respect to the box size $L$ and to the conduction length, respectively.   } %
	\label{fig:2D_energy_spectra}
\end{figure}

The differences between the velocity and density fluctuations in the snapshots are also visible in the corresponding shell-integrated energy spectra, which we plot in Fig.~\ref{fig:2D_energy_spectra}. The two spectra display contrasting features in $k$ space: the kinetic energy spectrum presents a peak at a relatively large scale at around $kL\approx 60-70$, or $k l_{\chi} \approx 1$ (where $l_{\chi}=\sqrt{\chi/\omega_T}$), which is followed by a steep power-law decay towards larger wavenumbers; the potential energy spectrum, on the other hand, is characterized by a plateau extending from low to high wavenumbers (ending at $kL \approx 300-400$), beyond which it gradually decreases until the dissipative cutoff. Large scales are dominated by kinetic fluctuations, while the reverse is true at smaller scales, confirming the visual impression given by Fig.~\ref{fig:2D_velocity_density}. Nevertheless, at the very smallest wavenumbers (largest scales) the kinetic and potential energies are roughly comparable, which we interpret as a sign of g-modes because they obey $u^2 \sim \theta^2 / N^2$. This estimate can be obtained either from linear theory, or by balancing the large-scale eddy turnover time with the typical gravity wave timescale. We explore g-modes in more detail in Section \ref{sec:g-modes}.

The kinetic spectral energy density in Fig.~\ref{fig:2D_energy_spectra} is characterized by an inertial range at intermediate wavenumbers, with a slope that follows rather closely $E_K \sim k^{-3}$ (black dotted line). In incompressible, two-dimensional turbulence, this slope \citep[with logarithmic corrections, see:][]{Kraichnan1971}  is indicative of a direct enstrophy cascade between the forcing scale and the viscous cutoff \citep{Kraichnan1967}. Classically, 
 a stationary state can only be reached if there is an energy sink at large scales. The origin of this sink, or drag, in our simulations was touched on in Sec.~\ref{sec:theoretical-scalings} and will be taken up in detail in Sec.~\ref{sec:detailed_balance}. But such a drag force will modify the enstrophy cascade at small scales also, leading to a steepening of the $k^{-3}$ slope \citep{Nam2000,Boffetta2005}, as we observe.

\begin{figure*}
		\begin{subfigure}{.5\linewidth}
		\centering
		\includegraphics[width=0.9\columnwidth]{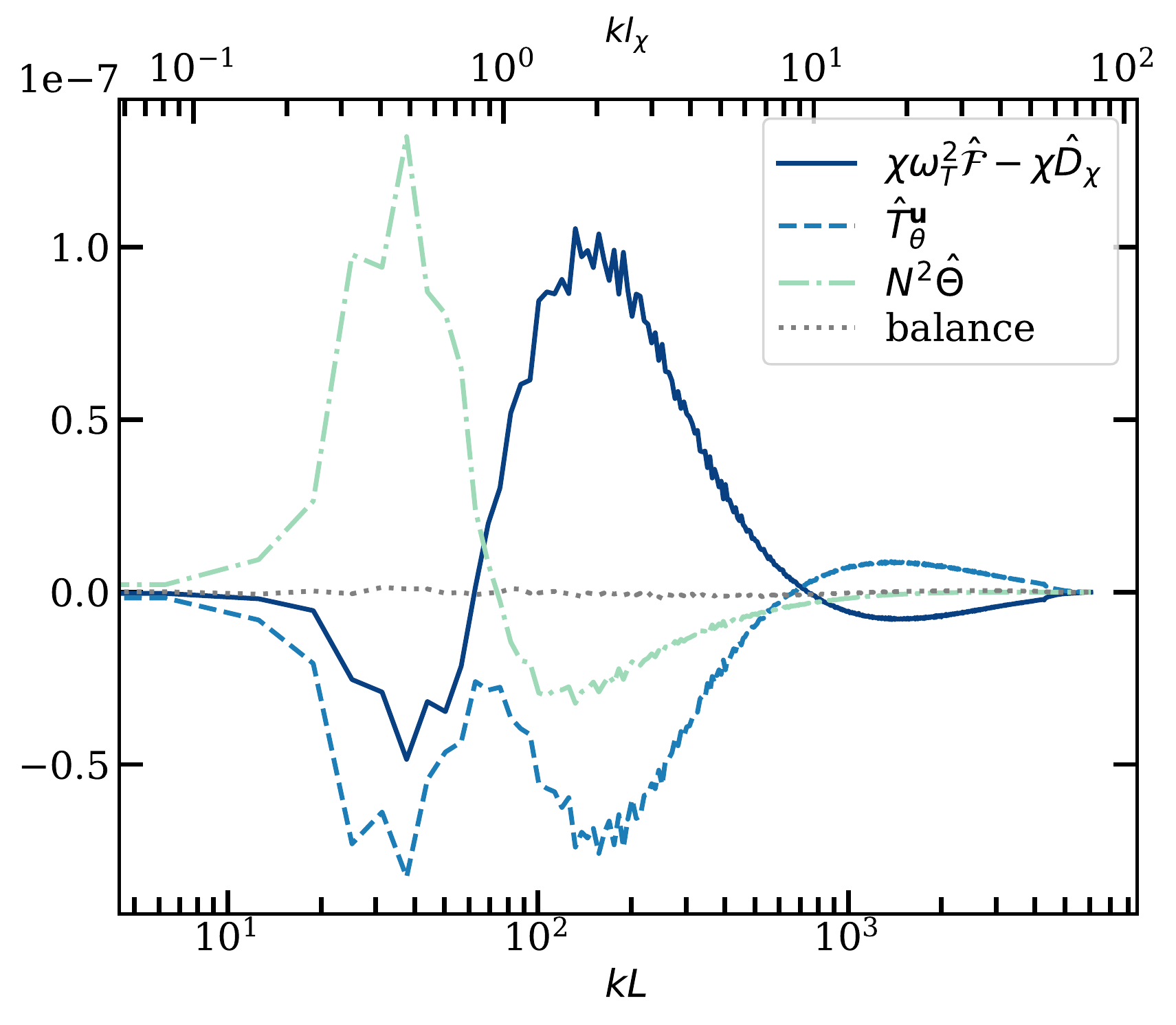}
		\caption{}
		\label{fig:2D_thermal_detail_energy}
	\end{subfigure}%
	\begin{subfigure}{.5\linewidth}
		\centering
		\includegraphics[width=0.9\columnwidth]{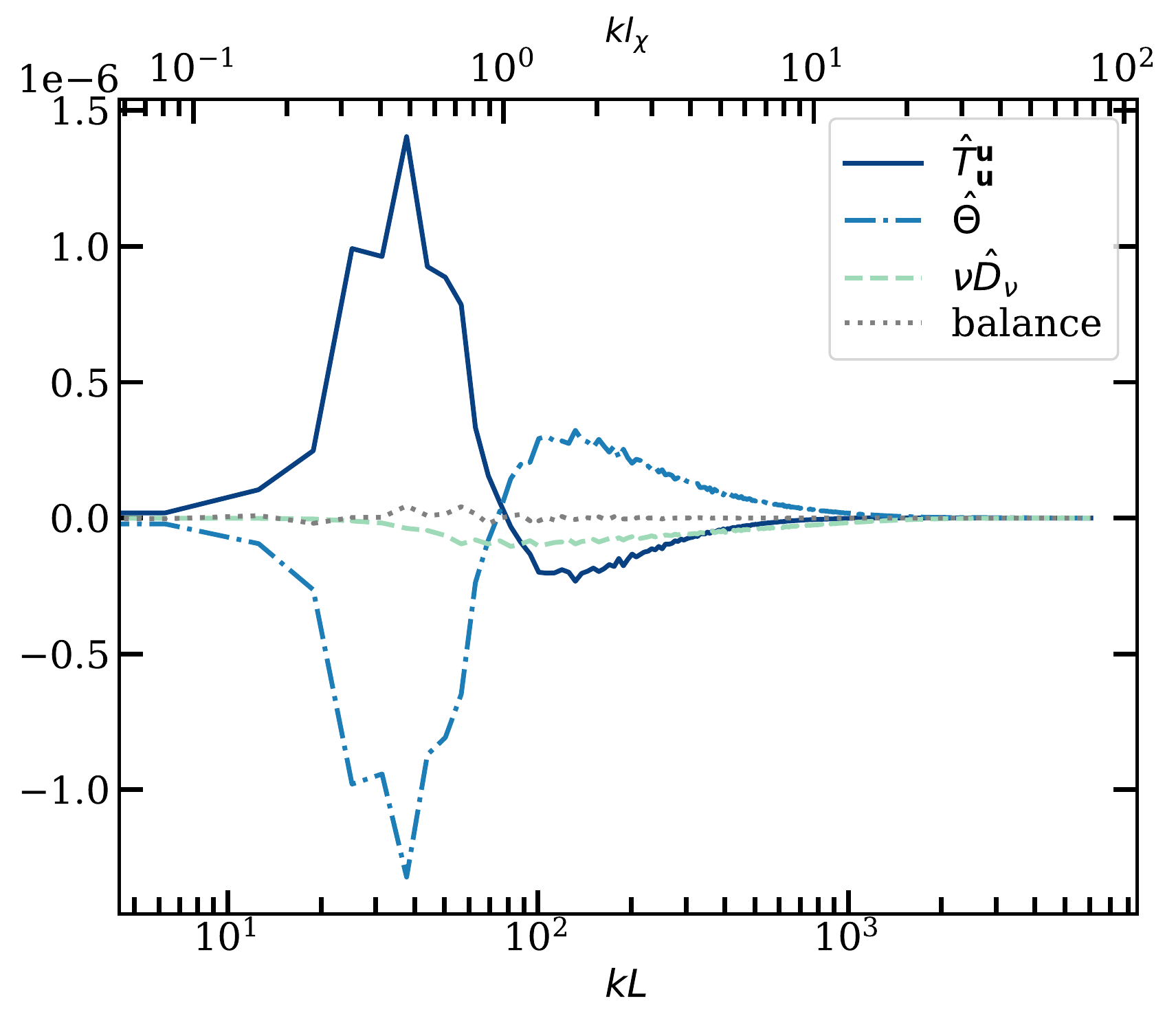}
		\caption{}
		\label{fig:2D_kin_detail_energy_loop}
	\end{subfigure}%
	\caption{Detailed thermal (left) and kinetic (right) energy balance of run R0 at saturation. The various terms in the balances are described in the legend.
	The dotted line in both panels represents the sum of these terms. 
	A large number of samples is necessary in order to obtain converged results and thus in both panels the spectra have been averaged over a time interval of $\Delta t = 100$ for a total of $10^4$ samples.
	}
	\label{fig:Pe3e3_N21e-1}
\end{figure*}

\subsection{Saturation Mechanism}
\label{sec:detailed_balance}

\subsubsection{The Energy `Flux Loop'} 

As is well established, in unstratified, 2D turbulence the conservation of two inviscid invariants --  the kinetic energy and the enstrophy (the volume integral of the squared vorticity) -- precipitates a direct cascade of enstrophy towards smaller scales and an inverse cascade of energy to large scales \citep{Kraichnan1967,Batchelor1969,Kraichnan1971,Boffetta2012}. As a result, energy accumulates at the largest scales (lowest wavenumbers) and, as mentioned earlier, a quasi-stationary state can only be reached in the presence of an energy sink on these scales, often modelled by an ad hoc drag term. This approach is typical of the atmospheric and oceanographic sciences, where a drag force stands in for the effects of mountains or a rough seabed \citep{Salmon1998}.

In stably stratified 2D turbulence, however, enstrophy is no longer conserved and the velocity field is coupled to the evolution of the buoyancy variable \citep{Bolgiano1959,Obukhov1959}. Buoyancy leaves small scales alone for the most part, and instead works most effectively on long scales: below a critical large scale (often taken to be the Ozmidov scale), buoyancy is negligible, and there can exist the usual inertial range of forced 2D turbulence; but above this critical scale, buoyancy holds sway and fluid motions are constrained to undergo buoyancy oscillations -- thus g-modes predominate, possibly excited by the said turbulence.
Moreover, buoyancy can arrest the inverse energy cascade at this critical scale, by imposing an effective energy sink or drag. More precisely, it establishes an energy \textit{flux-loop}: kinetic energy is sent up to large-scales by the inverse cascade; it is then converted into density fluctuations via the excitation of g-modes\footnote{This is similar to the excitation of Rossby waves in turbulent rotating shallow flows. There the Rhines length defines the transition between regimes in which turbulence or Rossby waves dominate \citep{Rhines1975}.}; and finally these fluctuations are nonlinearly advected back down to smaller scales, where they are dissipated \citep{Boffetta2011}.

Our 2D MTI simulations exhibit similar behaviour. The stable entropy stratification splits the $k$ space into two distinct ranges: (a) small scales, characterised by injection of energy via the MTI, the absence of g-modes, and a turbulent inverse cascade that transports kinetic energy up to (b) the large scales, upon which g-modes are excited and by absorbing kinetic energy acts as a sink, independent of the box size.

\subsubsection{Scale-by-scale Energy Balance}

In order to verify this picture of 2D MTI saturation, we track how energy is transferred at each scale through the scale-by-scale  energy balances of Eqs.~\eqref{eq:spectral_vel}-\eqref{eq:spectral_temp}. We plot the various component terms of these balances in Fig.~\ref{fig:Pe3e3_N21e-1} using data from run R0.

Turning to the thermal energy balance in Fig.~\ref{fig:2D_thermal_detail_energy} first, we find the net effect of the forcing term and of the anisotropic conduction (solid blue curve) is to inject energy into density modes at large wavenumbers $\gtrsim 10^2$ (near the maximum wavenumber of linear theory $k_{max}L\approx 220$). The action of these fluctuations is to alleviate the background temperature gradient, and drive the system toward isothermality (cf. Section \ref{sec:isothermality}). On the same small scales, the buoyancy term (dot-dashed light green) is negative, and thus transfers some of this energy into velocity fluctuations, which will subsequently be sent to larger scales by an inverse cascade (see below). On large scales, however, this is all reversed: the buoyancy term (dot-dashed light green), is positive and thus returns this energy to the density fluctuations, via g-mode excitation;
conduction and forcing (solid blue), on the other hand, act as an energy sink, damping the g-modes, in agreement with the strong damping rates derived from linear theory, see Eq.~\eqref{eq:g-modes-disp-rel}.
Both the anisotropic net energy injection and the buoyancy curves switch sign at roughly the same wavenumber $kL \approx 60-70$ (or $k l_{\chi} \approx 1$), which thus functions similarly to the Ozmidov scale in stratified hydrodynamic turbulence, and indeed we associate this transition scale with the buoyancy scale introduced in Section~\ref{sec:isothermality}. 
Finally, the advection term $\hat{T}_{\theta}^{\bs u}$ (dashed blue line) redistributes energy towards the smaller scales, and shows two distinct peaks where the buoyancy flux and the net forcing are maximum. 

The kinetic energy balance appears in Fig.~\ref{fig:2D_kin_detail_energy_loop}, and it too shows a distinct demarcation between large-scales and small-scales. In particular, we can recognize the shape of an "$\infty$" symbol (hence the label \textit{double-loop}): on smaller scales, the buoyancy term (dot-dashed blue curve) injects energy at a wavenumber comparable to -- but slightly smaller than -- the peak $k$ of the thermal forcing, while the negative nonlinear transfer term $\hat{T}_{\bs u}^{\bs u}$ (solid dark blue curve) sends that energy to larger scales via an inverse cascade; beyond a critical $k$ 
this energy is deposited (positive $\hat{T}_{\bs u}^{\bs u}$) and then converted back into thermal fluctuations via the negative buoyancy term, closing the loop.
As a final remark, we note that viscous dissipation removes only a small fraction of energy at all scales, and is thus a subdominant process, as is reasonable to expect at low $\mathrm{Pr}$.

\subsubsection{Large-Scale g-mode Excitation}\label{sec:g-modes}

A key ingredient of the flux-loop mechanism outlined above is the presence of large-scale g-modes
whose existence can only be inferred indirectly from the detailed energy balance. In this short subsection, we show direct evidence that these oscillations are present and dominant at large scales. 

\begin{figure}
	\centering
	\includegraphics[width=0.9\columnwidth]{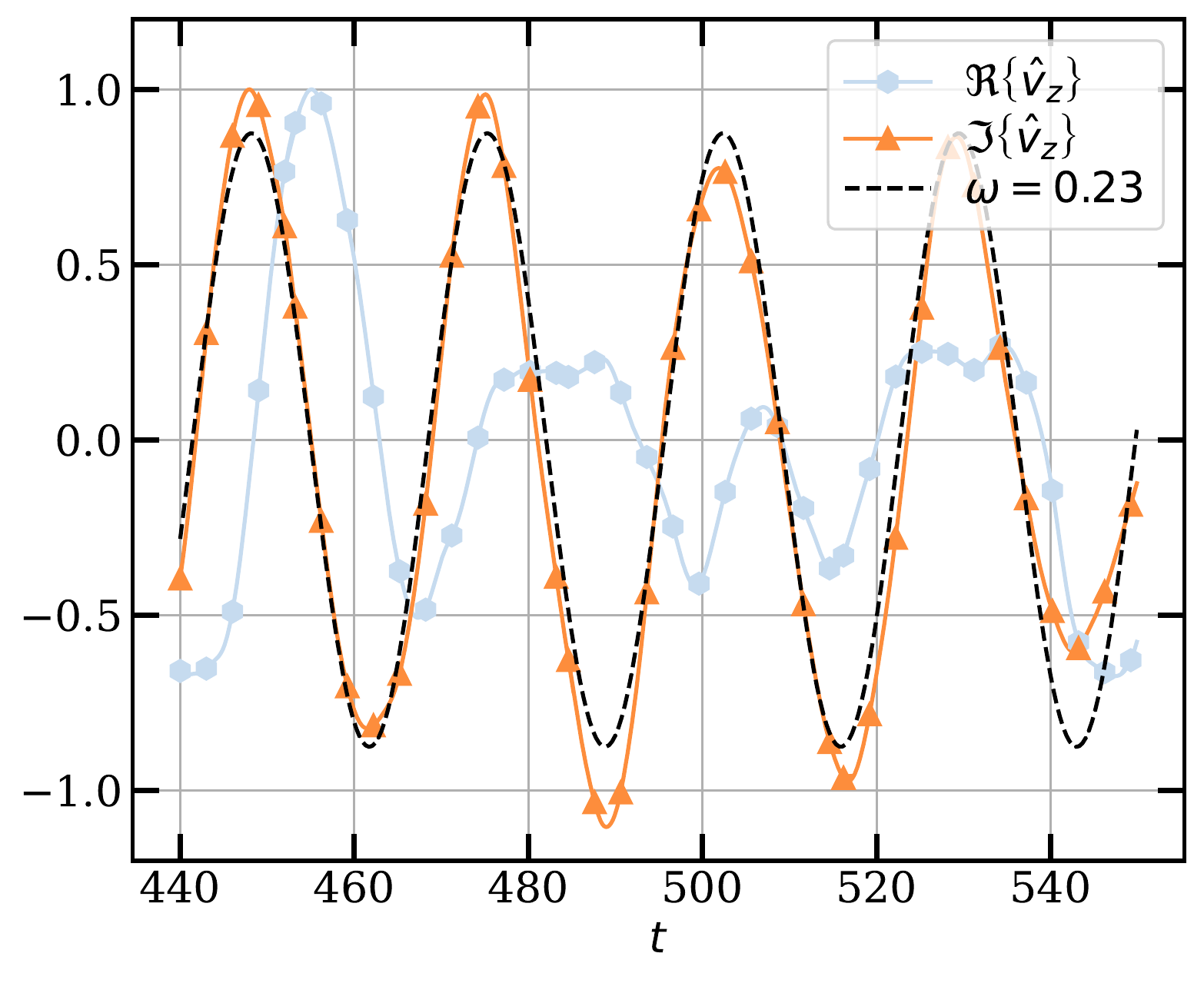}
	\caption{Large-scale g-mode excited at saturation for run R0. In the plot, we show the real and imaginary part of the $(1,1)$ mode in $v_{z}$. The dashed line is a best-fit of the imaginary part of the mode.}
	\label{fig:2D_gmodes}
\end{figure}

We plot in Fig.~\ref{fig:2D_gmodes} the time evolution of a large-scale Fourier mode $(n_x,n_z) = (1,1)$ of the vertical velocity taken from run R0. It thus possesses $k_x=k_z=2\pi/L$. We can see that the real and imaginary parts of the complex amplitude behave like a forced harmonic oscillator, with a time-dependent amplitude and with a frequency $\omega=0.23$. This is close to that expected from the dispersion relation of g-modes, namely $\omega= N/\sqrt{2}\approx 0.22$, and thus cements the identification. Similarly, we find good agreement between the theoretical g-modes frequency and the oscillation frequency of other large-scale Fourier modes, e.g. $(1,2),(2,1),(2,2),(3,1)$ and $(1,3)$. 
\subsubsection{Weakly sub-adiabatic cases}

The claim that it is buoyancy that arrests the inverse cascade is further supported by the results of a simulation with very weak entropy stratification $\tilde{N}^2 = 10^{-4}$ (and with otherwise the same numerical parameters of run R0), which we dub R0Nlow. In this run, after the MTI growth phase, the kinetic energy keeps growing with time without ever reaching saturation. Moreover, looking at the power spectrum, we notice a pile-up of energy at the lowest wavenumbers and the manifestation of coherent flow structure on the box scale. This striking behaviour is a consequence of the very low $\tilde{ N}$, which pushes the buoyancy scale beyond the size of the box; in other words, the entropy stratification is too weak to arrest the inverse cascade. We find analogous results for simulations where $\tilde{ N}$ is exactly zero.

\subsection{Dependence on Physical Parameters, and Scaling Laws}

To understand how the saturated state depends on thermal conductivity and stratification we perform a parameter sweep over a wide range of values of $\mathrm{Pe}$ (series $RsP$) and $\tilde{N}^2$ (series $RsN$). Our focus is on box-averaged quantities and on spectra. These are calculated over a time window of $200$ dynamical times so that our results are well-converged. The parameters and the results of the runs discussed in this section are summarized in Table~\ref{tab:table_runs}.

\subsubsection{Energetics}

\begin{figure}
	\centering
	\includegraphics[width=1.0\columnwidth]{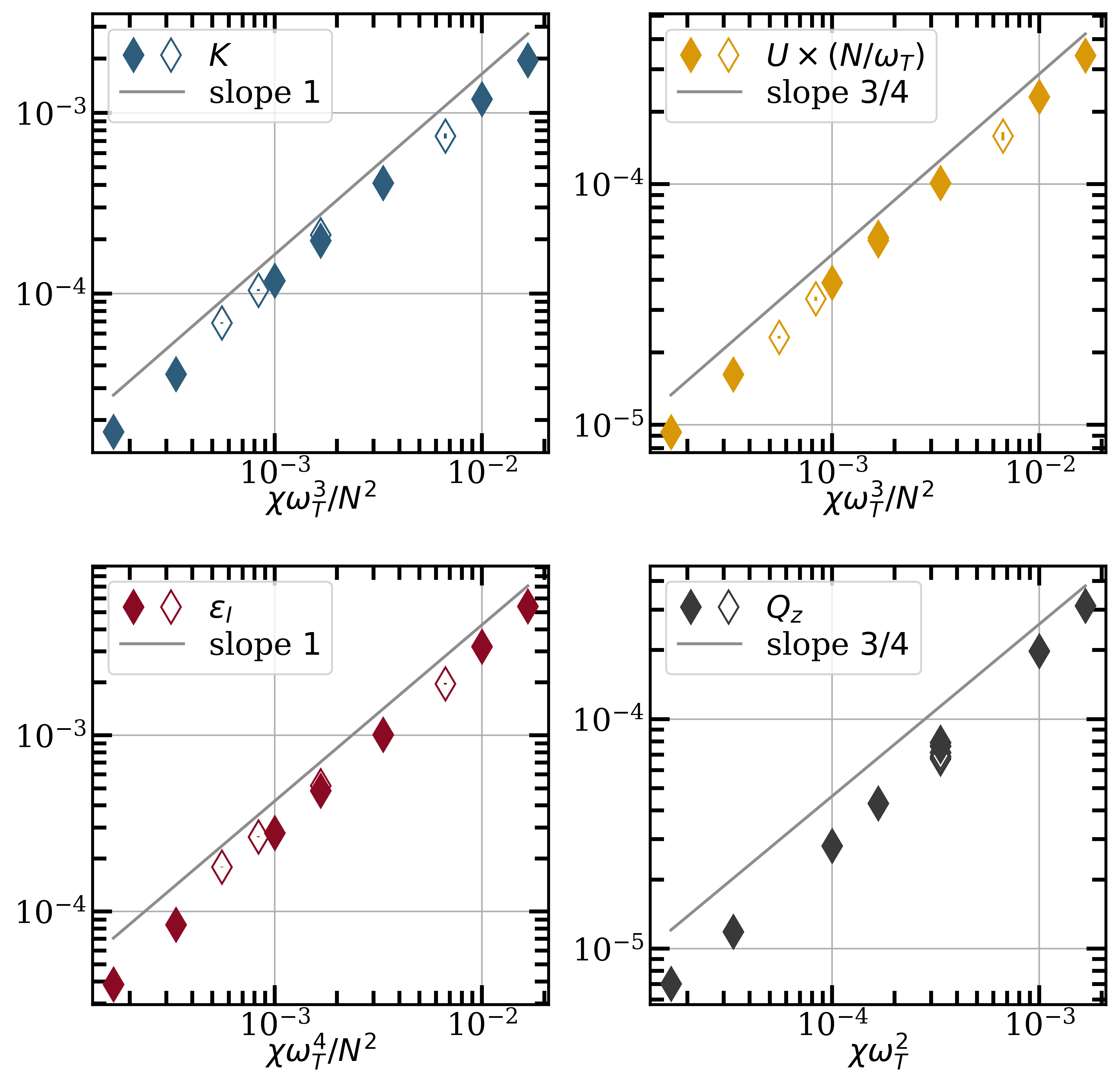}
	\caption{Plot of the turbulent quantities of the MTI at saturation for runs $RsP$ (full diamonds) and $RsN$ (empty diamonds). Top left: kinetic energy; top right: potential energy; bottom left: energy injection rate; bottom right: vertical heat flux. The grey solid lines are plotted for reference and, for the kinetic energy and the injection rate, represent the theoretical scalings derived in Section \ref{sec:theoretical-scalings}. Note that in the plot of $Q_z$ the empty diamonds collapse on top of each other.}
	\label{fig:2D_energetics_scalings}
\end{figure}

We plot in Fig.~\ref{fig:2D_energetics_scalings} the kinetic and potential energies at saturation (top left and top right panel, respectively), as well as the energy injection rate and the vertical heat flux (bottom left and bottom right panels) for runs $RsP$ and $RsN$. The most striking feature is that these mean quantities clearly possess power law dependencies on the thermal conductivity (through $\mathrm{Pe}$) and the entropy stratification ($\tilde{N}^2$), which we trace out with solid grey lines.
In particular, the saturated kinetic and potential energies increase as $\chi$ increases ($\mathrm{Pe}$ decreases), but decrease with increasing stratification $\tilde{N}^2$. By performing a best-fit of the kinetic energy as a function of the dimensionless combination $\mathrm{Pe}^{-1} \tilde{N}^{-2}$, we find that the variation of $K$ is consistent with a power-law exponent of $1$ -- which in dimensional units implies $K \sim \chi \omega_T^3 N^{-2}$, in agreement with our prediction in Section \ref{sec:theoretical-scalings}. On the other hand, the potential energy scales roughly as $\sim \mathrm{Pe}^{-3/4} \tilde{N}^{-5/2}$, and thus $U \propto \chi^{3/4}  N^{-5/2}$ in dimensional units. On dimensional grounds, it is clear at least one other physical process must be controlling the saturation of the thermal energy, such as resistivity or viscosity, but the dependency should be weak.

Moving on to the energy injection rate $\epsilon_I$, we also see both run series $RsP$ and $RsN$ manifest a scaling consistent with the theoretical prediction in Eq.~\eqref{eq:energy_input_mti}. The vertical heat flux also increases with increasing $\chi$, albeit at a shallower rate, while it is largely unaffected by changes in the entropy stratification. Finally, we confirm that the magnetic energy follows the approximate scaling $M \sim (\chi / \eta) (\omega_T / N)^2 B_0^2$ (not plotted), derived in Eq.~\eqref{eq:mag_energy_saturation}.

\subsubsection{Buoyancy Scale}

\begin{figure}
	\includegraphics[width=1.0\columnwidth]{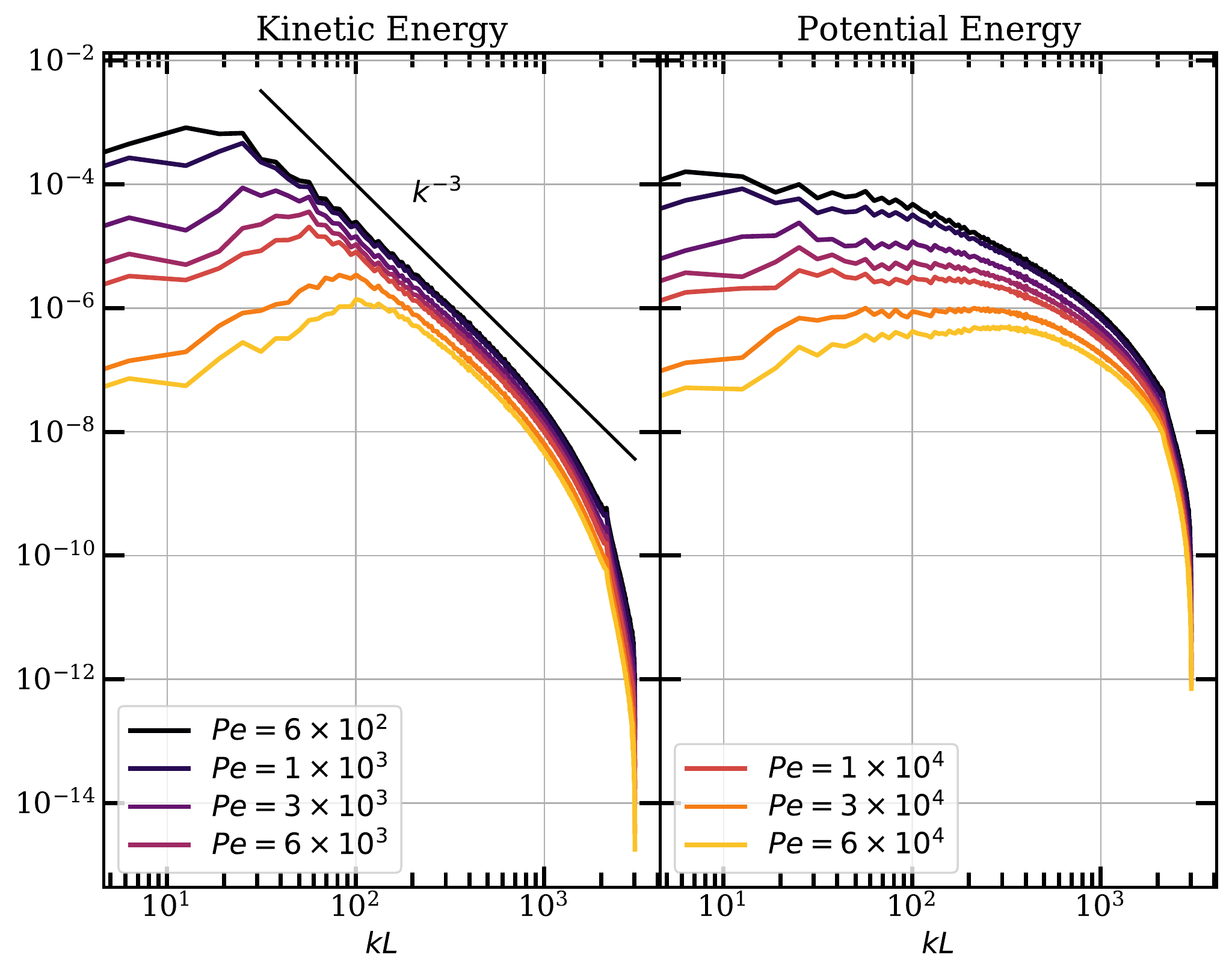}
\includegraphics[width=1.0\columnwidth]{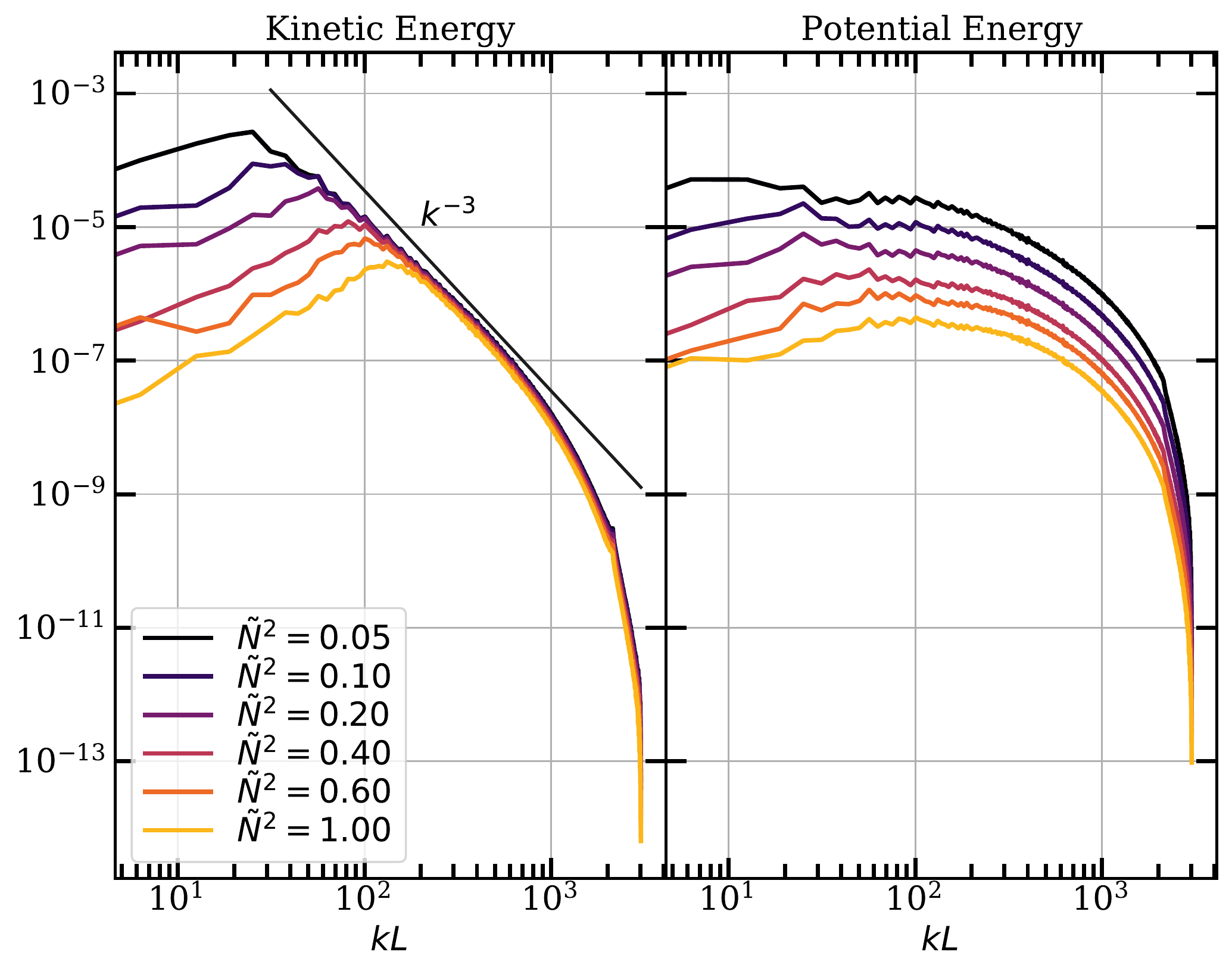}
\caption{Kinetic energy (left) and potential energy (right) power spectra at saturation for various $\mathrm{Pe}$ (top panels) and $\tilde{N}$ (bottom panels).}
\label{fig:2D_all_kin_th_spectrum}
\end{figure}

The dependency of the turbulent energies on $\mathrm{Pe}$ and $\tilde{ N}$ is tied to significant variation in the characteristic scales of the turbulence across the different runs. In Fig.~\ref{fig:2D_all_kin_th_spectrum} we plot the energy spectra for runs $RsP$ and $RsN$, respectively. We observe that the peak of the spectrum varies with the Peclet number and with the entropy stratification, moving towards lower wavenumbers as both $\mathrm{Pe}$ and $\tilde{ N}$ decrease: in other words, larger turbulent eddies result from greater values of thermal conductivity and for weaker entropy stratifications. The bottom left panel of Fig.~\ref{fig:2D_all_kin_th_spectrum} clearly demonstrates that on decreasing $\tilde{N}$ the kinetic energy peak moves to longer scales, while the spectrum at large wavenumbers is left practically unchanged. Meanwhile, in the top left panel, we find that increasing $\mathrm{Pe}$ drives the spectrum to shorter scales.

Contrary to the kinetic power spectra, those of the potential energy fail to demonstrate as clear a variation as the parameters vary (cf. right panels in Fig.~\ref{fig:2D_all_kin_th_spectrum}). Generally, they show larger amplitude density fluctuations at all wavelengths for lower values of $\mathrm{Pe}$ and of $\tilde{ N}$. This might be due to the fact that the MTI forcing term injects energy directly into the density fluctuations at intermediate-to-large scales, which dominate the dynamics and potentially explains the absence of an inertial range. 

The existence of a buoyancy scale, which can halt the inverse cascade of energy and prevent it from reaching the box scale is a remarkable property of 2D MTI. Its existence generally prohibits structures on the size of the box (e.g. elevator flows and vortices), unless the physical parameters of the simulation are such that the buoyancy (or integral) scale itself is larger than the box, such as in run R0Nlow, see Section~\ref{sec:g-modes}. To shed light on the physical mechanism that determines the buoyancy scale, we estimate numerically the wavenumber at which the kinetic power spectrum peaks, which we denote $k_{i}$, as follows:
\begin{equation}
k_{i}^{-1} = \frac{\int_{k_{min}}^{k_{max}} k^{-1} E_k d k}{\int_{k_{min}}^{k_{max}}  E_k d k},
\end{equation}
where the integrals are computed using the method of quadratures, and where the limits of integration are $k_{min} = 2 \pi /L$ and  $k_{max} = 2 \pi \times (2 \sqrt{2} N/3) /L$ (which is the largest, non-dealiased wavenumber of the 2D domain). Strictly speaking $k_i$ is the integral scale, but we identify this also with our buoyancy scale. 

\begin{figure}
    \centering
    	\includegraphics[width=1.0\columnwidth]{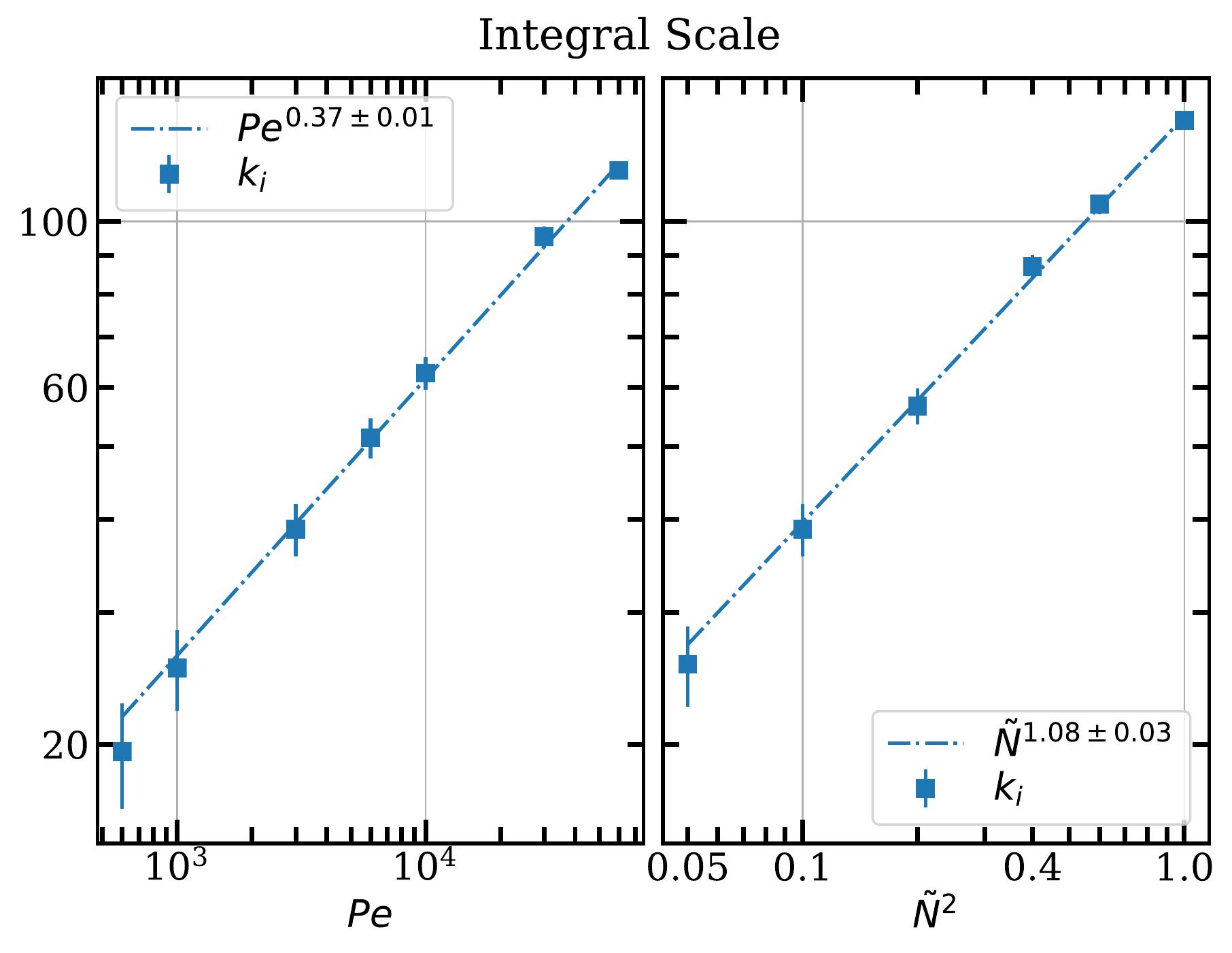}
    	\includegraphics[width=1.0\columnwidth]{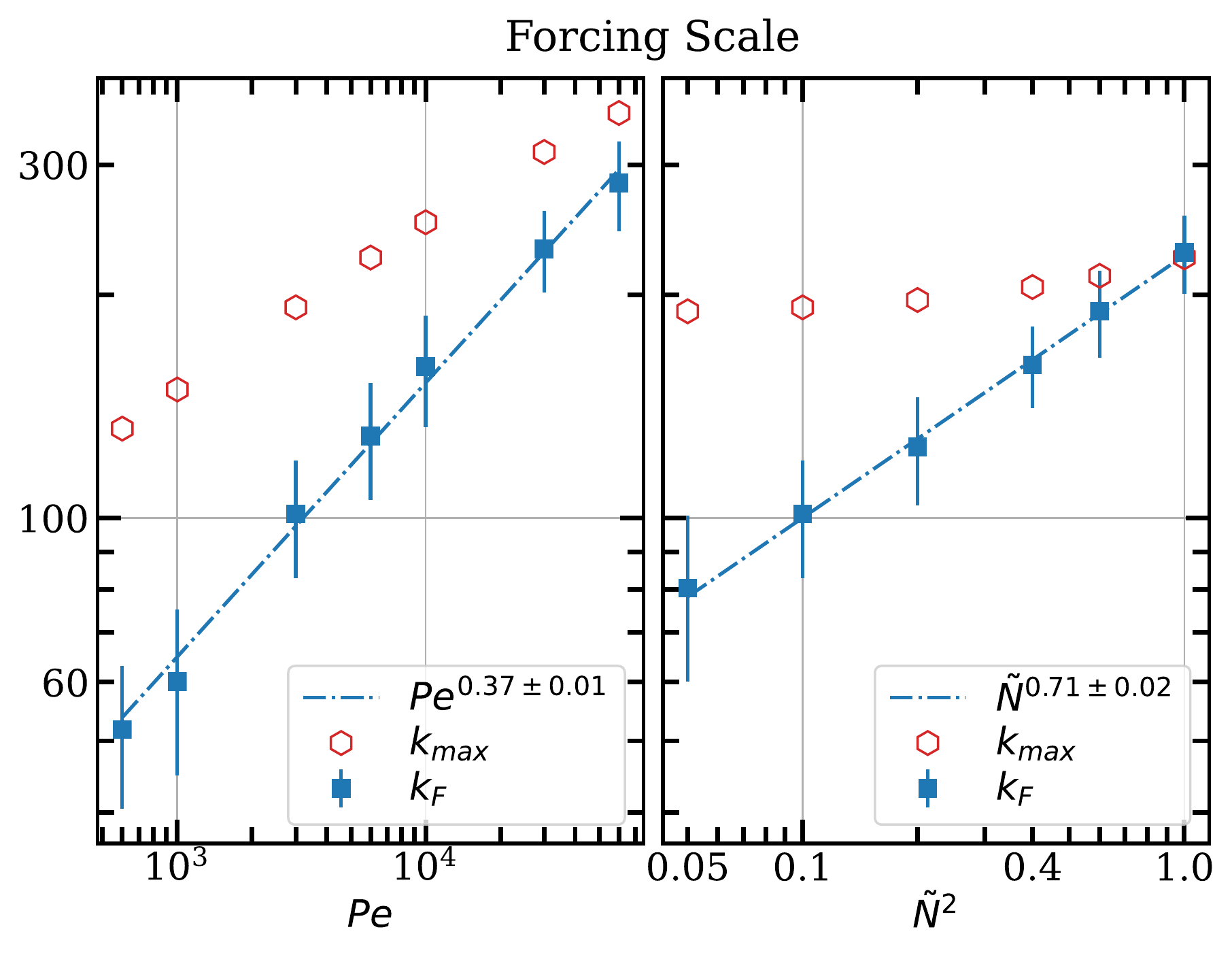}
    \caption{The buoyancy/integral (top) and forcing (bottom) wavenumbers, $k_i$ and $k_F$, as functions of $\mathrm{Pe}$ (left) and $\tilde{N}$ (right). In addition, we plot the wavenumber of maximum linear growth $k_{max}$ in the bottom panels, for comparison. }
    \label{fig:integralandforcing}
\end{figure}

We plot $k_i$ in the top two panels of Fig.~\ref{fig:integralandforcing} as a function of $\mathrm{Pe}$ and $\tilde{ N}$. Evidently, the buoyancy scale follows a power-law in both $\mathrm{Pe}$ and $\tilde{ N}$, and performing a best-fit we find that it scales as $k_i \sim \mathrm{Pe}^{0.37} \tilde{N}^{1.08}$. In comparison, the buoyancy scale proposed in Section \ref{sec:theoretical-scalings} yields a scaling $\sim \mathrm{Pe}^{0.5} \tilde{N}$. The dependency on the buoyancy frequency is in agreement, and the $\mathrm{Pe}$ is close, with the discrepancy arising from the residual influence of viscosity and resistivity. To check on this, we perform a series of runs where we keep $\mathrm{Pe}=10^3$ and $\tilde{ N}^2 = 0.1$ fixed, and vary either the resistivity or the viscosity, both over two orders of magnitude, in the range $\mathrm{Pr}=0.1-0.001$ and $q=0.1-0.001$ (but making sure to stay in the $\mathrm{Pr} \lesssim 1$, $q \lesssim 1$ regime), we find $k_B$ depends rather weakly on viscosity and even more so on resistivity, with a best-fit of $k_B$ on the Reynolds number ($=\mathrm{Pr}\, \mathrm{Pe} $) and on the magnetic Reynolds number ($=q\, \mathrm{Pe}$) returning slopes of $0.13(4)$ and $0.02(1)$, respectively. In dimensional units, this then yields roughly $k_B l_\chi \sim  \tilde{N}\,\mathrm{Pr}^{0.13} q^{0.02}$, which agrees with the theoretical prediction of Eq.~\eqref{bscale}, but with additional negligible dependencies on Pr and $q$.

\subsubsection{Forcing Scale}

We now focus our attention on the energy injected at the large wavenumbers, and how it varies with $\mathrm{Pe}$ and $\tilde{ N}$. In particular, we calculate the net input of energy, thus summing the forcing term $\chi \omega_T^2 \hat{\mathcal{F}}$ and of the dissipative term $\chi \hat{D}_{\chi}$ in the detailed energy balance of Eq.~\eqref{eq:spectral_temp}. 

We define the forcing scale $k_F$ to be the wavenumber for which the net input energy flux is largest, and we plot it in the bottom panels Fig.~\ref{fig:integralandforcing} as a function of $\mathrm{Pe}$ and $\tilde{ N}$. The forcing scale obeys a power-law with respect to these parameters, with energy being injected at larger wavenumbers as $\mathrm{Pe}$ and $\tilde{ N}$ increase. In Fig.~\ref{fig:integralandforcing} we also plot the wavenumber of maximum linear growth of the MTI, cf. Eq.~\eqref{eq:k_max_visc}. While the two estimates are of the same order of magnitude, their scaling laws differ, especially with respect to $\tilde{N}$. Evidently, at saturation the forcing mechanism does not simply follow from linear theory.

Similarly to the buoyancy scale, we find that $k_F$ depends very weakly on resistivity and viscosity, and a best-fit of $k_F$ on the Reynolds number and on the magnetic Reynolds number returns slopes of $0.06(3)$ and $0.05(2)$, respectively. Though weak, these scalings ensure that in dimensional units $k_F l_\chi \sim \tilde{N}^{7/10} \mathrm{Pr}^{0.06} q^{0.05}$, with a negligible dependence on $\mathrm{Pr}$ and $q$.

\subsection{Other Field Geometries}

All the simulations presented so far were initialized with a weak, uniform magnetic field aligned in the horizontal direction. To complement our analysis, we perform a series of runs with different magnetic initial conditions, namely with no net-flux, net vertical flux, and a very strong horizontal field. The results of these runs are presented in this section.

\subsubsection{No Net-Flux}

We perform a simulation where, instead of a uniform initial magnetic field, we initialize a sinusoidal horizontal field -- with wavelength equal to the box size -- such that no net-flux threads the domain. The absence of a net flux removes one constraint from the system and allows it to rearrange the magnetic field geometry freely. Because of the non-zero resistivity, we expect the magnetic field strength to eventually decay over time, but one question we wish to answer is whether the properties of the MTI at saturation are affected by the absence of a net flux.

Our zero-net flux results appear in Figs \ref{fig:2D_other_compare_Bz_kin_pot_energy}-\ref{fig:2D_other_compare_Bz_mag_energy} and Table~\ref{tab:table_runs}. We find that, while the magnetic energy decays, all the other saturated quantities are practically unchanged; the turbulent kinetic and potential energy levels lie almost the same as in run R0. This behaviour confirms that in the weak-field limit the MTI only cares about the orientation of the unit vector $\bs b$, and not on the actual strength of the field. Rather amusingly, as the magnetic field reaches negligible values, the MTI forcing of kinetic and density fluctuations continues unabated.

\subsubsection{Vertical Magnetic Field}

As pointed out by \citet{McCourt2011}, while a purely vertical equilibrium field is linearly stable to the MTI, it nonetheless can be destabilized by a sufficiently large perturbation, and is thus nonlinearly unstable. To explore this, we show the results of a run, dubbed R0Bz, with the same parameters as R0, but with a vertical $\bs B$. 

\begin{figure}
	\centering
	\includegraphics[width=0.95\columnwidth]{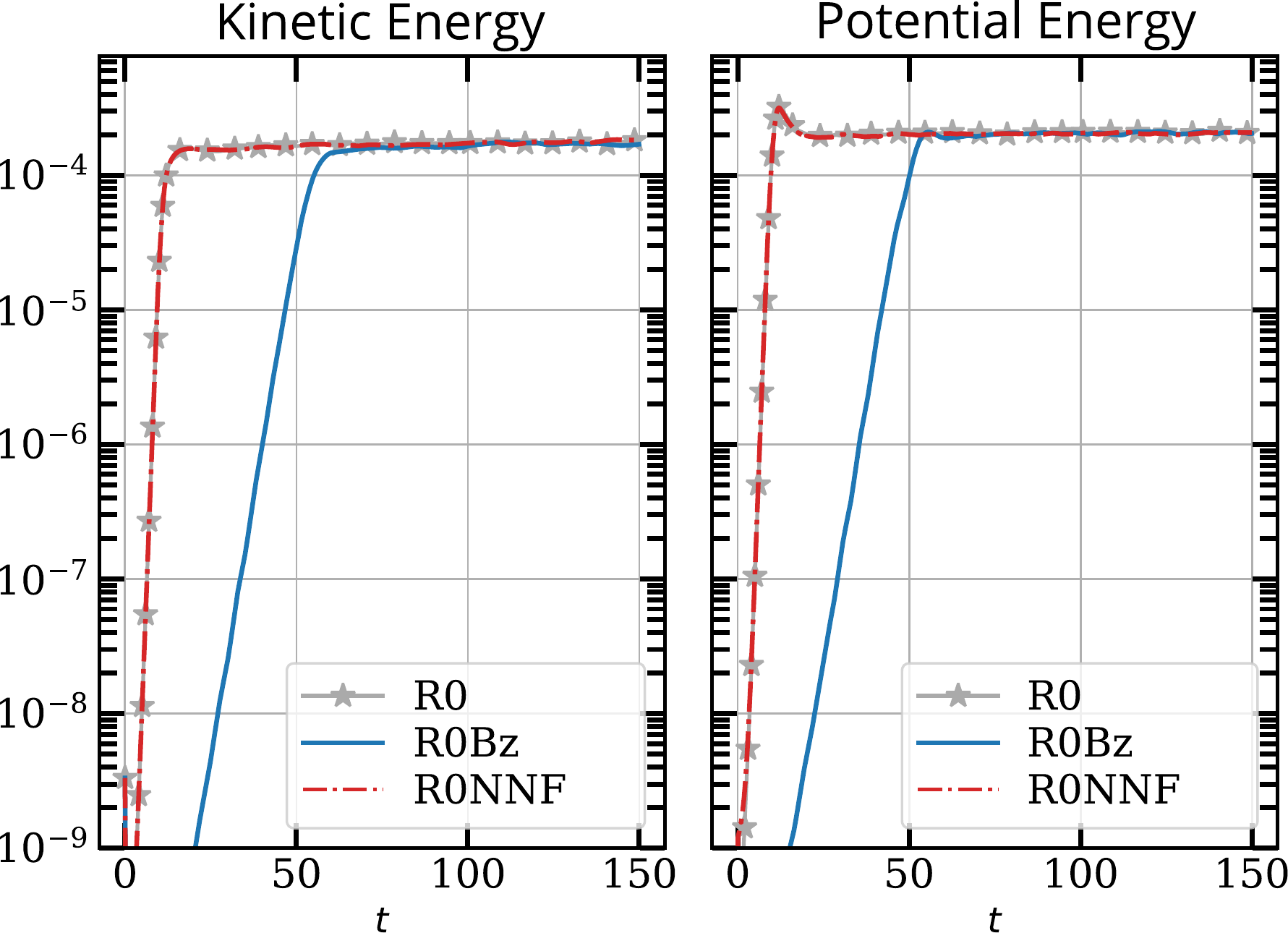}
	\caption{Left panel: comparison between the total kinetic energy of run R0 (starred line), run R0Bz (solid blue line), and run R0NNF (dash-dotted line). Right panel: the potential energy for the same runs.}
	\label{fig:2D_other_compare_Bz_kin_pot_energy}
\end{figure}

\begin{figure}
	\centering
	\includegraphics[width=1.0\columnwidth]{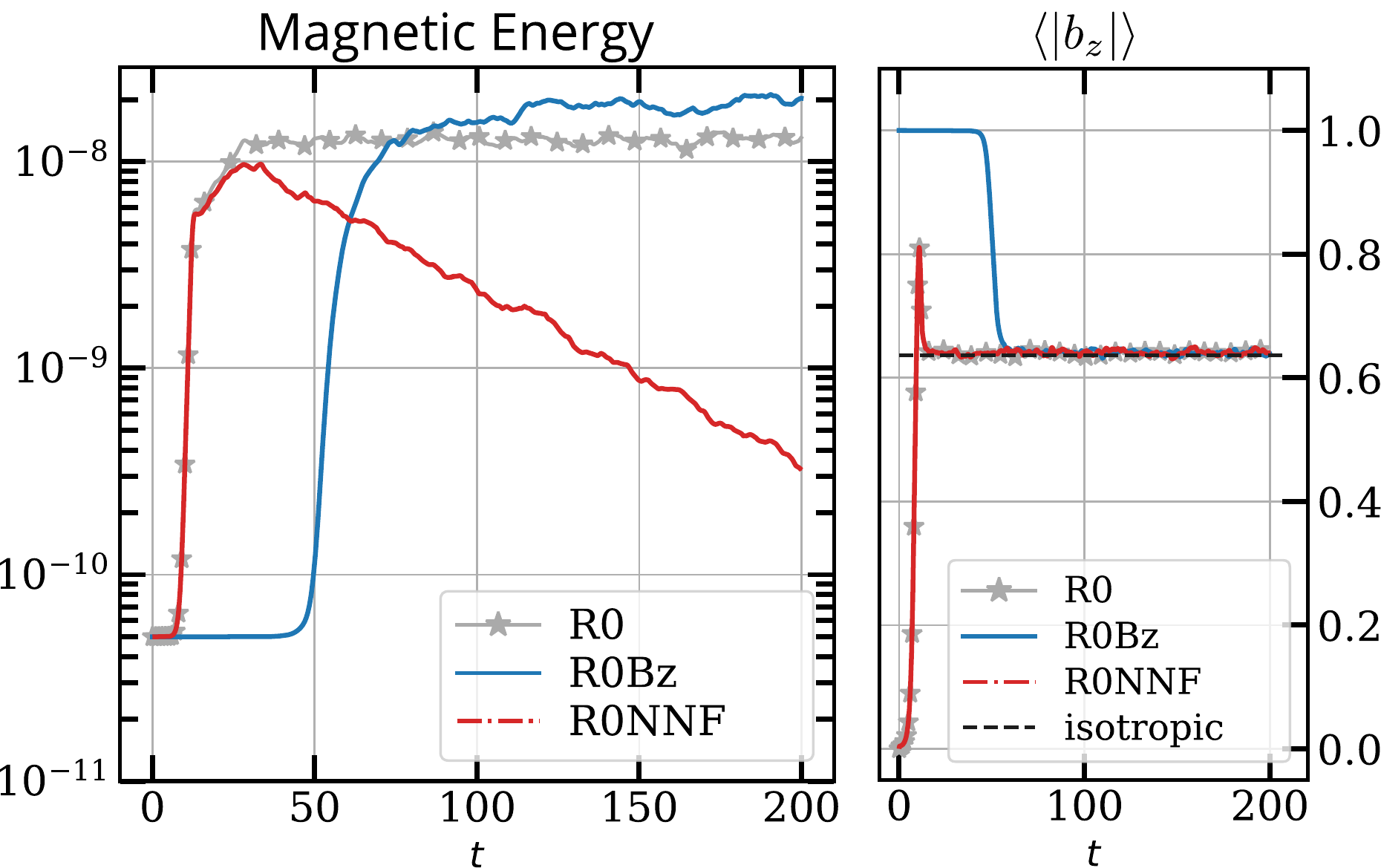}
	\caption{Left panel: comparison between the total magnetic energy of run R0 (starred line), run R0Bz (solid blue line), and run R0NNF (dash-dotted line). Right panel: average orientation of the magnetic field $b_z$.}
	\label{fig:2D_other_compare_Bz_mag_energy}
\end{figure}

In Fig.~\ref{fig:2D_other_compare_Bz_kin_pot_energy} we compare the kinetic and potential energies of runs R0 and R0Bz, while in Fig.~\ref{fig:2D_other_compare_Bz_mag_energy} we plot the magnetic energy and the average inclination of $\bs b$. After the initial growth phase, the magnetic field in R0Bz becomes isotropic and the saturated kinetic and thermal energy levels settle on the same values as in run R0. One small deviation manifests in the magnetic energy which is slightly larger than in run R0. This we attribute to the different amount of stretching of the field lines during the growth phase. Finally, we note that the Nusselt numbers of the two simulations at saturation are equivalent, as reported in Table~\ref{tab:table_runs}, suggesting that the relatively weak background vertical magnetic field does not impact on the heat conduction significantly.

\subsubsection{Strong Horizontal Magnetic Field}

The last simulation we discuss is run R0B1e-3, which is initialized with a strong, uniform, horizontal magnetic field of amplitude $B_0 = 10^{-3}$: this choice ensures that the magnetic field strength remains larger than the amplitude of the velocity fluctuations throughout the saturated stage, and should therefore be considered diametrically opposed to the weak-field cases we have mostly explored. 

\begin{figure}
	\centering
	\includegraphics[width=1.0\columnwidth]{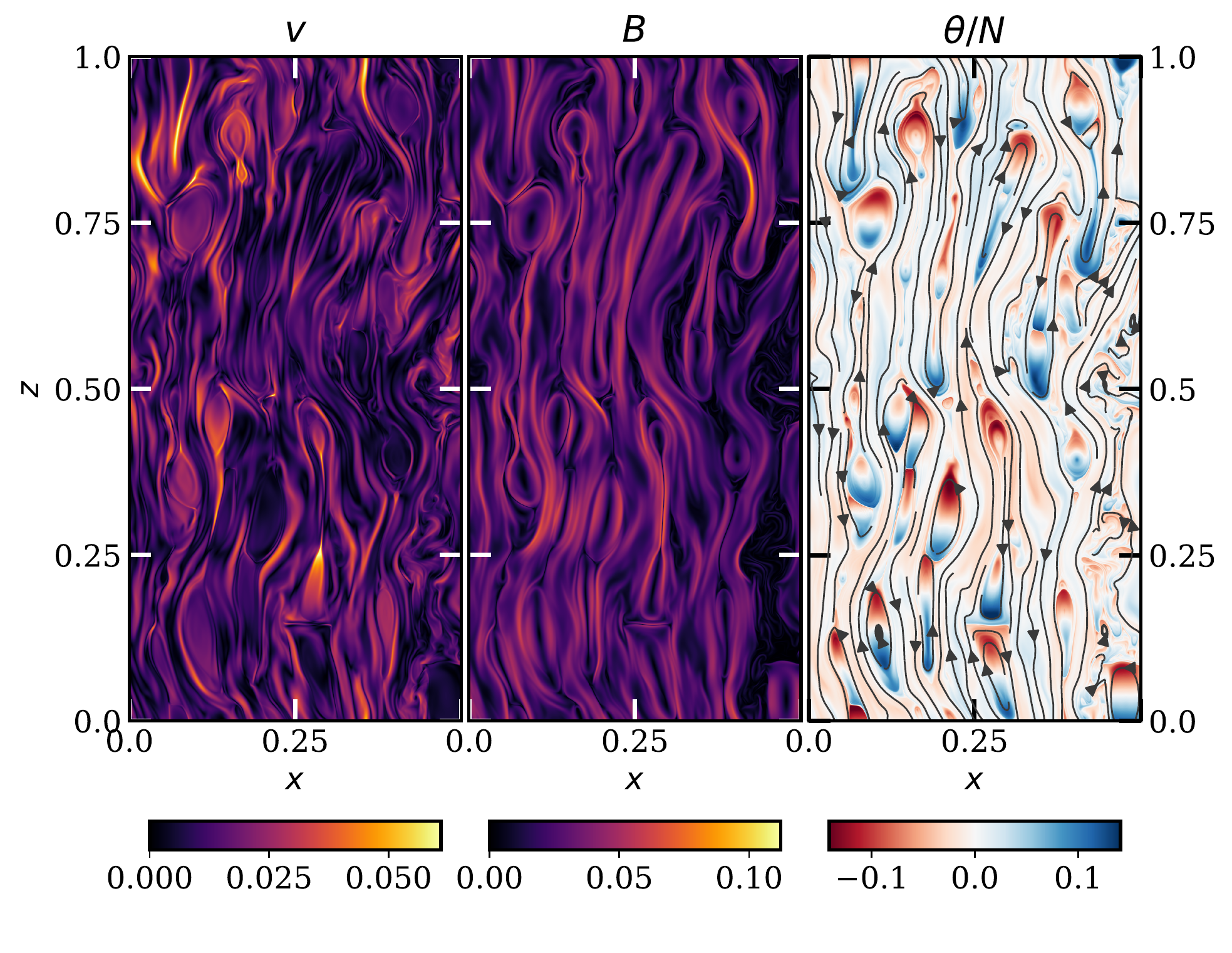}
	\caption{From left to right: magnitude of the velocity vector, amplitude of the magnetic field and density fluctuations for run R0B1e-3 at $t=295 \omega_T^{-1}$. Magnetic field lines have been overlaid on the density fluctuations. }
	\label{fig:2D_other_b_v_temp_strongB}
\end{figure}

In Fig.~\ref{fig:2D_other_b_v_temp_strongB} we show a snapshot at late times from run R0B1e-3: the three panels plot the velocity field strength, the magnetic field strength, and the density fluctuations, respectively. The stronger magnetic tension arising from the initial magnetic field sets a lower cutoff on the scale of the turbulence at saturation, so that only relatively large structures can form. In addition, we can clearly distinguish the presence of plasmoid-like structures, with magnetically insulated regions of space ("bubbles") that shear and move across the domain. Within each bubble, we recognize the action of thermal conduction along magnetic field lines, which heats the upper half of the bubble, and cools the lower half. 

The stronger magnetic field enhances the levels of potential energy by a factor of $2-3$, as one can see from Table~\ref{tab:table_runs}, while the total kinetic energy levels are relatively unchanged. However, by looking at the vertical and horizontal kinetic energies separately, we find the horizontal motions are significantly suppressed and thus most of the energy is contained in the vertical fluctuations. This marked anisotropy is also visible in the magnetic energy, with the horizontal component one order of magnitude less than the vertical. It is also echoed in the vertical elongation of the structures, clearly evident in Fig.~\ref{fig:2D_other_b_v_temp_strongB}. Similar vertical anisotropies have been observed by \citet{Kunz2012} in simulations of strong field MTI, and by \citet{Avara2013} in simulations of the HBI.

\begin{figure}
	\centering
	\includegraphics[width=0.9\columnwidth]{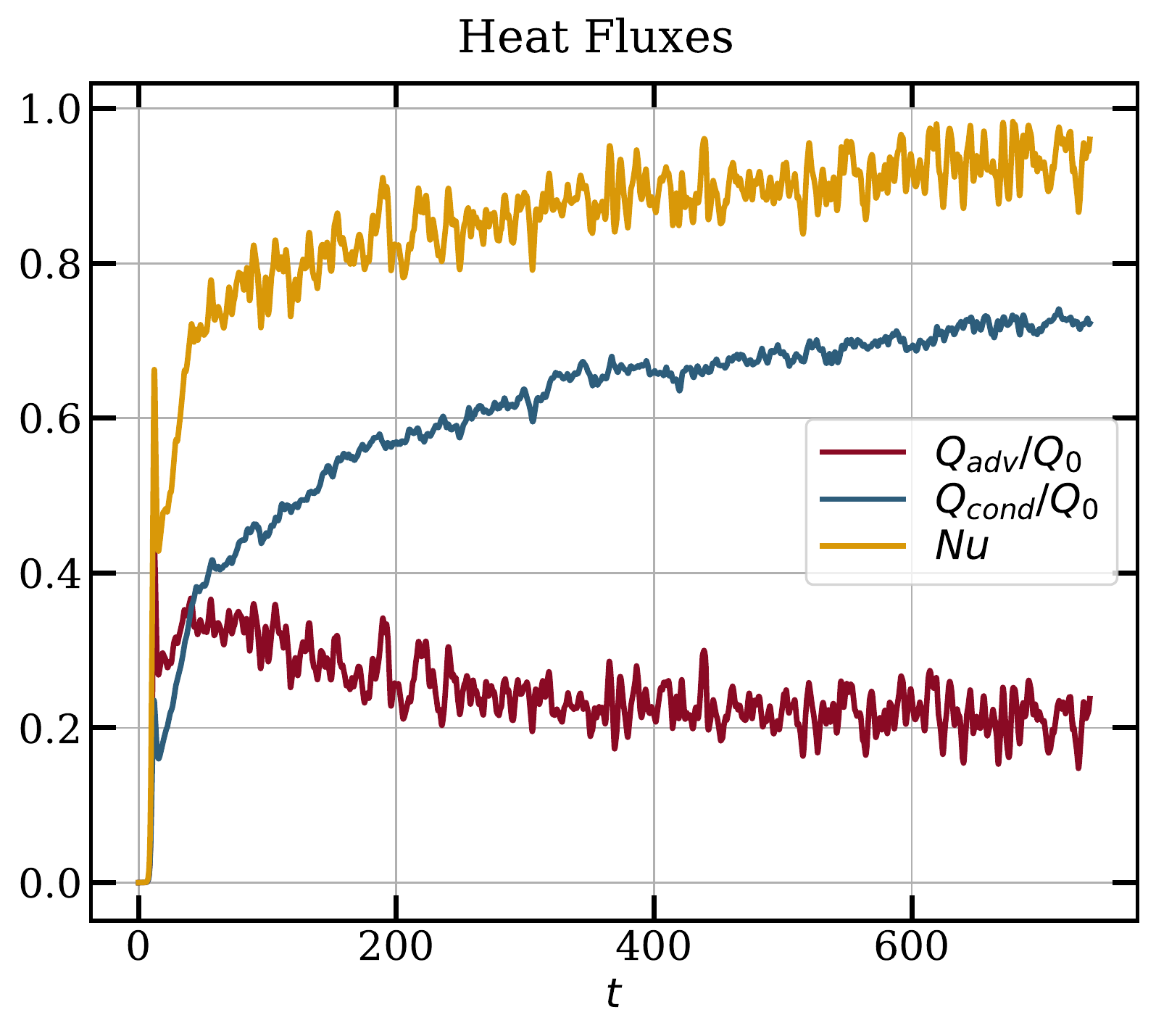}
	\caption{Time evolution of the advective (red) and conductive heat fluxes (blue) for run R0B1e-3, normalized by $Q_0 = \chi \omega_T^2$. We also show the Nusselt number (solid gold line), given by the sum of $Q_{adv}$ and $Q_{cond}$.}
	\label{fig:2D_other_nusselt_strongB}
\end{figure}

 In Fig.~\ref{fig:2D_other_nusselt_strongB},  we see that heat transport in R0B1e-3 is much more vigorous than in run R0: both conduction and turbulent advection are enhanced by the stronger magnetic fields. Nevertheless, the largest contribution comes from the conductive flux and is due to the fact that magnetic field lines are strongly biased in the vertical directions. As a result, run R0B1e-3 shows a Nusselt number that approaches unity. It is important to stress that, even after $\sim 700 \omega_T^{-1}$, run R0B1e-3 has not reached saturation. In particular, the vertical magnetic and kinetic energies continue to grow and the system becomes progressively dominated by vertically extended structures that may eventually reach the size of the box.


\section{Summary and Discussion}\label{sec:conclusions}

This paper presents a theoretical and numerical study of the MTI in 2D. In its first part, we presented the MHD equations with anisotropic heat conduction in the Boussinesq approximation, and obtained an important estimate of the energy injection rate at saturation in the limit of $\mathrm{Pr} \ll 1$ and weak magnetic fields. Modelling energy losses on large-scales as a form of Epstein drag, we arrived at approximate expressions for the kinetic energy at saturation, and for the scale, which we call "buoyancy scale", where the entropy stratification begins to suppress the MTI turbulence. Next, we studied the linear theory of the instability, deriving a dispersion relation that includes both (isotropic) viscosity and resistivity, from which we obtained an expression of the maximum growth rate in the asymptotic limit of low viscosity, low resistivity and weak magnetic fields. Finally, we analysed the evolution of an MTI eigenmode into the nonlinear regime. The "open field" linear MTI solution, where the perturbed magnetic field is purely vertical, is not an exact solution of the nonlinear Boussinesq equations (in contrast to the analogous MRI channel modes): the nonzero nonlinearity in the anisotropic thermal conduction instead forces the solution into quasi-steady buoyancy oscillations. These solutions, however, are vulnerable to parasitic Kelvin-Helmholtz instabilities, which disrupt the open field solution and initiate the MTI's breakdown into 2D turbulence. They thus prohibit MTI saturation via a simple formation of strong purely vertical field.

In the second part of the paper, we shared results of 2D numerical simulations performed with the SNOOPY code, modified to account for anisotropic heat conduction. The use of a Boussinesq code allowed us to perform an extensive sweep of the physical parameters: thermal conductivity, stratification, viscosity, and resistivity. We find that the MTI saturation mechanism consists of (a) energy input at small scales by the MTI, (b) its transfer to large-scales by a 2D turbulent inverse cascade, where (c) it is dumped into buoyancy oscillations at the critical buoyancy scale, which (d) are then damped. The properties of turbulence at saturation closely depend on the values of $\mathrm{Pe}$ and $N^2$, and follow convenient power laws. At saturation the energy injection rate and the kinetic energy scale as $\epsilon_I \sim \chi \omega_T^4 / N^2$ and $K \sim \chi \omega_T^3 / N^2$, respectively (in agreement with theory). The buoyancy scale, on the other hand, is close to the theoretical estimate of $\sim \chi^{-1/2} N$. 

These novel scaling laws are perhaps the main achievements of the paper. Previous compressible simulations were unable to observe these trends because they were too tightly constrained by a three-layered `sandwich' configuration \citep[see][]{Parrish2007}, which prohibited an exploration of the physical parameters, the entropy structure especially. In fact, the influence of stable stratification on the MTI has been mostly neglected, but a key theme of this paper has been exposing its critical role. Secondly, most local MTI studies of the MTI have adopted the "rapid conduction" regime, in which $\mathrm{Pe}\lesssim 1$ \citep[e.g.,][]{McCourt2011}. Local simulations in this regime, inevitably, produce flows limited by the box size, rather than the natural buoyancy length or the cluster system size, leading to suspect saturation properties. By adopting moderate conductivities, with $\mathrm{Pe}\gg 1$, we bypass this problem.
While we leave to Paper II an in-depth confrontation with cluster observations, we point out that a regime of moderate conduction is justified by the thermodynamic profiles reported by \citet{Ghirardini2019}, which we analyze in Sec.~\ref{sec:galaxy_clusters}. We find that the conduction length is typically less than the size of the cluster, especially if its magnitude is partially suppressed by microturbulence, for example.

While a Boussinesq model has facilitated a wider exploration of the MTI dynamics, we concede that its major limitation is the impossibility of studying the global dynamics of the MTI, i.e. lengthscales comparable to a cluster's pressure scale-height. To do so necessarily requires global simulations, such as in compressible (or anelastic) vertically global boxes, or in spheres or spherical shells. We also omit the important element of pressure anisotropy \citep{Kunz2011,Kunz2012}, as well as the suppression of heat conductivity by microinstabilities \citep[as in, e.g.,][for suppression by the mirror instability]{Berlok2020}, but aim to tackle this physics in future work. Finally, our simulations are two-dimensional and thus cannot support a magnetic dynamo, which may then feedback non-trivially on the MTI saturation. This and related processes we do cover in our companion paper, in which we present the results of three-dimensional MTI Boussinesq simulations that revise and extend the various findings advertised here.

\section*{Acknowledgements}

The authors thank Fran\c{c}ois Rincon and Thomas Berlok for helpful and generous advice on a draft version of the manuscript.

\section*{Data Availability}

The data underlying this article will be shared on reasonable request to the corresponding author.
%%%%%%%%%%%%%%%%%%%%%%%%%%%%%%%%%%%%%%%%%%%%%%%%%%



\bibliographystyle{mnras}
\bibliography{MTI_GalaxyClusters} % if your bibtex file is called example.bib

\begin{thebibliography}{}
\makeatletter
\relax
\def\mn@urlcharsother{\let\do\@makeother \do\$\do\&\do\#\do\^\do\_\do\%\do\~}
\def\mn@doi{\begingroup\mn@urlcharsother \@ifnextchar [ {\mn@doi@}
  {\mn@doi@[]}}
\def\mn@doi@[#1]#2{\def\@tempa{#1}\ifx\@tempa\@empty \href
  {http://dx.doi.org/#2} {doi:#2}\else \href {http://dx.doi.org/#2} {#1}\fi
  \endgroup}
\def\mn@eprint#1#2{\mn@eprint@#1:#2::\@nil}
\def\mn@eprint@arXiv#1{\href {http://arxiv.org/abs/#1} {{\tt arXiv:#1}}}
\def\mn@eprint@dblp#1{\href {http://dblp.uni-trier.de/rec/bibtex/#1.xml}
  {dblp:#1}}
\def\mn@eprint@#1:#2:#3:#4\@nil{\def\@tempa {#1}\def\@tempb {#2}\def\@tempc
  {#3}\ifx \@tempc \@empty \let \@tempc \@tempb \let \@tempb \@tempa \fi \ifx
  \@tempb \@empty \def\@tempb {arXiv}\fi \@ifundefined
  {mn@eprint@\@tempb}{\@tempb:\@tempc}{\expandafter \expandafter \csname
  mn@eprint@\@tempb\endcsname \expandafter{\@tempc}}}

\bibitem[\protect\citeauthoryear{Alexiades, Amiez  \& Gremaud}{Alexiades
  et~al.}{1996}]{Alexiades1996}
Alexiades V.,  Amiez G.,   Gremaud P.-A.,  1996, \mn@doi [Communications in
  Numerical Methods in Engineering]
  {10.1002/(SICI)1099-0887(199601)12:1<31::AID-CNM950>3.0.CO;2-5}, 12, 31

\bibitem[\protect\citeauthoryear{{Allen}, {Rapetti}, {Schmidt}, {Ebeling},
  {Morris}  \& {Fabian}}{{Allen} et~al.}{2008}]{Allen2008}
{Allen} S.~W.,  {Rapetti} D.~A.,  {Schmidt} R.~W.,  {Ebeling} H.,  {Morris}
  R.~G.,   {Fabian} A.~C.,  2008, \mn@doi [\mnras]
  {10.1111/j.1365-2966.2007.12610.x}, \href
  {https://ui.adsabs.harvard.edu/abs/2008MNRAS.383..879A} {383, 879}

\bibitem[\protect\citeauthoryear{{Allen}, {Evrard}  \& {Mantz}}{{Allen}
  et~al.}{2011}]{Allen2011}
{Allen} S.~W.,  {Evrard} A.~E.,   {Mantz} A.~B.,  2011, \mn@doi [\araa]
  {10.1146/annurev-astro-081710-102514}, \href
  {https://ui.adsabs.harvard.edu/abs/2011ARA&A..49..409A} {49, 409}

\bibitem[\protect\citeauthoryear{{Avara}, {Reynolds}  \&
  {Bogdanovi{\'c}}}{{Avara} et~al.}{2013}]{Avara2013}
{Avara} M.~J.,  {Reynolds} C.~S.,   {Bogdanovi{\'c}} T.,  2013, \mn@doi [\apj]
  {10.1088/0004-637X/773/2/171}, \href
  {https://ui.adsabs.harvard.edu/abs/2013ApJ...773..171A} {773, 171}

\bibitem[\protect\citeauthoryear{Balbus}{Balbus}{2000}]{Balbus2000}
Balbus S.~A.,  2000, \mn@doi [\apj] {10.1086/308732}, 534, 420

\bibitem[\protect\citeauthoryear{Balbus}{Balbus}{2001}]{Balbus2001}
Balbus S.~A.,  2001, \mn@doi [\apj] {10.1086/323875}, 562, 909

\bibitem[\protect\citeauthoryear{{Balbus} \& {Hawley}}{{Balbus} \&
  {Hawley}}{1991}]{Balbus1991}
{Balbus} S.~A.,  {Hawley} J.~F.,  1991, \mn@doi [\apj] {10.1086/170270}, \href
  {https://ui.adsabs.harvard.edu/abs/1991ApJ...376..214B} {376, 214}

\bibitem[\protect\citeauthoryear{{Bale}, {Pulupa}, {Salem}, {Chen}  \&
  {Quataert}}{{Bale} et~al.}{2013}]{Bale2013}
{Bale} S.~D.,  {Pulupa} M.,  {Salem} C.,  {Chen} C.~H.~K.,   {Quataert} E.,
  2013, \mn@doi [\apjl] {10.1088/2041-8205/769/2/L22}, \href
  {https://ui.adsabs.harvard.edu/abs/2013ApJ...769L..22B} {769, L22}

\bibitem[\protect\citeauthoryear{{Batchelor}}{{Batchelor}}{1969}]{Batchelor1969}
{Batchelor} G.~K.,  1969, \mn@doi [Physics of Fluids] {10.1063/1.1692443},
  \href {https://ui.adsabs.harvard.edu/abs/1969PhFl...12C.233B} {12, II}

\bibitem[\protect\citeauthoryear{{Berlok}, {Quataert}, {Pessah}  \&
  {Pfrommer}}{{Berlok} et~al.}{2020}]{Berlok2020}
{Berlok} T.,  {Quataert} E.,  {Pessah} M.~E.,   {Pfrommer} C.,  2020, arXiv
  e-prints, \href {https://ui.adsabs.harvard.edu/abs/2020arXiv200700018B} {p.
  arXiv:2007.00018}

\bibitem[\protect\citeauthoryear{{B{\^\i}rzan}, {Rafferty}, {McNamara}, {Wise}
  \& {Nulsen}}{{B{\^\i}rzan} et~al.}{2004}]{Birzan2004}
{B{\^\i}rzan} L.,  {Rafferty} D.~A.,  {McNamara} B.~R.,  {Wise} M.~W.,
  {Nulsen} P.~E.~J.,  2004, \mn@doi [\apj] {10.1086/383519}, \href
  {https://ui.adsabs.harvard.edu/abs/2004ApJ...607..800B} {607, 800}

\bibitem[\protect\citeauthoryear{Boffetta \& Ecke}{Boffetta \&
  Ecke}{2012}]{Boffetta2012}
Boffetta G.,  Ecke R.~E.,  2012, \mn@doi [Annual Review of Fluid Mechanics]
  {10.1146/annurev-fluid-120710-101240}, 44, 427

\bibitem[\protect\citeauthoryear{Boffetta, Cenedese, Espa  \&
  Musacchio}{Boffetta et~al.}{2005}]{Boffetta2005}
Boffetta G.,  Cenedese A.,  Espa S.,   Musacchio S.,  2005, \mn@doi
  [Europhysics Letters ({EPL})] {10.1209/epl/i2005-10111-6}, 71, 590

\bibitem[\protect\citeauthoryear{Boffetta, Lillo, Mazzino  \&
  Musacchio}{Boffetta et~al.}{2011}]{Boffetta2011}
Boffetta G.,  Lillo F.~D.,  Mazzino A.,   Musacchio S.,  2011, \mn@doi [EPL
  (Europhysics Letters)] {10.1209/0295-5075/95/34001}, 95, 34001

\bibitem[\protect\citeauthoryear{{B{\"o}hringer} \& {Werner}}{{B{\"o}hringer}
  \& {Werner}}{2010}]{Boehringer2010}
{B{\"o}hringer} H.,  {Werner} N.,  2010, \mn@doi [\aapr]
  {10.1007/s00159-009-0023-3}, \href
  {https://ui.adsabs.harvard.edu/abs/2010A&ARv..18..127B} {18, 127}

\bibitem[\protect\citeauthoryear{{Bolgiano}}{{Bolgiano}}{1959}]{Bolgiano1959}
{Bolgiano} R. J.,  1959, \mn@doi [\jgr] {10.1029/JZ064i012p02226}, \href
  {https://ui.adsabs.harvard.edu/abs/1959JGR....64.2226B} {64, 2226}

\bibitem[\protect\citeauthoryear{{Brada{\v{c}}}, {Allen}, {Treu}, {Ebeling},
  {Massey}, {Morris}, {von der Linden}  \& {Applegate}}{{Brada{\v{c}}}
  et~al.}{2008}]{Bradac2008}
{Brada{\v{c}}} M.,  {Allen} S.~W.,  {Treu} T.,  {Ebeling} H.,  {Massey} R.,
  {Morris} R.~G.,  {von der Linden} A.,   {Applegate} D.,  2008, \mn@doi [\apj]
  {10.1086/591246}, \href
  {https://ui.adsabs.harvard.edu/abs/2008ApJ...687..959B} {687, 959}

\bibitem[\protect\citeauthoryear{Braginskii}{Braginskii}{1965}]{Braginskii1965a}
Braginskii S.~I.,  1965, Reviews of Plasma Physics, 1, 205

\bibitem[\protect\citeauthoryear{{Bu}, {Yuan}  \& {Stone}}{{Bu}
  et~al.}{2011}]{Bu2011}
{Bu} D.-F.,  {Yuan} F.,   {Stone} J.~M.,  2011, \mn@doi [\mnras]
  {10.1111/j.1365-2966.2011.18354.x}, \href
  {https://ui.adsabs.harvard.edu/abs/2011MNRAS.413.2808B} {413, 2808}

\bibitem[\protect\citeauthoryear{{Calzavarini}, {Doering}, {Gibbon}, {Lohse},
  {Tanabe}  \& {Toschi}}{{Calzavarini} et~al.}{2006}]{calzone2006}
{Calzavarini} E.,  {Doering} C.~R.,  {Gibbon} J.~D.,  {Lohse} D.,  {Tanabe} A.,
    {Toschi} F.,  2006, \mn@doi [\pre] {10.1103/PhysRevE.73.035301}, \href
  {https://ui.adsabs.harvard.edu/abs/2006PhRvE..73c5301C} {73, 035301}

\bibitem[\protect\citeauthoryear{{Churazov}, {Forman}, {Jones}  \&
  {B{\"o}hringer}}{{Churazov} et~al.}{2003}]{Churazov2003}
{Churazov} E.,  {Forman} W.,  {Jones} C.,   {B{\"o}hringer} H.,  2003, \mn@doi
  [\apj] {10.1086/374923}, \href
  {https://ui.adsabs.harvard.edu/abs/2003ApJ...590..225C} {590, 225}

\bibitem[\protect\citeauthoryear{{Churazov} et~al.,}{{Churazov}
  et~al.}{2012}]{Churazov2012}
{Churazov} E.,  et~al., 2012, \mn@doi [\mnras]
  {10.1111/j.1365-2966.2011.20372.x}, \href
  {https://ui.adsabs.harvard.edu/abs/2012MNRAS.421.1123C} {421, 1123}

\bibitem[\protect\citeauthoryear{Corfield}{Corfield}{1984}]{Corfield1984}
Corfield C.~N.,  1984, \mn@doi [Geophysical \& Astrophysical Fluid Dynamics]
  {10.1080/03091928408248181}, 29, 19

\bibitem[\protect\citeauthoryear{Davidson}{Davidson}{2013}]{Davidson2013}
Davidson P.~A.,  2013, Turbulence in Rotating, Stratified and Electrically
  Conducting Fluids.
Cambridge University Press, \mn@doi{10.1017/CBO9781139208673}

\bibitem[\protect\citeauthoryear{{Drake}, {Pfrommer}, {Reynolds}, {Ruszkowski},
  {Swisdak}, {Einarsson}, {Hassam}  \& {Roberg-Clark}}{{Drake}
  et~al.}{2020}]{Drake2020}
{Drake} J.~F.,  {Pfrommer} C.,  {Reynolds} C.~S.,  {Ruszkowski} M.,  {Swisdak}
  M.,  {Einarsson} A.,  {Hassam} A.~B.,   {Roberg-Clark} G.~T.,  2020, arXiv
  e-prints, \href {https://ui.adsabs.harvard.edu/abs/2020arXiv200707931D} {p.
  arXiv:2007.07931}

\bibitem[\protect\citeauthoryear{{Fabian}}{{Fabian}}{1994}]{Fabian1994}
{Fabian} A.~C.,  1994, \mn@doi [\araa] {10.1146/annurev.aa.32.090194.001425},
  \href {https://ui.adsabs.harvard.edu/abs/1994ARA&A..32..277F} {32, 277}

\bibitem[\protect\citeauthoryear{{Fabian}}{{Fabian}}{2012}]{Fabian2012}
{Fabian} A.~C.,  2012, \mn@doi [\araa] {10.1146/annurev-astro-081811-125521},
  \href {https://ui.adsabs.harvard.edu/abs/2012ARA&A..50..455F} {50, 455}

\bibitem[\protect\citeauthoryear{{Fabian}, {Sanders}, {Allen}, {Crawford},
  {Iwasawa}, {Johnstone}, {Schmidt}  \& {Taylor}}{{Fabian}
  et~al.}{2003}]{Fabian2003}
{Fabian} A.~C.,  {Sanders} J.~S.,  {Allen} S.~W.,  {Crawford} C.~S.,  {Iwasawa}
  K.,  {Johnstone} R.~M.,  {Schmidt} R.~W.,   {Taylor} G.~B.,  2003, \mn@doi
  [\mnras] {10.1046/j.1365-8711.2003.06902.x}, \href
  {https://ui.adsabs.harvard.edu/abs/2003MNRAS.344L..43F} {344, L43}

\bibitem[\protect\citeauthoryear{{Foucart}, {Chandra}, {Gammie}  \&
  {Quataert}}{{Foucart} et~al.}{2016}]{Foucart2016}
{Foucart} F.,  {Chandra} M.,  {Gammie} C.~F.,   {Quataert} E.,  2016, \mn@doi
  [\mnras] {10.1093/mnras/stv2687}, \href
  {https://ui.adsabs.harvard.edu/abs/2016MNRAS.456.1332F} {456, 1332}

\bibitem[\protect\citeauthoryear{{Foucart}, {Chandra}, {Gammie}, {Quataert}  \&
  {Tchekhovskoy}}{{Foucart} et~al.}{2017}]{Foucart2017}
{Foucart} F.,  {Chandra} M.,  {Gammie} C.~F.,  {Quataert} E.,   {Tchekhovskoy}
  A.,  2017, \mn@doi [\mnras] {10.1093/mnras/stx1368}, \href
  {https://ui.adsabs.harvard.edu/abs/2017MNRAS.470.2240F} {470, 2240}

\bibitem[\protect\citeauthoryear{{Garaud}}{{Garaud}}{2018}]{Garaud2018}
{Garaud} P.,  2018, \mn@doi [Annual Review of Fluid Mechanics]
  {10.1146/annurev-fluid-122316-045234}, \href
  {https://ui.adsabs.harvard.edu/abs/2018AnRFM..50..275G} {50, 275}

\bibitem[\protect\citeauthoryear{{Ghirardini} et~al.,}{{Ghirardini}
  et~al.}{2019}]{Ghirardini2019}
{Ghirardini} V.,  et~al., 2019, \mn@doi [\aap] {10.1051/0004-6361/201833325},
  \href {https://ui.adsabs.harvard.edu/abs/2019A&A...621A..41G} {621, A41}

\bibitem[\protect\citeauthoryear{{Goodman} \& {Xu}}{{Goodman} \&
  {Xu}}{1994}]{Goodman1994}
{Goodman} J.,  {Xu} G.,  1994, \mn@doi [\apj] {10.1086/174562}, \href
  {https://ui.adsabs.harvard.edu/abs/1994ApJ...432..213G} {432, 213}

\bibitem[\protect\citeauthoryear{{Kannan}, {Vogelsberger}, {Pfrommer},
  {Weinberger}, {Springel}, {Hernquist}, {Puchwein}  \& {Pakmor}}{{Kannan}
  et~al.}{2017}]{Kannan2017}
{Kannan} R.,  {Vogelsberger} M.,  {Pfrommer} C.,  {Weinberger} R.,  {Springel}
  V.,  {Hernquist} L.,  {Puchwein} E.,   {Pakmor} R.,  2017, \mn@doi [\apjl]
  {10.3847/2041-8213/aa624b}, \href
  {https://ui.adsabs.harvard.edu/abs/2017ApJ...837L..18K} {837, L18}

\bibitem[\protect\citeauthoryear{Komarov, Schekochihin, Churazov  \&
  Spitkovsky}{Komarov et~al.}{2018}]{Komarov2018}
Komarov S.,  Schekochihin A.~A.,  Churazov E.,   Spitkovsky A.,  2018, \mn@doi
  [Journal of Plasma Physics] {10.1017/S0022377818000399}, 84, 905840305

\bibitem[\protect\citeauthoryear{Kraichnan}{Kraichnan}{1967}]{Kraichnan1967}
Kraichnan R.~H.,  1967, \mn@doi [The Physics of Fluids] {10.1063/1.1762301},
  10, 1417

\bibitem[\protect\citeauthoryear{Kraichnan}{Kraichnan}{1971}]{Kraichnan1971}
Kraichnan R.~H.,  1971, \mn@doi [Journal of Fluid Mechanics]
  {10.1017/S0022112071001216}, 47, 525\u2013535

\bibitem[\protect\citeauthoryear{{Kulsrud}}{{Kulsrud}}{2005}]{Kulsrud2005}
{Kulsrud} R.~M.,  2005, {Plasma physics for astrophysics}

\bibitem[\protect\citeauthoryear{Kunz}{Kunz}{2011}]{Kunz2011}
Kunz M.~W.,  2011, \mn@doi [\mnras] {10.1111/j.1365-2966.2011.19303.x}, 417,
  602

\bibitem[\protect\citeauthoryear{Kunz, Bogdanović, Reynolds  \& Stone}{Kunz
  et~al.}{2012}]{Kunz2012}
Kunz M.~W.,  Bogdanović T.,  Reynolds C.~S.,   Stone J.~M.,  2012, \mn@doi
  [\apj] {10.1088/0004-637X/754/2/122}, 754, 122

\bibitem[\protect\citeauthoryear{Latter}{Latter}{2016}]{Latter2016}
Latter H.~N.,  2016, \mn@doi [\mnras] {10.1093/mnras/stv2449}, 455, 2608

\bibitem[\protect\citeauthoryear{{Leccardi} \& {Molendi}}{{Leccardi} \&
  {Molendi}}{2008}]{Leccardi2008}
{Leccardi} A.,  {Molendi} S.,  2008, \mn@doi [\aap]
  {10.1051/0004-6361:200809538}, \href
  {https://ui.adsabs.harvard.edu/abs/2008A&A...486..359L} {486, 359}

\bibitem[\protect\citeauthoryear{{Lesur}}{{Lesur}}{2015}]{Lesur2015}
{Lesur} G.,  2015, {Snoopy: General purpose spectral solver} (\mn@eprint {ascl}
  {1505.022})

\bibitem[\protect\citeauthoryear{Lesur \& Longaretti}{Lesur \&
  Longaretti}{2007}]{Lesur2007}
Lesur G.,  Longaretti P.-Y.,  2007, \mn@doi [Monthly Notices of the Royal
  Astronomical Society] {10.1111/j.1365-2966.2007.11888.x}, 378, 1471

\bibitem[\protect\citeauthoryear{{Markevitch} \& {Vikhlinin}}{{Markevitch} \&
  {Vikhlinin}}{2007}]{Markevitch2007}
{Markevitch} M.,  {Vikhlinin} A.,  2007, \mn@doi [\physrep]
  {10.1016/j.physrep.2007.01.001}, \href
  {https://ui.adsabs.harvard.edu/abs/2007PhR...443....1M} {443, 1}

\bibitem[\protect\citeauthoryear{{Markevitch}, {Vikhlinin}  \&
  {Mazzotta}}{{Markevitch} et~al.}{2001}]{Markevitch2001}
{Markevitch} M.,  {Vikhlinin} A.,   {Mazzotta} P.,  2001, \mn@doi [\apjl]
  {10.1086/337973}, \href
  {https://ui.adsabs.harvard.edu/abs/2001ApJ...562L.153M} {562, L153}

\bibitem[\protect\citeauthoryear{{Markevitch}, {Gonzalez}, {David},
  {Vikhlinin}, {Murray}, {Forman}, {Jones}  \& {Tucker}}{{Markevitch}
  et~al.}{2002}]{Markevitch2002}
{Markevitch} M.,  {Gonzalez} A.~H.,  {David} L.,  {Vikhlinin} A.,  {Murray} S.,
   {Forman} W.,  {Jones} C.,   {Tucker} W.,  2002, \mn@doi [\apjl]
  {10.1086/339619}, \href
  {https://ui.adsabs.harvard.edu/abs/2002ApJ...567L..27M} {567, L27}

\bibitem[\protect\citeauthoryear{McCourt, Parrish, Sharma  \& Quataert}{McCourt
  et~al.}{2011}]{McCourt2011}
McCourt M.,  Parrish I.~J.,  Sharma P.,   Quataert E.,  2011, \mn@doi [\mnras]
  {10.1111/j.1365-2966.2011.18216.x}, 413, 1295

\bibitem[\protect\citeauthoryear{{McNamara} et~al.,}{{McNamara}
  et~al.}{2000}]{McNamara2000}
{McNamara} B.~R.,  et~al., 2000, \mn@doi [\apjl] {10.1086/312662}, \href
  {https://ui.adsabs.harvard.edu/abs/2000ApJ...534L.135M} {534, L135}

\bibitem[\protect\citeauthoryear{{Mignone}, {Bodo}, {Massaglia}, {Matsakos},
  {Tesileanu}, {Zanni}  \& {Ferrari}}{{Mignone} et~al.}{2007}]{Mignone2007}
{Mignone} A.,  {Bodo} G.,  {Massaglia} S.,  {Matsakos} T.,  {Tesileanu} O.,
  {Zanni} C.,   {Ferrari} A.,  2007, \mn@doi [\apjs] {10.1086/513316}, \href
  {https://ui.adsabs.harvard.edu/abs/2007ApJS..170..228M} {170, 228}

\bibitem[\protect\citeauthoryear{Mohr, Mathiesen  \& Evrard}{Mohr
  et~al.}{1999}]{Mohr1999}
Mohr J.~J.,  Mathiesen B.,   Evrard A.~E.,  1999, \mn@doi [The Astrophysical
  Journal] {10.1086/307227}, 517, 627

\bibitem[\protect\citeauthoryear{Nam, Ott, Antonsen  \& Guzdar}{Nam
  et~al.}{2000}]{Nam2000}
Nam K.,  Ott E.,  Antonsen T.~M.,   Guzdar P.~N.,  2000, \mn@doi [Phys. Rev.
  Lett.] {10.1103/PhysRevLett.84.5134}, 84, 5134

\bibitem[\protect\citeauthoryear{{Narayan} \& {Medvedev}}{{Narayan} \&
  {Medvedev}}{2001}]{Narayan2001}
{Narayan} R.,  {Medvedev} M.~V.,  2001, \mn@doi [\apjl] {10.1086/338325}, \href
  {https://ui.adsabs.harvard.edu/abs/2001ApJ...562L.129N} {562, L129}

\bibitem[\protect\citeauthoryear{Obukhov}{Obukhov}{1959}]{Obukhov1959}
Obukhov A.,  1959, in Dokl. Akad. Nauk SSSR. pp 1246--1248

\bibitem[\protect\citeauthoryear{Parrish \& Stone}{Parrish \&
  Stone}{2005}]{Parrish2005}
Parrish I.~J.,  Stone J.~M.,  2005, \mn@doi [\apj] {10.1086/444589}, 633, 334

\bibitem[\protect\citeauthoryear{Parrish \& Stone}{Parrish \&
  Stone}{2007}]{Parrish2007}
Parrish I.~J.,  Stone J.~M.,  2007, \mn@doi [\apj] {10.1086/518881}, 664, 135

\bibitem[\protect\citeauthoryear{Parrish, Stone  \& Lemaster}{Parrish
  et~al.}{2008}]{Parrish2008}
Parrish I.~J.,  Stone J.~M.,   Lemaster N.,  2008, \mn@doi [\apj]
  {10.1086/592380}, 688, 905

\bibitem[\protect\citeauthoryear{Parrish, McCourt, Quataert  \& Sharma}{Parrish
  et~al.}{2012a}]{Parrish2012}
Parrish I.~J.,  McCourt M.,  Quataert E.,   Sharma P.,  2012a, \mn@doi [\mnras]
  {10.1111/j.1745-3933.2011.01171.x}, 419, L29

\bibitem[\protect\citeauthoryear{Parrish, McCourt, Quataert  \& Sharma}{Parrish
  et~al.}{2012b}]{Parrish2012a}
Parrish I.~J.,  McCourt M.,  Quataert E.,   Sharma P.,  2012b, \mn@doi [\mnras]
  {10.1111/j.1365-2966.2012.20650.x}, 422, 704

\bibitem[\protect\citeauthoryear{{Pratt}, {B{\"o}hringer}, {Croston}, {Arnaud},
  {Borgani}, {Finoguenov}  \& {Temple}}{{Pratt} et~al.}{2007}]{Pratt2007}
{Pratt} G.~W.,  {B{\"o}hringer} H.,  {Croston} J.~H.,  {Arnaud} M.,  {Borgani}
  S.,  {Finoguenov} A.,   {Temple} R.~F.,  2007, \mn@doi [\aap]
  {10.1051/0004-6361:20065676}, \href
  {https://ui.adsabs.harvard.edu/abs/2007A&A...461...71P} {461, 71}

\bibitem[\protect\citeauthoryear{Quataert}{Quataert}{2008}]{Quataert2008}
Quataert E.,  2008, \mn@doi [\apj] {10.1086/525248}, \href
  {https://ui.adsabs.harvard.edu/abs/2008ApJ...673..758Q} {673, 758}

\bibitem[\protect\citeauthoryear{{Rhines}}{{Rhines}}{1975}]{Rhines1975}
{Rhines} P.~B.,  1975, \mn@doi [Journal of Fluid Mechanics]
  {10.1017/S0022112075001504}, \href
  {https://ui.adsabs.harvard.edu/abs/1975JFM....69..417R} {69, 417}

\bibitem[\protect\citeauthoryear{{Riquelme}, {Quataert}  \&
  {Verscharen}}{{Riquelme} et~al.}{2016}]{Riquelme2016}
{Riquelme} M.~A.,  {Quataert} E.,   {Verscharen} D.,  2016, \mn@doi [\apj]
  {10.3847/0004-637X/824/2/123}, \href
  {https://ui.adsabs.harvard.edu/abs/2016ApJ...824..123R} {824, 123}

\bibitem[\protect\citeauthoryear{{Roberg-Clark}, {Drake}, {Reynolds}  \&
  {Swisdak}}{{Roberg-Clark} et~al.}{2016}]{Roberg2016}
{Roberg-Clark} G.~T.,  {Drake} J.~F.,  {Reynolds} C.~S.,   {Swisdak} M.,  2016,
  \mn@doi [\apjl] {10.3847/2041-8205/830/1/L9}, \href
  {https://ui.adsabs.harvard.edu/abs/2016ApJ...830L...9R} {830, L9}

\bibitem[\protect\citeauthoryear{{Roberg-Clark}, {Drake}, {Reynolds}  \&
  {Swisdak}}{{Roberg-Clark} et~al.}{2018}]{Roberg2018}
{Roberg-Clark} G.~T.,  {Drake} J.~F.,  {Reynolds} C.~S.,   {Swisdak} M.,  2018,
  \mn@doi [\prl] {10.1103/PhysRevLett.120.035101}, \href
  {https://ui.adsabs.harvard.edu/abs/2018PhRvL.120c5101R} {120, 035101}

\bibitem[\protect\citeauthoryear{Rosati, Borgani  \& Norman}{Rosati
  et~al.}{2002}]{Rosati2002}
Rosati P.,  Borgani S.,   Norman C.,  2002, \mn@doi [Annual Review of Astronomy
  and Astrophysics] {10.1146/annurev.astro.40.120401.150547}, 40, 539

\bibitem[\protect\citeauthoryear{{Ruszkowski} \& {Oh}}{{Ruszkowski} \&
  {Oh}}{2010}]{Ruszkowski2010}
{Ruszkowski} M.,  {Oh} S.~P.,  2010, \mn@doi [\apj]
  {10.1088/0004-637X/713/2/1332}, \href
  {https://ui.adsabs.harvard.edu/abs/2010ApJ...713.1332R} {713, 1332}

\bibitem[\protect\citeauthoryear{{Ruszkowski}, {Lee}, {Br{\"u}ggen}, {Parrish}
  \& {Oh}}{{Ruszkowski} et~al.}{2011}]{Ruszkowski2011}
{Ruszkowski} M.,  {Lee} D.,  {Br{\"u}ggen} M.,  {Parrish} I.,   {Oh} S.~P.,
  2011, \mn@doi [\apj] {10.1088/0004-637X/740/2/81}, \href
  {https://ui.adsabs.harvard.edu/abs/2011ApJ...740...81R} {740, 81}

\bibitem[\protect\citeauthoryear{Salmon}{Salmon}{1998}]{Salmon1998}
Salmon R.,  1998, Lectures on geophysical fluid dynamics.
Oxford University Press

\bibitem[\protect\citeauthoryear{{Sarazin}}{{Sarazin}}{1988}]{Sarazin1988}
{Sarazin} C.~L.,  1988, {X-ray emission from clusters of galaxies}

\bibitem[\protect\citeauthoryear{{Schekochihin} \& {Cowley}}{{Schekochihin} \&
  {Cowley}}{2006}]{Schekochihin2006}
{Schekochihin} A.~A.,  {Cowley} S.~C.,  2006, \mn@doi [Physics of Plasmas]
  {10.1063/1.2179053}, \href
  {https://ui.adsabs.harvard.edu/abs/2006PhPl...13e6501S} {13, 056501}

\bibitem[\protect\citeauthoryear{{Schekochihin}, {Cowley}, {Kulsrud}, {Hammett}
   \& {Sharma}}{{Schekochihin} et~al.}{2005}]{Schekochihin2005}
{Schekochihin} A.~A.,  {Cowley} S.~C.,  {Kulsrud} R.~M.,  {Hammett} G.~W.,
  {Sharma} P.,  2005, \mn@doi [\apj] {10.1086/431202}, \href
  {https://ui.adsabs.harvard.edu/abs/2005ApJ...629..139S} {629, 139}

\bibitem[\protect\citeauthoryear{{Schekochihin}, {Iskakov}, {Cowley},
  {McWilliams}, {Proctor}  \& {Yousef}}{{Schekochihin}
  et~al.}{2007}]{Schekochihin2007}
{Schekochihin} A.~A.,  {Iskakov} A.~B.,  {Cowley} S.~C.,  {McWilliams} J.~C.,
  {Proctor} M.~R.~E.,   {Yousef} T.~A.,  2007, \mn@doi [New Journal of Physics]
  {10.1088/1367-2630/9/8/300}, \href
  {https://ui.adsabs.harvard.edu/abs/2007NJPh....9..300S} {9, 300}

\bibitem[\protect\citeauthoryear{{Schuecker}, {Finoguenov}, {Miniati},
  {B{\"o}hringer}  \& {Briel}}{{Schuecker} et~al.}{2004}]{Schuecker2004}
{Schuecker} P.,  {Finoguenov} A.,  {Miniati} F.,  {B{\"o}hringer} H.,   {Briel}
  U.~G.,  2004, \mn@doi [\aap] {10.1051/0004-6361:20041039}, \href
  {https://ui.adsabs.harvard.edu/abs/2004A&A...426..387S} {426, 387}

\bibitem[\protect\citeauthoryear{Schwarzschild}{Schwarzschild}{1906}]{Schwarzschild1906}
Schwarzschild K.,  1906, Nachrichten von der Gesellschaft der Wissenschaften zu
  Göttingen, Mathematisch-Physikalische Klasse, 1906, 41

\bibitem[\protect\citeauthoryear{{Sharma}, {Quataert}  \& {Stone}}{{Sharma}
  et~al.}{2008}]{Sharma2008}
{Sharma} P.,  {Quataert} E.,   {Stone} J.~M.,  2008, \mn@doi [\mnras]
  {10.1111/j.1365-2966.2008.13686.x}, \href
  {https://ui.adsabs.harvard.edu/abs/2008MNRAS.389.1815S} {389, 1815}

\bibitem[\protect\citeauthoryear{Spiegel}{Spiegel}{1971}]{Spiegel1971}
Spiegel E.~A.,  1971, \mn@doi [\araa] {10.1146/annurev.aa.09.090171.001543},
  \href {https://ui.adsabs.harvard.edu/abs/1971ARA&A...9..323S} {9, 323}

\bibitem[\protect\citeauthoryear{{Spiegel} \& {Weiss}}{{Spiegel} \&
  {Weiss}}{1982}]{Spiegel1982}
{Spiegel} E.~A.,  {Weiss} N.~O.,  1982, \mn@doi [Geophysical and Astrophysical
  Fluid Dynamics] {10.1080/03091928208209028}, \href
  {https://ui.adsabs.harvard.edu/abs/1982GApFD..22..219S} {22, 219}

\bibitem[\protect\citeauthoryear{Spitzer}{Spitzer}{1962}]{Spitzer1962}
Spitzer L.,  1962, Physics of Fully Ionized Gases

\bibitem[\protect\citeauthoryear{{Strang}}{{Strang}}{1968}]{Strang1968}
{Strang} G.,  1968, \mn@doi [SIAM Journal on Numerical Analysis]
  {10.1137/0705041}, \href
  {https://ui.adsabs.harvard.edu/abs/1968SJNA....5..506S} {5, 506}

\bibitem[\protect\citeauthoryear{{Vikhlinin}, {Kravtsov}, {Forman}, {Jones},
  {Markevitch}, {Murray}  \& {Van Speybroeck}}{{Vikhlinin}
  et~al.}{2006}]{Vikhlinin2006}
{Vikhlinin} A.,  {Kravtsov} A.,  {Forman} W.,  {Jones} C.,  {Markevitch} M.,
  {Murray} S.~S.,   {Van Speybroeck} L.,  2006, \mn@doi [\apj]
  {10.1086/500288}, \href
  {https://ui.adsabs.harvard.edu/abs/2006ApJ...640..691V} {640, 691}

\bibitem[\protect\citeauthoryear{{Vogt} \& {En{\ss}lin}}{{Vogt} \&
  {En{\ss}lin}}{2005}]{Vogt2005}
{Vogt} C.,  {En{\ss}lin} T.~A.,  2005, \mn@doi [\aap]
  {10.1051/0004-6361:20041839}, \href
  {https://ui.adsabs.harvard.edu/abs/2005A&A...434...67V} {434, 67}

\bibitem[\protect\citeauthoryear{{Voigt} \& {Fabian}}{{Voigt} \&
  {Fabian}}{2004}]{Voigt2004}
{Voigt} L.~M.,  {Fabian} A.~C.,  2004, \mn@doi [\mnras]
  {10.1111/j.1365-2966.2004.07285.x}, \href
  {https://ui.adsabs.harvard.edu/abs/2004MNRAS.347.1130V} {347, 1130}

\bibitem[\protect\citeauthoryear{Voit}{Voit}{2005}]{Voit2005}
Voit G.~M.,  2005, \mn@doi [Rev. Mod. Phys.] {10.1103/RevModPhys.77.207}, 77,
  207

\bibitem[\protect\citeauthoryear{{Yang} \& {Reynolds}}{{Yang} \&
  {Reynolds}}{2016}]{Yang2016}
{Yang} H. Y.~K.,  {Reynolds} C.~S.,  2016, \mn@doi [\apj]
  {10.3847/0004-637X/818/2/181}, \href
  {https://ui.adsabs.harvard.edu/abs/2016ApJ...818..181Y} {818, 181}

\bibitem[\protect\citeauthoryear{Yuan \& Narayan}{Yuan \&
  Narayan}{2014}]{Yuan2014}
Yuan F.,  Narayan R.,  2014, \mn@doi [Annual Review of Astronomy and
  Astrophysics] {10.1146/annurev-astro-082812-141003}, 52, 529

\bibitem[\protect\citeauthoryear{{Zakamska} \& {Narayan}}{{Zakamska} \&
  {Narayan}}{2003}]{Zakamska2003}
{Zakamska} N.~L.,  {Narayan} R.,  2003, \mn@doi [\apj] {10.1086/344641}, \href
  {https://ui.adsabs.harvard.edu/abs/2003ApJ...582..162Z} {582, 162}

\bibitem[\protect\citeauthoryear{{Zhuravleva} et~al.,}{{Zhuravleva}
  et~al.}{2014}]{Zhuravleva2014b}
{Zhuravleva} I.,  et~al., 2014, \mn@doi [\nat] {10.1038/nature13830}, \href
  {https://ui.adsabs.harvard.edu/abs/2014Natur.515...85Z} {515, 85}

\bibitem[\protect\citeauthoryear{{Zhuravleva}, {Churazov}, {Schekochihin},
  {Allen}, {Vikhlinin}  \& {Werner}}{{Zhuravleva}
  et~al.}{2019}]{Zhuravleva2019}
{Zhuravleva} I.,  {Churazov} E.,  {Schekochihin} A.~A.,  {Allen} S.~W.,
  {Vikhlinin} A.,   {Werner} N.,  2019, \mn@doi [Nature Astronomy]
  {10.1038/s41550-019-0794-z}, \href
  {https://ui.adsabs.harvard.edu/abs/2019NatAs...3..832Z} {3, 832}

\bibitem[\protect\citeauthoryear{{ZuHone}, {Markevitch}, {Ruszkowski}  \&
  {Lee}}{{ZuHone} et~al.}{2013}]{ZuHone2013a}
{ZuHone} J.~A.,  {Markevitch} M.,  {Ruszkowski} M.,   {Lee} D.,  2013, \mn@doi
  [\apj] {10.1088/0004-637X/762/2/69}, \href
  {https://ui.adsabs.harvard.edu/abs/2013ApJ...762...69Z} {762, 69}

\makeatother
\end{thebibliography}

%\begin{thebibliography}{99}
%\end{thebibliography}

%%%%%%%%%%%%%%%%%%%%%%%%%%%%%%%%%%%%%%%%%%%%%%%%%%

%%%%%%%%%%%%%%%%% APPENDICES %%%%%%%%%%%%%%%%%%%%%

\appendix

\section{Derivation of Boussinesq Equations}\label{app:boussinesq_derivation}

The compressible MHD equations with anisotropic heat conduction take the following form:
\begin{gather}
\left( \partial_t + \bs u \cdot \nabla \right) \rho = -\rho \Div \bs u, \label{app:compressible_MHD_1} \\
\left( \partial_t + \bs u \cdot \nabla \right) \bs u = -\frac{1}{\rho} \nabla p_{tot} - \bs g  + \frac{\bs B \cdot \nabla \bs B}{4 \pi \rho} + \nu \Lap \bs u, \label{app:compressible_MHD_2} \\
\left( \partial_t + \bs u \cdot \nabla \right) \bs B = \bs B \cdot \nabla \bs u + \eta \Lap \bs B, \label{app:compressible_MHD_3} \\
\left( \partial_t + \bs u \cdot \nabla \right) s = \frac{\gamma -1}{p} \nabla \cdot \left[ \kappa \bs b \bs b \cdot \nabla T \right], \label{app:compressible_MHD_4}
\end{gather}
where $\bs g $ is the gravitational acceleration, $s=\ln(p \rho^{-\gamma})$ is the fluid entropy, and $	\kappa$ is the Spitzer conductivity. 

To derive the Boussinesq MTI equations, we follow closely the work of \citet{Spiegel1982} and \citet{Corfield1984}. For simplicity we consider a spherically symmetric cluster, which is in hydrostatic equilibrium with $\bs u = 0$ and supports no conductive flux, such that
\begin{equation}
\partial_R \bar{p}_{tot}  = \partial_R \bar{p} = - \bar{\rho} g,
\end{equation}
where we have assumed the background magnetic field is weak, and where the overbar represents an equilibrium variable. Next, we define a typical scale-height $H$ as $H \sim \left( \partial_R \ln \bar{p} \right)^{-1}$. From dimensional analysis it follows that $\bar{p} \sim  \rho g H$ and that the squared sound speed is $\bar{c}_S^2 \sim gH$.

We perturb Eqs~\eqref{app:compressible_MHD_1}-\eqref{app:compressible_MHD_4} around the background equilibrium so that $f = \bar{f} + f'$, where the primed quantity represents the fluctuating part, and assume that the perturbations vary on a short length-scale $\lambda \sim \left( \nabla \ln \lvert f' \rvert \right)^{-1}\ll H$, and the perturbed velocities are of order $w \sim \lvert \bs u' \rvert$. The two small fluctuation scales can be then combined to yield a typical timescale of $ \sim \lambda / w$. 
We next focus on a fixed location in the cluster, given in spherical coordinates by $(R_0,\vartheta_0,\varphi_0)$, and introduce local Cartesian coordinates centred on that location:
\begin{equation}
 x=R_0(\varphi-\varphi_0), \qquad y=R_0(\vartheta-\vartheta_0), \qquad z= R-R_0, 
 \end{equation}
so that $x,y,z\sim \lambda $. We define the small parameter $\epsilon=\lambda/R_0$, and if we restrict ourselves to distances $\lambda$ from our chosen point, to leading order in $\epsilon$ the del operator becomes 
$\nabla = \bs e_x\partial_x+ \bs e_y\partial_y + \bs e_z\partial_z $, 
and the various spherical terms in the differential operators are subdominant. We next Taylor expand the background fields in small $z$. To leading order in $\epsilon$ we have
\begin{align*}
g = g_0 + \dots, \quad
\nabla \bar{s} = \bs{e}_z \left( \partial_r \bar{s} \right)_0 + \dots, \quad
\nabla \bar{T} = \bs{e}_z \left( \partial_r \bar{T} \right)_0 + \dots,
\end{align*}
where the 0 subscript denotes evaluation at $R=R_0$.
To make further progress the density and total pressure perturbations are ordered as $\rho'/\bar{\rho}_0 \sim \mathcal{M}^2 / \epsilon$ and $p'_{tot}/\bar{p}_{tot,0} \sim \mathcal{M}^2 $, respectively, where $\mathcal{M} \sim w/c_S$ is the Mach number. In what follows we take $\mathcal{M}\sim \epsilon$. Assuming that magnetic fields remain small, at first order in $\epsilon$, the linearized equation of state becomes	$\rho'/\rho_0 = -T'/T_0$.
By nondimensionaling the fluid variables as $\bs u ' = w \bs u^*$, $\bs B' = w \sqrt{4 \pi \bar{\rho}_0} \bs B^*$, $\rho' = \epsilon \bar{\rho}_0 \rho^*$ and $p' = \epsilon^2 \bar{\rho}_0 (g_0 H) p^*$ and replacing them in Eqs~\eqref{app:compressible_MHD_1}-\eqref{app:compressible_MHD_4} we can then expand at leading order in $\epsilon$ to obtain the Boussinesq equations. Apart from the presence of the anisotropic conduction term in the entropy equation, this procedure is entirely analogous to other derivations of the Boussinesq MHD equations in the literature. Therefore, we only show in detail the treatment of the entropy equation below.

Upon expanding the fluid quantities in Eq.~\eqref{app:compressible_MHD_4} into equilibria and perturbations, we rescale all time and spatial derivatives as $\nabla = \lambda^{-1} \nabla^* $ and $\partial_t = (w/\lambda) \partial_t^*$, and note that, at first order in $\epsilon$, the entropy and temperature fluctuations are $s' = -\gamma \rho' / \bar{\rho}_0 $ and $T' = - \rho' / \bar{\rho}_0 \bar{T}_0$, respectively. The entropy equation at $\bigO (\epsilon)$ then reads:
\begin{gather}
- \frac{w}{\lambda} \gamma \epsilon \left( \partial_t^* + \bs u^* \cdot \nabla^* \right) \rho^* = - \frac{w}{H} u_z^* \left(\partial_z^* \bar{s} \right)_0 + \\
-  \frac{(\gamma-1)\bar{T}_0 \bar{\kappa}_0}{\bar{p}_0} \left\{ \frac{\epsilon}{\lambda^2} \nabla^* \cdot \left[  \bs b \bs b \cdot \nabla^* \rho^* \right] - \frac{1}{\lambda H}  \left(\partial_z^* \ln \bar{T} \right)_0 \nabla^* \cdot \left( \bs b  b_z \right) \right\} \nonumber.
\end{gather}
Defining a thermal \textit{diffusivity} $\chi$ -- which is non-dimensionalized as $\chi^* = \chi /(\lambda w)$ -- where
\begin{equation}
\chi = \left( \frac{\gamma-1}{\gamma} \right) \frac{\bar{T}_0}{\bar{p}_0}\bar{\kappa}_0,
\end{equation}
we finally obtain the Boussinesq equation for the density fluctuations,
\begin{align}
\left( \partial_t^* + \bs u^* \cdot \nabla^* \right) \rho^* &= \frac{1}{\gamma} u_z^* \left(\partial_z^* \bar{s} \right)_0 + \chi^* \nabla^* \cdot \left[  \bs b \bs b \cdot \nabla^* \rho^* \right] \\ 
& - \chi^* \left(\partial_z^* \ln \bar{T} \right)_0 \nabla^* \cdot \left( \bs b  b_z \right), \nonumber
\end{align}
which reduces to Eq.~\eqref{eq:buoyancy_eq} after reintroducing dimensional units and defining the buoyancy variable $\theta = g_0 \rho'/\rho_0$.

\begin{figure}
    \centering
    	\includegraphics[width=0.9\columnwidth]{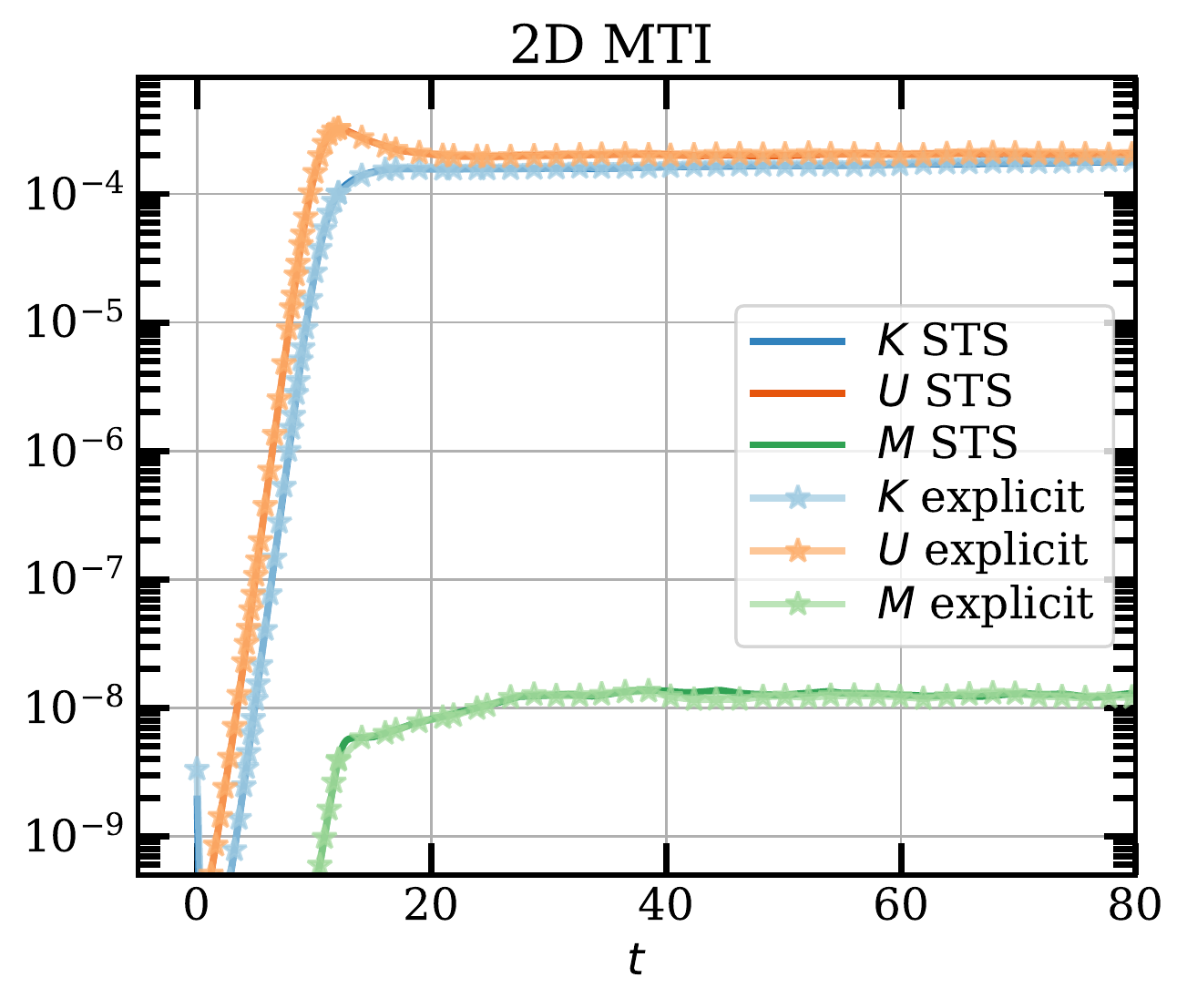}
    \caption{Comparison of the total kinetic, magnetic and potential energies for a run in 2D with explicit integration of the anisotropic diffusion term and without. %Bottom panel: comparison of the total kinetic, and magnetic energies for a run in 3D with explicit integration of the anisotropic diffusion term and without.
    %Note that for both cases, the two simulations share the same initial conditions.
    }
    \label{fig:Appendix_comparisons}
\end{figure}

\section{Benchmark of Super-Time Stepping in SNOOPY}\label{app:STS_benchmark}

In this Appendix we validate our implementation of the Super-Time Stepping algorithm in SNOOPY by comparing the results of two sets of simulations: one in 2D and one in 3D, where we focus on magnetic dynamo growth (see Paper II).
As we can see from Fig.~\ref{fig:Appendix_comparisons}, volume averaged quantities are well described by STS methods, with a typical speed-up of the calculations by a factor of $\sim 10$. Super-time stepping performs well also in 3D simulations (not shown), where -- in addition to the saturated energy levels --  the rate of growth of the kinematic dynamo is also well approximated.

\section{Spectral Fluxes}\label{app:spectral_fluxes}

In the following Appendix, we show the expression of the spectral fluxes that enter in the scale-by-scale energy balance of the Boussinesq MTI equations, see Eq.~\eqref{eq:spectral_vel}-\eqref{eq:spectral_temp}, and the text thereafter: 
\begin{align}
\hat{T}_{\bs u}^{\bs u} (\bs k, t) &= \int k_i \Im \left\{ \hat{u}_j^* (\bs k, t) \hat{u}_i (\bs q, t) \hat{u}_j (\bs k- \bs q, t) \right\} d\bs q ,\\
\hat{T}_{\theta}^{\bs u} (\bs k, t) &= \int k_i \Im \left\{   \hat{\theta}^* (\bs k, t) \hat{u}_i (\bs q, t) \hat{\theta} (\bs k- \bs q, t) \right\} d\bs q,\\
\hat{T}_{\bs B}^{\bs u} (\bs k, t) &= \int k_i \Im \left\{ \hat{B}_j^* (\bs k, t) \hat{u}_i (\bs q, t) \hat{B}_j (\bs k- \bs q, t) \right\} d\bs q ,\\
\hat{\Theta} (\bs k, t) &= \Re \left\{ \hat{\theta}^*  (\bs k, t) \hat{u}_z  (\bs k, t)  \right\}, \\
D_{\nu} (\bs k, t) &=  k^2 \lvert \hat{u} (\bs k, t) \rvert^2, \\
D_{\eta} (\bs k, t) &= k^2 \lvert \hat{B} (\bs k, t) \rvert^2, \\
D_{\chi} (\bs k, t) &= -  \int \int \left[  k_j (\bs k - \bs p - \bs q)_m  \right. \nonumber \\
&\left. \Re  \left\{ \hat{\theta}^* (\bs k, t) \hat{b}_j (\bs q, t) \hat{b}_m (\bs p, t)  \hat{\theta} (\bs k - \bs p - \bs q, t) \right\} \right] d\bs q d\bs p, \\
\hat{\mathcal{F}}(\bs k, t)  &=  \int k_i \Im \left\{ \hat{\theta}^* (\bs k, t) \hat{b}_i (\bs q, t) \hat{b}_z (\bs k- \bs q, t) \right\} d\bs q, \\
L (\bs k, t) &=  \int k_i \Im \left\{ \hat{B}_j^* (\bs k, t) \hat{B}_i (\bs q, t) \hat{u}_j (\bs k- \bs q, t) \right\} d\bs q .
\end{align}

%%%%%%%%%%%%%%%%%%%%%%%%%%%%%%%%%%%%%%%%%%%%%%%%%%

% Don't change these lines
\bsp	% typesetting comment
\label{lastpage}
\end{document}